\documentclass[structabstract]{aa}  
\usepackage{graphicx}
\usepackage{txfonts}
\usepackage{longtable}
\usepackage{appendix}

\begin{document}
   \title{The hot core towards the intermediate mass protostar NGC~7129~FIRS~2  }
   \subtitle{Chemical similarities with Orion KL\footnotemark}
   \author{A. Fuente\inst{1}
       \and 
       J. Cernicharo\inst{2}
      \and
         P. Caselli\inst{3}
       \and
        C. M$^{\textrm{c}}$Coey\inst{4}
       \and   
       D. Johnstone\inst{5,6,7}
       \and
       M. Fich\inst{4}
      \and
   	T. van Kempen\inst{8}
   \and
   Aina Palau\inst{9}
       \and
        U.A. Y{\i}ld{\i}z\inst{10}
         \and
        B. Tercero\inst{2}
       \and
       A. L\'opez\inst{2}
   }

   \institute{Observatorio Astron\'omico Nacional (OAN,IGN), Apdo 112, E-28803 Alcal\'a de Henares (Spain)
   \email{a.fuente@oan.es}
   \and
   Instituto de Ciencia de Materiales de Madrid (ICMM) C/Sor Juana In\'es de la Cruz N3, Cantoblanco, 28049 Madrid, Spain
   \and            
   Max Planck Institute for Extraterrestrial Physics, Postfach 1312, 85741 Garching, Germany
   \and
   Department of Physics and Astronomy, University of Waterloo, Waterloo, Ontario, N2L 3G1, Canada
   \and
   Department of Physics \& Astronomy, University of Victoria, Victoria, BC, V8P 1A1, Canada
   \and
   National Research Council of Canada, Herzberg Institute of Astrophysics, 5071 West Saanich Road,
   Victoria, BC, V9E 2E7, Canada
   \and
   Joint Astronomy Centre, 660 North A'ohoku Place, University Park, Hilo, HI 96720, USA
   \and
   Leiden Observatory, Leiden University, P.O. Box 9513, 2300 RA Leiden, Netherlands
   \and
   Centro de Radioastronom\'ia y Astrof\'isica, Universidad Nacional 
   Aut\'onoma de M\'exico, P.O. Box 3-72, 58090 Morelia, Michoac\'an, M\'exico
   \and
Jet Propulsion Laboratory, California Institute of Technology, 4800 Oak
Grove Drive, Pasadena CA, 91109, USA
}


 
  \abstract
  %
  {This paper is dedicated to the study of the chemistry of the intermediate-mass (IM) hot core NGC~7129~FIRS~2, probably the most
   compact warm core found in the 2 to 8 M$_{\odot}$ stellar mass range. }
 {Our aim is to determine the chemical composition of the IM hot core NGC~7129~FIRS~2, providing new insights on the chemistry of hot cores in a more general context.}
    {NGC~7129~FIRS~2  (hereafter, FIRS~2) is located at a distance of 1250 pc and high spatial resolution observations are required to resolve the hot core at its center. We present 
a molecular survey from 218200~MHz to 221800~MHz carried out with the IRAM Plateau de Bure Interferometer (PdBI). These observations were complemented with a 
long integration single-dish
spectrum taken with the IRAM 30m telescope in Pico de Veleta (Spain). We used a Local Thermodynamic Equilibrium (LTE) single 
temperature code to model the whole dataset.}
  {The interferometric spectrum is crowded with a total of $\approx$300 lines from which a few dozens remain unidentified yet. 
  The spectrum has been modeled with a 
  total of  20 species and their isomers, isotopologues and deuterated compounds. Complex molecules like methyl formate (CH$_3$OCHO), ethanol (CH$_3$CH$_2$OH),
  glycolaldehyde (CH$_2$OHCHO), acetone (CH$_3$COCH$_3$), dimethyl ether (CH$_3$OCH$_3$), ethyl cyanide (CH$_3$CH$_2$CN) and the G conformer of         
  ethylene glycol (aGg'-(CH$_2$OH)$_2$) are  among the detected species. The detection of vibrationally excited lines of
  CH$_3$CN, CH$_3$OCHO, CH$_3$OH, OCS, HC$_3$N and CH$_3$CHO proves the existence of gas and dust at high 
   temperatures. In fact, the gas kinetic temperature estimated from the vibrational lines of CH$_3$CN, $\sim$405$_{-67}^{+100}$~K, is similar to that measured in
massive hot cores. Our data allow an extensive comparison of the chemistry in FIRS~2 and the Orion hot core.}
   {We find a quite similar chemistry in FIRS~2 and Orion. Most of the studied fractional molecular abundances agree within a factor of 5. Larger differences 
  are only found for the deuterated compounds D$_2$CO and CH$_2$DOH and a few 
  molecules (CH$_3$CH$_2$CN, SO$_2$, HNCO and CH$_3$CHO). Since the physical conditions are similar in both hot cores, only different initial conditions 
 (warmer pre-collapse and collapse phase in the case of Orion) and/or different crossing time of the gas in the hot core can explain this behavior. We discuss these two
 scenarios.
}

   \keywords{astrochemistry --
                stars:formation -- ISM: molecules --
                ISM: individual (NGC~7129~FIRS~2) }

   \maketitle
\footnotetext{Based on observations carried out with the IRAM Plateau de Bure Interferometer. IRAM is supported by INSU/CNRS (France), MPG (Germany), and IGN (Spain).}

\section{Introduction}

Intermediate-mass young stellar objects (IMs) are the precursors of stars in the 2$-$8~M$_\odot$ mass range (Herbig Ae/Be stars). 
They share some characteristics with massive young stellar objects
(clustering, PDR) but they are more common than the more massive stars and therefore they can be found closer to the Sun 
(d$<$1~kpc), which allows determination of the physical and chemical structure of their envelopes at similar spatial scales as
can be done for low mass protostars. From a chemical point of view, IMs are interesting because they constitute the link between 
the low-mass and high-mass ranges, covering an
intermediate range of luminosities, densities and temperatures. 

Hot molecular cores are compact ($<$0.05~pc) objects with high temperatures ($>$100 K) and densities 
(n(H$_2$)$>$ 10$^6$ cm$^{-3}$), which are characterized by a very rich chemistry of complex organic molecules 
(COMs). In fact, the prototype of massive hot core, that associated with Orion KL, hosts 
one of the richest molecular chemistry observed in the interstellar
medium (Blake et al. 1987; Beuther et al. 2005, 2006; Tercero et al. 2010. 2011, 2013; Motiyenko et al. 2012, Cernicharo et al. 2013).
COMs have also been detected, coming from the inner regions of a few low-mass protostars, 
the so called hot corinos (IRAS~16293-2422: Cazaux et al. 2003, 
Bottinelli et al. 2004a, Pineda et al. 2012; NGC 1333 IRAS 2A: J{\o}rgensen et al.2005, 2007; 
NGC1333-IRAS 4AB: Bottinelli et al. 2004b, Bottinelli et al. 2007). 
These regions have smaller sizes ($\sim$150~AU) and lower temperatures ($\sim$100~K).
Although high sensitivity interferometric data on hot cores and corinos are still scarce and the comparison 
is difficult, there is some general consensus 
that hot corinos are richer in O-bearing molecules like CH$_3$OCHO, CH$_3$CHO, 
CH$_3$OCH$_3$ or HCOOH, and poorer in N-bearing compounds. Furthermore, large abundances of deuterated molecules and
large deuterium fractions are only associated with hot corinos (Vastel et al. 2003; Parise et al. 2004; Demyk et al. 2010). 

A handful of well studied hot cores exist in the IM range: 
NGC~7129~FIRS~2 (Fuente et al. 2005a, hereafter FU05), IC~1396~N (Neri et al. 2007, Fuente et al. 2009),  
IRAS~22198+6336 (S\'anchez-Monge et al. 2010, Palau et al. 2011), AFGL 5142 (Palau et al., 2011). The sizes of these 
IM hot cores range from $\sim$130 AU (IRAS~22198) to $\sim$900 AU (NGC~7129) and their chemistries 
present some differences. FU05 proposed that IM hot cores, like hot corinos, are richer 
in H$_2$CO and HCOOH relative to CH$_3$OH than massive hot cores, 
and they did not find any variation in the CH$_3$CN/CH$_3$OH abundance ratio across the stellar range.
Palau et al. (2011) studied the chemistry of IRAS~22198+6336  and AFGL~5142                                                                                      
and concluded that these IM hot cores are
richer in oxygenated molecules with two or more CH$_{2/3}$-groups than hot corinos but poorer in N-bearing molecules
than massive hot cores. As yet, the number of studied objects is very low and the derived column
densities have large uncertainties. 

In this paper, we present a 4~GHz interferometric frequency survey towards the hot core NGC~7129~FIRS~2 (hereafter, FIRS~2)
that, with a luminosity of $\sim$500 L$_\odot$, lies near the middle of the IM luminosity range.
FIRS~2 is a prototypical young IM protostar, it is associated with energetic outflows 
(Fuente et al. 2001) and presents clear signs of CO depletion and enhanced deuterium fractionation when observed
with single-dish telescopes (scales of $\sim$10000~AU) (Fuente et al. 2005b; 
Alonso-Albi et al. 2010). It is located at a distance of 1250$\pm$50~pc from the Sun (Shevchenko \& Yakubov 1989) and FU05 detected a
compact hot core at its center. FIRS~2 was observed with Herschel within the Water In Star-forming regions (WISH) key program 
(van Dishoeck et al. 2011) and it is one of the best known IM protostars with a quite complete set of data at both
millimeter and far-IR wavelengths  (Johnstone et al. 2010, Fich et al. 2010, Fuente et al. 2012). This source and IRAS~22198+6336 
are the only hot cores without clear signs of clustering which suggests that their luminosity could come
from a single binary system (Palau et al. 2013). 

\section{Observations}
The interferometric observations were carried out with the Plateau de Bure Interferometer (PdBI) 
in its CD configuration during August and November, 2012. This configuration provided an angular 
resolution of 1.43$\arcsec$$\times$1.26$\arcsec$ PA 144$^\circ$  ($\sim$1870~AU$\times$1647~AU at the distance of 
FIRS~2) at the central frequency. The 1~mm receivers were tuned at 219.360~GHz which allowed the simultaneous imaging of
the C$^{18}$O 2$\rightarrow$1 and $^{13}$CO 2$\rightarrow$1 lines, both within the $\sim$4~GHz receiver's
band. The narrow  40~MHz correlator units were used to sample the C$^{18}$O and $^{13}$CO lines at the high spectral 
resolution of 40~kHz. The wide band correlator WideX  sampled the whole 4~GHz bandwidth 
with a spectral resolution of $\sim$2~MHz. MWC 349 was used as flux calibrator (1.92 Jy) and the rms in
each Widex channel, 4$-$6 mJy/beam ($\sim$ 0.058-0.087 K). 

The interferometric observations were complemented with single-dish observations using the 30m telescope and the 
same frequency setup. These observations were done in December 2012 and the achieved rms(T$_{\rm a}^{*}$) was 
$\sim$0.012~K (0.09~Jy) in a channel of $\sim$1.953~MHz. Therefore, the 30m observations are
$\sim$15 times less sensitive than the PdB images. The telescope was pointed towards the phase center of the 
interferometric observations: $21^{\mathrm{h}}43^{\mathrm{m}}01\fs7$, $+66\degr03\arcmin23\farcs6$
(J2000). Forward and main beam efficiencies are 0.92 and 0.63, respectively.
The selected intensity scale is main beam temperature. 

This paper is dedicated to the analysis of the chemical complexity revealed 
by the WideX data towards the mm continuum source.

\section{Results}
\subsection{Continuum maps}
The high density of lines detected made it impossible to accurately measure the continuum flux
from the low spectral resolution WideX data. Instead, we used the empty emission channels around the C$^{18}$O and $^{13}$CO lines measured at 
higher spectral resolution using the 40~MHz correlator units to determine the continuum flux. As expected, the continuum flux was 
slightly different at 219.560~GHz and 220.398~GHz. 
We derived two continuum maps at 219.560~GHz and 220.299~GHz. Fitting the visibilities we obtained
the position of the compact source:  $21^{\mathrm{h}}43^{\mathrm{m}}01\fs67$, $+66\degr03\arcmin23\farcs7$, i.e,
offset by $-$0.17$\arcsec$ in right ascension from our phase center. The measured fluxes are 0.37(0.01)~Jy at 219.560~GHz and 0.38(0.01) 
at 220.398~GHz,
respectively. In FU05, we determined a mm emission spectral index of 2.56 based on the continuum images at 86 GHz and
230~GHz. 
The new fluxes at 219.6~GHz and 220.4~GHz are consistent with these results taking into account the
uncertainty of 10\% in the absolute flux calibration.

\subsection{Molecular Lines}
In Figs.~A.1-4 we show the WideX spectra towards the continuum peak.
The continuum image at 219.560~GHz was subtracted from the spectral 
maps before the cleaning process. This subtraction is not perfect since the continuum presents a smooth slope across the
observed bandwidth, but  it is good enough for our current detection goals.

The spectrum towards the compact source is crowded with lines, typical of those found in
massive hot cores. In order to estimate the flux that the interferometer is missing, we compared the 
PdBI spectrum with that obtained with the 30m telescope. 
In Table~1 we show the list of all the lines detected with both telescopes
and the fraction of flux recovered using the PdBI. Note that only the H$_2$CO, CH$_3$OH, SO, C$^{18}$O and
$^{13}$CO lines are missing a significant fraction of their fluxes when observed with the PdBI. The emission of OCS, HNCO,
CH$_3$CN and HC$_3$N comes mainly from the hot core although 20\% of the flux of the lowest energy HNCO line is missing. 
This is also consistent with the spatial distribution as observed
with the PdBI. In Fig. 1, we show the integrated intensity images of some of the more intense lines. Only $^{13}$CO,
C$^{18}$O and SO present large scale structure. In all the other molecules the emission is point-like at the
angular resolution of our observations.

\begin{table*}
\caption{30m/PdBI lines$^1$}
\begin{tabular}{lccccc} \hline \hline 
Line            &                                    &    Freq    &  30m Flux$^2$               &  PdBI Flux$^3$                   & PdBI/30m  \\ 
                &                                    &    (MHz)   &  (Jy$\times$km s$^{-1}$) &  Flux (Jy/beam$\times$km s$^{-1}$) &            \\ 
\hline
O$^ {13}$CS     & 18$\rightarrow$17                  &    218199       &  1.07 (0.40)       &  0.63 (0.04)                  & $\sim$60 \%       \\
H$_2$CO         & 3$_{0,3}$$\rightarrow$2$_{0,2}$    &    218222       &  41.11 (1.13)      &  4.51 (0.19)                  & $\sim$10\%        \\
HC$_3$N         & 24$\rightarrow$23                  &    218325       &  2.40 (0.38)       &  1.84 (0.34)                  & $\sim$100\%       \\
CH$_3$OH        & 4$_{2,0}$$\rightarrow$3$_{1,0}$    &    218440       &  12.54 (0.49)      &  2.80 (0.03)                  & $\sim$20\%  \\
H$_2$CO         & 3$_{2,2}$$\rightarrow$2$_{2,1}$    &    218476       &  10.80 (0.42)      &  3.26 (0.56)                  & $\sim$30\%  \\
H$_2$CO         & 3$_{2,1}$$\rightarrow$2$_{2,0}$    &    218760       &  10.54 (0.42)      &  2.98 (0.11)                  & $\sim$30\%  \\
OCS             & 18$\rightarrow$17                  &    218903       &  3.11 (0.44)       &  3.29 (0.05)                  & $\sim$100\%  \\
C$^{18}$O       & 2$\rightarrow$1                    &    219560       &  32.85 (0.20)      &  2.47 (0.10)                  & $<$10\%    \\
HNCO            & 10$_{3,8}$$\rightarrow$$_{3,7}$$^*$     &    219657    &   $<$ 1.13       &  1.04 (0.11)                  & $\sim$100\%   \\
                & 10$_{3,7}$$\rightarrow$$_{3,6}$$^*$     &              &                  &                               &  \\
HNCO            & 10$_{2,9}$$\rightarrow$9$_{2,8}$$^*$      &    219734    &   1.91 (0.53)  &  2.13 (0.12)                  & $\sim$100\%  \\
                & 10$_{2,8}$$\rightarrow$9$_{2,7}$$^*$      &    219737       &        \\
HNCO            & 10$_{0,10}$$\rightarrow$9$_{0,9}$     &    219798       & 3.14 (0.42)     & 2.12 (0.12)                   & $\sim$70\%    \\
H$_2$$^{13}$CO  & 3$_{1,2}$$\rightarrow$2$_{1,1}$       &    219909       & 1.96 (0.37)     & 1.18 (0.11)                   & $\sim$60\% \\
SO              & 5$_6$$\rightarrow$4$_5$               &    219949       & 43.20 (2.75)    & 8.65 (0.21)                   & $\sim$20\%  \\
E-CH$_3$OH      & 8$_0$$\rightarrow$7$_1$               &    220078       & 3.59 (0.95)     & 3.95 (0.14)                   & $\sim$100\% \\
$^{13}$CO       & 2$\rightarrow$1                       &    220399       & 144.49 (1.67)   & 5.54 (0.13)                   & $<$1\%  \\
CH$_3$CN        & 12$_{5}$$\rightarrow$11$_{5}$       &    220641         & 2.18 (0.64)     & 2.50 (0.13)                   & $\sim$100\% \\
CH$_3$CN        & 12$_{4}$$\rightarrow$11$_{4}$       &    220679         & 2.00 (0.25)     & 2.20 (0.08)                    & $\sim$100\% \\
CH$_3$CN        & 12$_{3}$$\rightarrow$11$_{3}$       &    220709         & 1.68 (0.65)     & 2.16 (0.70)                   & $\sim$100\% \\
CH$_3$CN        & 12$_{2}$$\rightarrow$11$_{2}$       &    220730         & 2.07 (0.58)     & 2.05 (0.46)                   & $\sim$100\% \\
CH$_3$CN        & 12$_{0}$$\rightarrow$11$_{0}$$^*$       &    220747         & 5.14 (0.71)     & 4.05 (0.52)                   & $\sim$100\% \\
                & 12$_{1}$$\rightarrow$11$_{1}$$^*$       &    220743         &                 &                               & $\sim$100\% \\
\hline \hline
\end{tabular}

\noindent
$^1$Comparison between the flux measured with the 30m and PdB telescopes for all the lines detected with the 30m single-dish telescope.\\
$^2$Velocity-integrated flux density in the 30m spectrum. \\
$^3$Velocity-integrated flux density towards the emission peak. \\
$^*$Blended lines.\\
\end{table*}

\begin{figure*}
\includegraphics{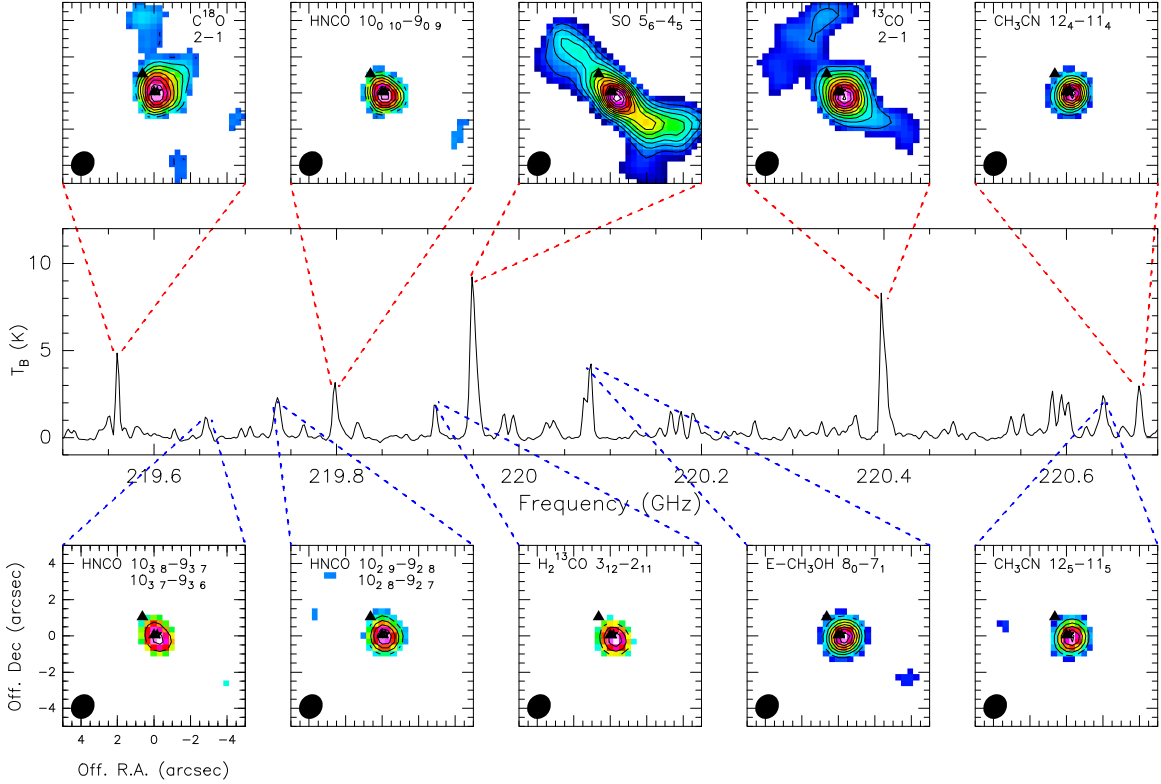}
\vspace{12.0cm}
      \caption{{\it Middle:} PdBI spectrum towards the emission peak. {\it Top and bottom:} Integrated intensity maps of the most intense 
lines in the portion of spectrum between 219.5~GHz and 220.7~GHz In all the panels the contours start and increases at steps of 3$\times$$\sigma$. 
Sigma and peak values in Jy/beam$\times$km~s$^{-1}$ are:
0.10, 2.47 (C$^{18}$O 2-1); 0.12, 2.12 (HNCO 10$_0$-9$_0$); 0.21, 8.65 (SO 5$_6$-4$_5$); 0.22, 7.28 ($^{13}$CO 2-1); 0.08, 2.20 (CH$_3$CN 12$_4$-11$_4$); 
0.11, 1.04 (HNCO 10$_3$-9$_3$); 0.12, 2.13 (HNCO 10$_2$-9$_2$); 0.11, 1.18 (H$_2$$^{13}$CO 3$_{1,2}$-2$_{1,1}$); 0.14, 3.95 (E-CH$_3$OH 8$_{0}$-7$_{1}$); 
0.13, 2.5 (CH$_3$CN 12$_5$-11$_5$). Filled triangles indicate the water masers as observed with the NRAO Very Large Array 
(Aina Palau, private communication). The masers trace the direction of the youngest outflow (Fuente et al. 2001) that is also clearly seen in the SO and
$^{13}$CO images. The (0,0) position is the phase center of the interferometric 
observations: $21^{\mathrm{h}}43^{\mathrm{m}}01\fs7$, $+66\degr03\arcmin23\farcs6$(J2000).}
         \label{Fig 1}
\end{figure*}

\section{Line identification and rotational diagrams}
For line identification we used Jos\'e Cernicharo's personal catalogue which is included in
the radiative transfer and molecular excitation code MADEX (Cernicharo 2012), the JPL (Pickett 1991,1998)
and the CDMS (M\"uller et al. 2001, 2005) catalogues. Our procedure can be summarized as follows: We started with the
identification of the lines of the most common species. The rotation temperatures, molecular column densities 
and source sizes were estimated by fitting the line intensities of the whole family of molecules (all isotopologues)
within the program MADEX. In these calculations, we assume Local Thermodynamic Equilibrium, v$_{\rm LSR}$=$-$10~km~s$^{-1}$, a
linewidth of 8~km s$^{-1}$ and a face-on disk as the source morphology. In the case of a good fit, we create a synthesized 
spectrum using the derived parameters and subtract it from the observations. 
Then, we go on with line identification in the residual spectrum. Sometimes, especially for the weak lines, the fit was not good enough. 
In these cases, we did not subtract the synthesized spectrum 
in order to avoid spurious features and just went on identifying the features in 
the original spectrum.
 
The number of parameters that can be determined for each species depends on the number of lines detected. For example, 
the source size can only be determined when the lines of the main isotopologue are optically thick. In the optically thick case, 
$T_{\rm B} \approx \eta_{\rm ff} \times J_\nu(T_{\rm rot})$, where $\eta_{\rm ff}$ is 
the beam filling factor, and  $J_\nu(T_{\rm rot})=h�\nu / kT_{\rm rot}  \times (1 - \exp(h \nu /kT_{\rm rot}))^{-1}$ with T$_{\rm rot}$, 
the rotation temperature. Providing that we know $T_{\rm rot}$, we can estimate 
the effective size of the emitting region from the line intensities. The emission sizes cannot be
constrained if the main isotopologue lines are optically thin.

For the less abundant molecules, the number of detected lines was too low or the line intensities were 
too uncertain to estimate the rotation temperature. In these cases, a rotation temperature of $\sim$200~K was assumed
and only the molecular column densities are calculated using MADEX. In Sect. 4.1 to 4.19, we give a 
more detailed description of the fitting for each species and the results of these calculations 
are shown in Tables~2 and 3.

Frequencies and line identifications are shown in Table A.1.
We compare our synthesized spectrum including all species with observations in Figs.~A.1-4.  

\begin{table}
\caption{Results of the rotational diagram analysis$^1$}
\begin{tabular}{llcc} \hline \hline 
Species                 &  T$_{\rm rot}$ (K)   &   N (cm$^{-2}$) & R$^2$ \\ \hline
HNCO                    & 260$^{+4}_{-5}$      &  1.6$^{+0.1}_{-0.1}$$\times$10$^{15}$   &  0.99 \\
CH$_3$CN $\nu$=0$^3$    & 905$^{+250}_{-161}$  &  1.6$^{+0.2}_{-0.2}$$\times$10$^{15}$   &  0.90 \\
CH$_3$CN $\nu_8$=1      & 405$^{+100}_{-67}$   &  9.4$^{-3.1}_{+4.6}$$\times$10$^{14}$   &  0.81 \\
CH$_3$OCHO  $\nu_t$=0,1  & 265$^{+139}_{-69}$    &  2.8$^{+1.9}_{-0.8}$$\times$10$^{16}$   &  0.60 \\
CH$_2$DOH               & 157$^{+15}_{-13}$    &  1.4$^{+0.3}_{-0.3}$$\times$10$^{16}$   & 0.99 \\
aGg'-(CH$_2$OH)$_2$     & 145$^{+37}_{-25}$    &  2.0$^{+0.3}_{-0.3}$$\times$10$^{15}$  & 0.92 \\  
\hline \hline
\end{tabular}

\noindent
$^1$Beam filling solution.
$^2$Correlation coefficient of the least square fitting.
$^3$This fit was not finally used in the model because the detection of the $^{13}$C isotopologue
showed that the lines are optically thick. \\
\end{table}

\begin{figure*}
\includegraphics{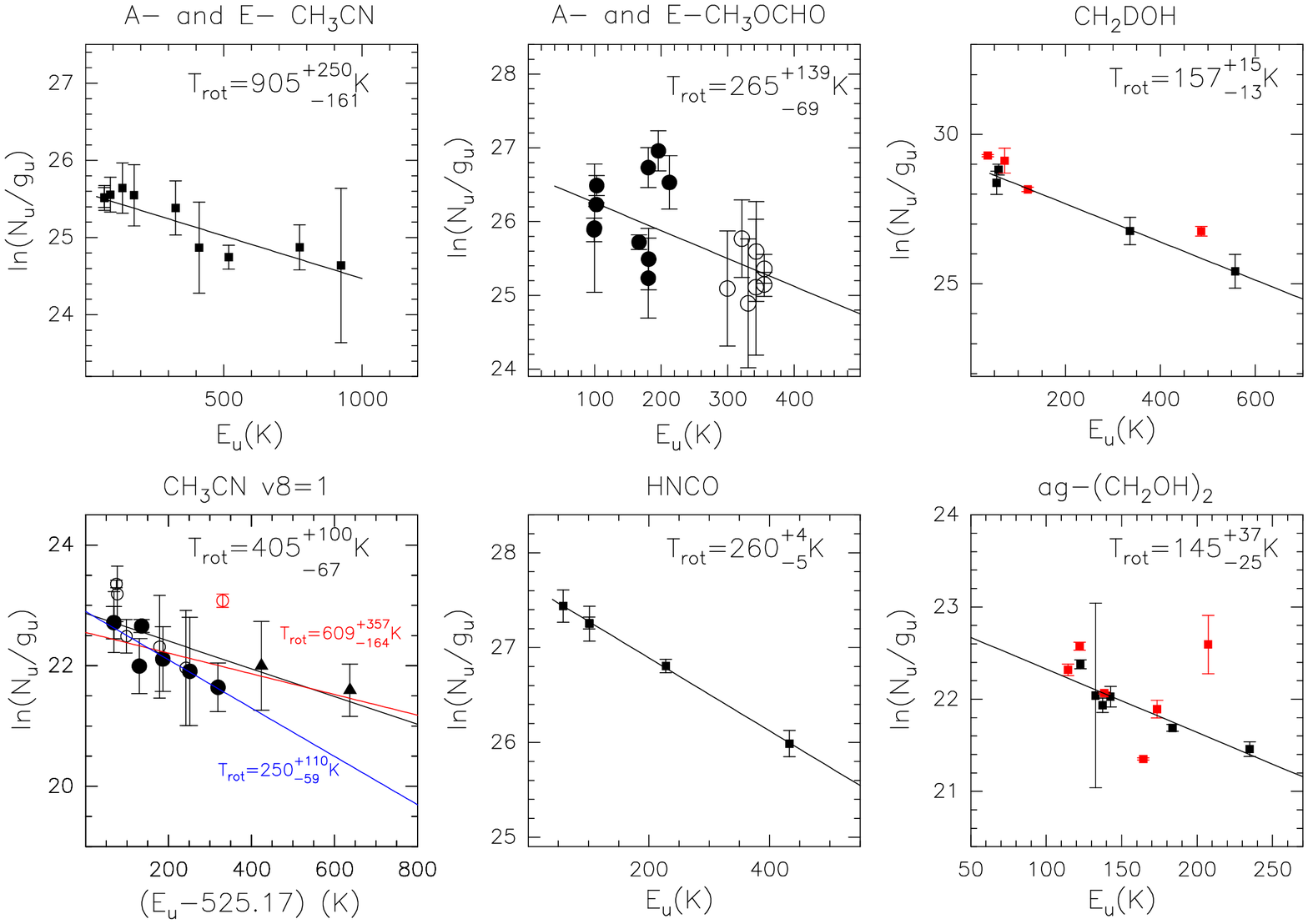}
\vspace{10.0cm}
      \caption{Results of the rotational diagrams for CH$_3$CN $\nu$=0,  CH$_3$CN $\nu_8$=1, CH$_3$OCHO , HNCO, CH$_2$DOH and aGg'-(CH$_2$OH)$_2$. The high rotation
      temperature fitted to the CH$_3$CN $\nu$=0 rotational diagram is the consequence of the lines being optically thick and is not considered in our modeling (see Sect. 4.1). 
      In the rotational diagram of the CH$_3$CN $\nu_8$=1 vibrational state we use different symbols for: blended lines (empty circles), isolated or partially blended lines (filled circles) and
      the points with E$_{\rm rot}$$>$400~K (triangles). The three straight lines correspond to the least square fittings with: all the points (black), 
      only unblended lines (red), unblended lines with E$_{\rm rot}$$<$400 K (blue). For
      CH$_3$OCHO, the points corresponding to the $\nu$=0 lines are indicated with filled squares and those of $\nu_t$=1 with empty circles. 
      The parameters of the transitions used in the rotation diagrams are listed in Tables A.3-8. Red points have not been used in the least square fitting because 
      they correspond to heavily blended lines.}
         \label{Fig 5}
 \end{figure*}

\subsection{Methyl cyanide: CH$_3$CN, CH$_3$NC}
Methyl cyanide (CH$_3$CN) is one of the best thermometers in hot cores. The symmetric rotor presents a $K$-ladder structure 
with transitions that are easily thermalized at the densities prevailing in hot cores. Besides, it is well known that this molecule is 
especially abundant in this kind of regions.
We have detected the whole 12$_K$$\rightarrow$11$_K$ ladder, from $K$=0 to $K$=11 in the $\nu$=0 ground vibrational state and in the $\nu_8$=1 
vibrational level at E$_{\rm vib}$=525.17~K (see~Table A.4). The high number of detected lines would allow us to determine the rotation and 
vibrational temperature of this molecule providing that the emission is optically thin (see Fig.~2).
As discussed below, this is not the case of the $\nu$=0 lines. Instead, we used the  $\nu_8$=1 lines to estimate
the rotation temperature. Unfortunately, most of the lines are blended or partially blended. This is the cause
of the large uncertainties in the integrated intensities shown in Table A.4. All the lines have T$_b$$>$3$\sigma$ but the
uncertainty in the linewidth is large for partially blended lines. 
As a first step we plotted all the detected lines in a rotational diagram and visually check if any of them were clearly above or below 
the straight line defined by the other points. We removed the line at 221265~MHz that was clealy above the fit and 
derived T$_{\rm rot}$=405$_{-67}^{+100}$~K using the rest of the points (the parameters of this fit are shown in Table~2). 
The fit was good (R=0.81) suggesting that the contribution of other species to the measured line integrated intensities were
within the errors. To corroborate this assumption, as a second step, we removed the lines that were severely blended (difference in frequency less than
4~MHz) and kept only those that were isolated or partially blended. At the end we had only 8 transitions and derived T$_{\rm rot}$=609$_{-167}^{+357}$~K.
The higher rotation temperature was due to the suppression of the lines at 221199, 221394, 221403 and 221422~MHz. 
The obtained value was still in agreement with our previous fit within the errors. We were worried because the velocity integrated line intensities of the transitions 
with E$_{rot}$$>$400~K (221059, 221210~MHz) present large uncertainties. In order to check the robustness of our temperture estimate, 
we make a new least square fitting without these lines. This
fit provided a lower limit to the rotation temperatue of T$_{\rm rot}$=250$_{-59}^{+110}$~K. Again the value obtained for the rotation temperature
is in agreement with that shown in Table 2 within the errors. Therefore, we decided to keep the estimate shown in Table~2 as a good value for
the average rotation temperature although we are aware that, more likely, we have strong temperature gradients accross the hot core with temperatures varying 
between $\sim$250~K and $\sim$600~K.

The results from the CH$_3$CN $\nu$=0 rotational diagram are shown in Table~2. We obtained a higher rotational temperature in $\nu$=0 than
in $\nu_8$=1. One could think that the high temperature in the ground vibrational state was an artifact due to the missing flux in the low $K$ components
of the CH$_3$CN 12$_K$$\rightarrow$11$_K$ ladder. However, comparison of the interferometric PdBI and single-dish 30m observations showed that the 
spatial filtering is not important and therefore this possibility was discarded. An alternative explanation is that the ground state vibrational 
lines are optically thick. The opacity can be derived by comparing the CH$_3$CN and CH$_3$$^{13}$CN lines (the lines of the other $^{13}$C isotopologue
$^{13}$CH$_3$CN are outside the observed frequency range). Unfortunately the CH$_3$$^{13}$CN 12$_0$$\rightarrow$11$_0$ and  
12$_1$$\rightarrow$11$_1$ lines are blended with those of the main isotopologue. The lowest $K$ transition without any obvious 
contamination is $K$=4 (220571~MHz) where we have a feature with T$_b$$\sim$0.39 K (with an uncertainty of 25\% because of
partial blending with other lines). Comparing with the main isotopologue, we have 
I(CH$_3$CN 12$_4$$\rightarrow$11$_4$)/I(CH$_3$$^{13}$CN 12$_4$$\rightarrow$11$_4$)$\sim$7, which implies opacities of $\sim$10 
in the main isotopologue line. Since the line opacities are very high, the rotational diagram analysis is not valid and
the resulting rotation temperature in the ground state is severely over-estimated. Instead, we 
derived the CH$_3$CN column density assuming T$_{\rm rot}$=405~K (the same as in the $\nu_8$=1 vibrational level)
and an opacity of $\sim$10 for the CH$_3$CN 12$_4$$\rightarrow$11$_4$ line and obtained  
N(E-CH$_3$CN)=N(A-CH$_3$CN)=2.6$\times$10$^{18}$~cm$^{-2}$, i.e., a total (A+E) CH$_3$CN column density 
of 5.2$\times$10$^{18}$~cm$^{-2}$. With these parameters, we fit the high energy rotational lines with a source 
size of $\approx$0.07$\arcsec$ (125~AU). The uncertainty in the source size comes from the assumed rotation temperature. By varying the rotation
temperature between 250 and 600~K we would obtain sizes of 100$-$160~AU. Note that these sizes are quite small, similar to typical scales 
of circumstellar disks.

This compact source model successfully predicted the intensities of the high excitation lines, but the intensities of the low energy transitions 
were severely underestimated. 
The only way to fit the whole spectrum was to consider a two-component model: (i) the compact one described above,
and (ii) a more extended second component that is filling the beam. We fit the $K$$<$4 lines with N(E-CH$_3$CN)=N(A-CH$_3$CN)=2.5$\times$10$^{14}$~cm$^{-2}$
and T$_{\rm rot}$=350~K for this extended component. We tried with a temperature of 250~K obtained from the fitting of the E$_{rot}$$<$400~K $\nu_8$=1 lines,
but this lower temperature would overestimate the $K$=0 and $K$=1 lines. Besides the assumed value is within the uncertaintly of this fit (=250$_{-59}^{+110}$~K).

Our estimate of the total methyl cyanide column density relies on the $^{13}$C isotopologue line. 
Of course there is also the possibility that the CH$_3$$^{13}$CN 12$_4$$\rightarrow$11$_4$ line is contaminated by another line and
its real intensity is lower. We did not find any good candidate at this frequency.
One line of EA-(CH$_3$)$_2$CO is very close in frequency but the emission of this line is very weak (see also Sect. 4.8). 
Furthermore, the small sizes derived for other species (see following sections) prompted us to consider the two-component model as the most 
likely one. We would also like to point out that we refer to a beam filling component at the
spatial resolution of our observations ($\sim$1900~AU) as $``$extended component". This component would not encompass the entire envelope.   

We also looked for the deuterated compounds in our spectrum. The most abundant species CH$_2$DCN and CHD$_2$CN have no intense 
lines in the observed frequency range. Only CD$_3$CN has three intense transitions at 219.965, 219.983 and 219.993~GHz. We have detected the
three lines but need a very low rotation temperature, T$_{\rm rot}$=40~K, to fit them. Because of
this surprisingly low rotation temperature, we consider
that a misidentification is possible and keep the derived column density as an upper value to the real one.
%
%
Several weak lines of CH$_3$NC were identified, a molecule so far only detected towards Sgr B2 (Cernicharo et al. 1988). 
Unfortunately, most of them are blended or partially blended. 
Assuming that CH$_3$NC is coming
from the CH$_3$CN compact core, we calculated a CH$_3$NC column density 
of 1.5$\times$10$^{16}$~cm$^{-2}$, i.e., about 350 times lower than that of CH$_3$CN. Since we have only one unblended line, we keep
this detection as tentative.
\subsection{Isocyanic acid: HNCO, HN$^{13}$CO}
The HNCO lines are well fitted with T$_{\rm rot}$=260$_{-5}^{+4}$~K and N(HNCO)=1.6$\times$10$^{15}$~cm$^{-2}$ (see Fig.~2 and Table~2). 
In our frequency range, most of the lines of HN$^{13}$CO are blended with those of the main isotopologue, 
therefore we cannot estimate the line opacities.
Comparing with the 30m spectrum, we calculated that the interferometer is missing about 20\% of the flux of the
low energy lines, showing that there is an extended component. For this reason, we adopt 
the beam filling solution.
We looked for the isomer HCNO and the deuterated compound DNCO but unfortunately they  do not have 
intense transitions in the sampled frequency range.
%
%
\subsection{Methyl formate: A-CH$_3$OCHO , E-CH$_3$OCHO }
We detected more than sixty lines of A-CH$_3$OCHO  and E-CH$_3$OCHO  in our spectrum, both in the $v_t$=0,1 vibrational states, 
the latter at E$_{\rm vib}$=188.40~K. We tried to fit 
a rotational diagram to each vibrational state separately in order to measure the rotation and vibrational temperatures independently
but this was not possible. The dynamical range of the E$_u$'s within each vibrational level is very small and the uncertainties in the integrated
intensities are too large to obtain a good fit. However we could get a reasonable fit
by considering the two vibrational levels together (see Table~2). We measured T$_{vib,rot}$$\sim$265$_{-69}^{+139}$~K, 
slightly lower than that of CH$_3$CN but still in good agreement taking into account the associated uncertainties. 
We fixed the temperature and varied the source size to fit the lines. The best fit was obtained with a size of $\sim$0.1$\arcsec$.  
Even with this small size, the opacities of the CH$_3$OCHO $\nu$=0 lines are always $\leq$1.3 suggesting that our estimate of the rotation temperature  
is reliable. Unfortunately, the lines of the $^{13}$C isotopologues are blended with other species and it was not possible to have
an estimate of their column densities. We used a weak feature at 128.952~GHz to derive an upper limit to the total column density 
of $^{13}$CH$_3$OCHO  of $<$3.0$\times$10$^{17}$ cm$^{-2}$ in the $\sim$0.1~$\arcsec$ core, which
would imply an upper limit to the total (A- +E-) column density of $<$2.0$\times$10$^{19}$~cm$^{-2}$ assuming $^{12}$C/$^{13}$C=65. 
This allow us to say that the methyl formate column density is accurate within a factor of 4.
About deuteration, we have three detected lines at 219242~MHz, 218730~MHz and 219132~MHz that would 
correspond to CH$_3$OCDO and  CH$_2$DOCHO. The first two lines are blended with lines of EE-(CH$_3$)$_2$CO 
which makes the derived deuterium fraction more uncertain. Taking into account the contribution of
acetone, we fitted the intensities of the three lines assuming a 
deuterium fraction of $\sim$0.06. 
%
%
%
\subsection{Ethyl cyanide: CH$_3$CH$_2$CN}
The detection of eight lines of CH$_3$CH$_2$CN strongly confirms the identification of this
species. Although three of them are heavily blended, the other five lines are enough to
identify this species.
The lines belong to the ground state and to the lowest energy vibrational states at E$_{\rm vib}$=297.12~K (Daly et al. 2013).
We tried to fit the rotational diagram but unfortunately the points did not follow a clear pattern and we could not
determine the rotation temperature. Hence, we fixed 
the rotation temperature to a value of 200~K (a reasonable value from the rotational temperatures measured with the other species, see Table~2) 
and fitted the column density. A beam averaged column density value of 5.2$\times$10$^{14}$~cm$^{-2}$ was derived from the intensity of the 
most intense lines. The fit improved assuming that the emission
comes from a compact source with a diameter of $\sim$0.1$\arcsec$, because some lines become optically thick.
This is the solution we finally adopted (see Table~3). We searched for the lines of the $^{13}$C isotopologues and the deuterated species 
but the lines were too weak to obtain a significant upper limit to their column densities.
%
%
\begin{table}
\caption{NGC~7129 FIRS~2 hot core model}
\begin{tabular}{lllll} \hline \hline 
{\tiny Species}   & {\tiny T$_{\rm rot}$(K)}   &   {\tiny N$_X$ (cm$^{-2}$)}$^1$    &   \multicolumn{1}{c}{\tiny D($\arcsec$)$^2$} & \\ \hline
{\tiny CH$_3$CN $\nu$=0}          &  {\tiny 405}             &  {\tiny 5.2$\times$10$^{18}$}   & {\tiny 0.07} & E$^4$  \\
                                  &  {\tiny 350}             &  {\tiny 5.0$\times$10$^{14}$}   & BF$^3$ & G$^5$            \\
{\tiny CH$_3$CN $\nu_8$=1}        &  {\tiny 405}             &  {\tiny 7.0$\times$10$^{16}$}   & {\tiny 0.07} & G   \\
{\tiny CH$_3$$^{13}$CN $\nu$=0}   &  {\tiny 405}             &  {\tiny 8.0$\times$10$^{16}$}   & {\tiny 0.07} & E  \\
                                  &  {\tiny 350}             &  {\tiny 7.7$\times$10$^{12}$}   &  BF & G \\
{\tiny CH$_3$NC}$^*$              &  {\tiny 405}             & {\tiny  3.0$\times$10$^{16}$}   & {\tiny 0.07}& G   \\
{\tiny CD$_3$CN}$^*$              &  {\tiny 40}              & {\tiny  $<$5.0$\times$10$^{13}$}   & BF& G   \\
{\tiny CH$_3$OH $\nu_t$=0,1}        &  {\tiny 238}               &  {\tiny 3.4$\times$10$^{20}$}  &  {\tiny 0.12} & G  \\
                                    &  {\tiny 157}$^6$               &  {\tiny 2.0$\times$10$^{16}$}$^6$  &  BF   & G         \\
{\tiny $^{13}$CH$_3$OH $\nu$=0}    & {\tiny 238}             &  {\tiny 4.2$\times$10$^{18}$}  &  {\tiny 0.12} & G  \\
{\tiny $^{13}$CH$_3$OH $\nu_t$=1}    & {\tiny 238}             &  {\tiny 1.1$\times$10$^{18}$}  &  {\tiny 0.12} & G  \\
{\tiny CH$_2$DOH}                    &  {\tiny 157}           &    {\tiny 1.4$\times$10$^{16}$} &  BF & G   \\ 

{\tiny CH$_3$OCHO  $\nu$=0}       & {\tiny 265}        &  {\tiny 3.4$\times$10$^{18}$}        &  {\tiny 0.12}  & E  \\
{\tiny CH$_3$OCHO  $\nu_t$=1}       & {\tiny 265}        &  {\tiny 1.7$\times$10$^{18}$}         &  {\tiny 0.12} & E   \\                             
{\tiny CH$_3$OCDO}                  & {\tiny 265}        &  {\tiny 1.6$\times$10$^{17}$}         &  {\tiny 0.12} & G  \\
{\tiny S-CH$_2$DOCHO}               & {\tiny 265}        &  {\tiny 8.0$\times$10$^{16}$}      &  {\tiny 0.12} & G \\
{\tiny A-CH$_2$DOCHO}               & {\tiny 265}        &  {\tiny 8.0$\times$10$^{16}$}      &  {\tiny 0.12} & G \\
{\tiny CH$_3$CH$_2$CN}             & {\tiny 200}        &  {\tiny 9.4$\times$10$^{16}$}      &  {\tiny 0.12} & E \\                                 
{\tiny CH$_3$CH$_2$CN $\nu_{13}$=1 /$\nu_{21}$=1}   & {\tiny 200}    &  {\tiny 2.9$\times$10$^{16}$}   &  {\tiny 0.12} & E \\
{\tiny CH$_2$CHCN $\nu_{11}$=0,1}                   &  {\tiny 200}   &  {\tiny $<$9.4$\times$10$^{17}$} &  {\tiny 0.12,} & G \\
{\tiny CH$_3$CH$_2$OH}           & {\tiny 200}    &  {\tiny 3.0$\times$10$^{18}$}   &  {\tiny 0.12}& E \\ 
{\tiny CH$_3$COCH$_3$}           &  {\tiny 200}   &  {\tiny 1.0$\times$10$^{15}$}   &  BF & G   \\
%
%
{\tiny (o+p)-H$_2$CO }          & {\tiny 200}    &  {\tiny 7.0$\times$10$^{17}$}   &  {\tiny 0.34} & E   \\
{\tiny (o+p)-H$_2$$^{13}$CO}    & {\tiny 200}    &  {\tiny 1.1$\times$10$^{16}$}   &  {\tiny 0.34} & E     \\
{\tiny (o+p)-D$_2$CO}           & {\tiny 200}    &  {\tiny 5.9$\times$10$^{15}$}   &  {\tiny 0.34} & E     \\
{\tiny OCS    }                       & {\tiny 200}    &  {\tiny 2.0$\times$10$^{18}$}   &  {\tiny 0.32} & E    \\ 
{\tiny OCS   $\nu_2$=1 }              & {\tiny 200}    &  {\tiny 2.4$\times$10$^{16}$}   &  {\tiny 0.32} & E     \\ 
{\tiny O$^{13}$CS        }            & {\tiny 200}    &  {\tiny 3.1$\times$10$^{16}$}   &  {\tiny 0.32} & E   \\ 
{\tiny HC$_3$N         }              & {\tiny 200}    &  {\tiny 2.0$\times$10$^{16}$}   &  {\tiny 0.20} & E     \\ 
{\tiny HC$_3$N  $\nu_7$=1}              & {\tiny 200}    &  {\tiny 2.0$\times$10$^{15}$}   &  {\tiny 0.20} & E  \\ 
%
%
{\tiny DC$_3$N}$^*$                       & {\tiny 200}    &  {\tiny 3.1$\times$10$^{14}$}   &  {\tiny 0.20} & G   \\   
{\tiny SO}                             &    {\tiny 200}    & {\tiny $>>$2.4$\times$10$^{15}$} & BF & E \\   
{\tiny S$^{18}$O}                     &  {\tiny 200}    &  {\tiny $<$8.0$\times$10$^{15}$}   &  {\tiny 0.6} & G  \\
{\tiny SO$_2$}                        & {\tiny 200}     &  {\tiny 6.0$\times$10$^{15}$}   & BF & G   \\
{\tiny $^{34}$SO$_2$}                 &  {\tiny 200}    &  {\tiny 2.7$\times$10$^{14}$}    & BF & G    \\
{\tiny HNCO}                          &  {\tiny 260}    &  {\tiny 1.6$\times$10$^{15}$}    & BF & E \\
{\tiny HN$^{13}$CO}                   & {\tiny 260}    &  {\tiny 2.5$\times$10$^{13}$}     & BF & E          \\
%
%
%
{\tiny (o+p)-H$_2$$^{13}$CS}          & {\tiny 200}    &  {\tiny 1.2$\times$10$^{15}$}   &  {\tiny 0.60} & E     \\
{\tiny (o+p)-HDCS}$^*$                & {\tiny 200}    &  {\tiny 1.0$\times$10$^{15}$}   &  {\tiny 0.60} & E     \\
{\tiny HCOOH}                         & {\tiny 100}    &  {\tiny 9.5$\times$10$^{18}$}   &  {\tiny 0.12} & E   \\   
{\tiny H$^{13}$COOH}                  & {\tiny 100}    &  {\tiny 1.5$\times$10$^{17}$}   &  {\tiny 0.12} & E    \\   
{\tiny HCOOD}$^*$                         & {\tiny 100}    &  {\tiny 1.0$\times$10$^{17}$}   &  {\tiny 0.12} & E    \\   
{\tiny CH$_3$OCH$_3$}                 & {\tiny 200}    &  {\tiny 4.0$\times$10$^{18}$}   &  {\tiny 0.12} & E   \\   
{\tiny CH$_3$CHO  $\nu_t$=0}          &  {\tiny 100}   &  {\tiny 4.0$\times$10$^{15}$}   & BF & G   \\  
{\tiny CH$_3$CHO  $\nu_t$=1}          &  {\tiny 100}   &  {\tiny 5.6$\times$10$^{14}$}   & BF & G   \\  
{\tiny CH$_3$CHO  $\nu_t$=2}          &  {\tiny 100}   &  {\tiny 1.1$\times$10$^{14}$}   & BF &  G   \\  
{\tiny aGg'-(CH$_2$OH)$_2$}           &  {\tiny 145}   &  {\tiny  2.0$\times$10$^{15}$}  &  BF & G  \\  
{\tiny (o+p)-H$_2$CCO}$^*$            &  {\tiny 200}   & {\tiny 6.7$\times$10$^{14}$}    &  BF & G \\    
{\tiny CH$_2$OHCHO}                   & {\tiny 265}     &  {\tiny  1.0$\times$10$^{15}$}  & BF & G  \\ 
{\tiny NH$_2$CHO}$^*$                     &  {\tiny 200}    &  {\tiny  3.0$\times$10$^{14}$}  & BF & G \\ 
\hline \hline
\end{tabular}

\noindent
$^1$Estimated column densities. $^2$Diameter of the emission region for the compact component assuming a disk morphology.
$^3$BF: Beam filling component.$^4$E: The disk diameter estimated from our modeling. $^5$ G: only a guess because the 
size of the disk cannot be constrained on basis of our data.$^6$The fit to this component is not reliable because the lines suffers from spatial
filtering and opacity effects (see text). $^*$Tentative detection
\end{table}

\subsection{Methanol: A- and E-CH$_3$OH, A- and E- $^{13}$CH$_3$OH, CH$_2$DOH}
Only five lines of methanol (including the A- and E- species and their vibrational states) were detected and their intensities
were not consistent with LTE. This is not unexpected since the emission of the 
low energy lines is optically thick and extended (see Table~1).  
We searched for the lines of $^{13}$CH$_3$OH and the deuterated species CH$_2$DOH. Ten lines of
CH$_2$DOH were detected. We removed from our fit the lines that could be blended with other more intense lines and we 
still kept 4 $``$clean$"$ lines that allowed us to derive the results shown in Fig.~2 and Table~2.
The rotation temperature of  CH$_2$DOH, $\sim$157$_{-13}^{+15}$~K, is lower than that of CH$_3$CN suggesting that it arises 
from a different region within the hot core. Based on the different spatial distribution of the high excitation and low excitation
lines of methanol, FU05 proposed the existence of two components: a hot component
towards the hot core and a colder component that extends along the outflow direction. Although the two components
are expected to contribute to the total CH$_2$DOH emission, its low rotation temperature suggests that it is dominated
by the extended one. 

For $^{13}$CH$_3$OH, we detected two lines of the $\nu_t$=0 vibrational state (220323 and 221282~MHz) and one
(221424~MHz) of the $\nu_t$=1 state. Although the emissions of the low energy lines of the main isotopologue
are  extended (see Table~1), we considered that the emissions of the 13-methanol lines come mainly from
the compact source. The line at 220323~MHz is heavily blended and cannot be used to estimate the 13-methanol column
density. Using the other two lines we fitted a vibrational/rotation temperature of $\sim$238~K
and a total $^{13}$CH$_3$OH $\nu_t=0,1$ column density of 5.3$\times$10$^{18}$~cm$^{-2}$.  
In order to estimate the uncertainty in our estimate of the 13-methanol
column density, we repeated the fit by fixing the rotation temperature to 200~K and 400~K. 
The derived column densities are 3.0$\times$10$^{18}$~cm$^{-2}$ and 6.1$\times$10$^{18}$~cm$^{-2}$, respectively, which suggests 
that our estimate of the $^{13}$C-methanol column density is accurate within a factor of 2.

Since the CH$_3$OH lines are optically thick, we derived the methanol column density from 
$^{13}$CH$_3$OH assuming that all the $^{13}$CH$_3$OH emission comes from the compact component and $^{12}$C/$^{13}$C=65. 
Then we added an extended component to fit the rest of lines. The value obtained for the CH$_3$OH column density in the extended component 
is not reliable because the intensities of the main isotopologue lines are affected by spatial filtering and opacity effects. 
Moreover, the separation of the methanol emission in these two components is model-dependent since we fixed the size of the compact core in our
calculations. The beam average CH$_3$OH column density derived from  $^{13}$CH$_3$OH is, however, reliable as long as the $^{13}$C-isotopolgue is 
optically thin regardless of the detailed spatial distribution of methanol within the observational beam. For this reason, we used 
this value to calculate the deuterium fraction, [CH$_2$DOH]/[CH$_3$OH]=0.02. This is an average value within the hot core.
%
%
\subsection{Trans- and Gauche- ethanol: T- and G-CH$_3$CH$_2$OH}
We detected six ethanol lines which includes four from T-CH$_3$CH$_2$OH and two from the
G conformer that is at 60~K above the T-state.
However, only three lines are clearly  detected (at 218554, 218655, and 220155~MHz) since
the other  are significantly blended with transitions from other species
or U-lines. Because of this, we could not apply the rotational diagram technique.
Instead, we assumed a rotation temperature of T$_{\rm rot}$=200~K,
and derived the column densities using the most intense lines. We assumed a size of $\sim$0.1$\arcsec$
to better fit the data published by FU05 (see Sect.5). The results are shown in Table~3.
%
%
\subsection{Acetone: AA-(CH$_3$)$_2$CO, AE-(CH$_3$)$_2$CO, EA-(CH$_3$)$_2$CO, EE-(CH$_3$)$_2$CO}
Several lines of acetone were detected in our spectrum (218773, 219076, 219220, 219242, 219264, 
220355, 220466, 220764, 220962~MHz) being the most intense 
at 219220, 219242 and 219264~MHz. Because many of them are blended, it is not possible to determine 
an accurate rotation temperature. We assumed T$_{\rm rot}$=200~K and obtained a total 
(AA+EA+AE+EE) beam average column density of 1.0$\times$10$^{15}$~cm$^{-2}$. 
%
%
\subsection{Formaldehyde: H$_2$CO}
Three lines of p-H$_2$CO and one of o-H$_2$$^{13}$CO were detected in the PdBI spectrum. The lines of the main isotopologue are optically 
thick and most of their emission has been filtered by the interferometer (see Table~1). Because of the large fraction (85\%) of missing flux in 
the main isotopologue line, we are surely missing an extended component. According with Fig. 1, the emission of the o-H$_2$$^{13}$CO
is unresolved by our observations. Besides, FU05 derived a size of 0.58$\pm$0.24\arcsec  $\times$ 0.40$\pm$0.24\arcsec for the D$_2$CO emission.
Based on these results, we were able to fit all the lines of this work and FU05 with an (ortho+para)-H$_2$$^{13}$CO column density of 
1.1$\times$10$^{16}$~cm$^{-2}$ and (ortho+para)- D$_2$CO column density of  5.9$\times$10$^{15}$~cm$^{-2}$, and a source with a
diameter of 0.34$\arcsec$ ( $\sim$426~AU ). We assume T$_{\rm rot}$=200K and an ortho-to-para ratio of~3. 

There are no H$_2$C$^{18}$O and HDCO lines in the frequency range of our observations that we could use to further constrain
our model. 
%
%
%
\subsection{Carbonyl sulfide: OCS, O$^{13}$CS}
The $J$= 19$\rightarrow$18 rotational lines of the ground vibrational states of OCS and O$^{13}$CS
lie in the range of wavelengths covered by our observations. 
We measured T$_b$[OCS 19$\rightarrow$18]/T$_b$[O$^{13}$CS 19$\rightarrow$18]$\sim$5.
Comparing the 30m and PdB spectra, we discarded 
the possibility that this low 
intensity ratio is a consequence of
the spatial filtering. The opacity of the OCS  19$\rightarrow$18 line must be very high, $\tau$$\sim$12 .
Assuming  T$_{\rm rot}$=200~K, we derived a OCS column density of $\sim$2$\times$10$^{18}$~cm$^{-2}$ and a source size 
of $\approx$0.32$\arcsec$ (400~AU), four times that derived for the compact component of CH$_3$CN. In our spectrum, we also detected two lines 
of the OCS $\nu_2$=1 vibrational state at E$_{\rm vib}$=748.8~K (see Table A.1). Assuming the same source size, the intensities of 
these lines imply a vibrational temperature, T$_{\rm vib}$=170~K, in agreement with the rotational temperature we assumed.
The same assumptions are used to fit the OC$^{33}$S J=19$\rightarrow$18 line in FU05.
%
%
%
%
%
\subsection{Sulfur monoxide: SO, S$^{18}$O}
Only the intense line of SO at 219949~MHz was detected. From its intensity and assuming T$_{\rm rot}$=200~K, we
derived a beam averaged SO column density of 2.4$\times$10$^{15}$~cm$^{-2}$. The emission of this line is 
extended (see Table~1 and Fig.~1) and, most likely, optically thick. There are no lines of $^{34}$SO, $^{33}$SO and/or S$^{18}$O in the 
spectrum that could be used to determine the line opacity. 
As commented in Sect. 5 a possible detection of S$^{18}$O in the spectrum published by FU05 suggests that the column density could 
be as much as a factor of $\sim$600 higher (see Sect.5 and Table~3). Unfortunately this line is blended with other intense
lines which makes the estimate uncertain. We consider our fitted value as a lower limit to the real one.  
%
%
\subsection{Sulfur dioxide: SO$_2$, $^{34}$SO$_2$}
One SO$_2$ line at  219276~MHz and two $^{34}$SO$_2$ lines at  219355 and 221115~MHz, respectively were detected.
These lines are reasonably fitted using T$_{\rm rot}$=200~K, a beam averaged SO$_2$ column density of 6$\times$10$^{15}$~cm$^{-2}$
and $^{32}$S/$^{34}$S=22. The SO$_2$ line is optically thin providing a good estimate of the total number of SO$_2$ molecules
within our beam. On the other hand, we could not estimate an accurate value of the opacity that would allow us to  
constrain the size of the emitting region. We checked that the SO$_2$ line would be optically thin even if all the emission arises from 
the 0.1$\arcsec$ compact core, however.
%
%
\subsection{Cyanoacetylene: HCCCN,HCCCN $\nu_7$=1, DCCCN, HCCNC}
We detected one HC$_3$N line at 218325~MHz and two lines of the  vibrational state $\nu_7$=1 at 218861~MHz 
and 219174~MHz, respectively, although the one at 219174~MHz is blended with an intense ethanol line. In addition, 
we tentatively detected the H$^{13}$CCCN 25$\rightarrow$24 line. Unfortunately the frequency of the 
H$^{13}$CCCN line is close to that of the  $^{13}$CO 2$\rightarrow$1 line and could have some contribution 
from its wing. Assuming that the isotopologue line is not blended with $^{13}$CO, we fit this line with T$_{\rm rot}$=200~K 
and N(H$^{13}$CCCN)=8.8$\times$10$^{14}$~cm$^{-2}$ and a source size of $\sim$0.20$\arcsec$ which implies an
upper limit to the HC$_3$N column density of $<$5.7$\times$10$^{16}$~cm$^{-2}$. This value, however,
overestimates the intensity of the lines of the main isotopologue, that are better fitted with N(HCCCN)=2.0$\times$10$^{16}$~cm$^{-2}$. 
We adopt the latter solution and conclude that our column density is accurate within a factor of 3.
There is a weak feature at  219490~MHz that could correspond to  DCCCN J=26$\rightarrow$25. 
Since this line is very weak and there is no other DCCCN line in the spectrum that could confirm the detection, 
we will consider it as a tentative detection and use its intensity to derive an upper limit to the DCCCN column density of
N(DCCCN)/N(HCCCN)$<$0.016. We could have
detected one line of HCCNC at 218558~MHz but it is blended with an intense ethanol line.
Therefore we can only derive a lower limit to the N(HCCCN)/N(HCCNC) column density ratio
of $>$13.

 
\subsection{Acetaldehyde: CH$_3$CHO}
Very few intense lines of CH$_3$CHO are found in the observed wavelength range and some of them are blended with 
others from abundant species. We detected without blending: the 220446~MHz line of 
E-CH$_3$CHO, and marginally the 221222~MHz line of A-CH$_3$CHO $\nu_t$=1, the 219780~MHz 
line of A-CH$_3$CHO $\nu_t$=2, and 221216~MHz line of E-CH$_3$CHO $\nu_t$=2. 
All these lines were fitted assuming
T$_{\rm vib}$=100~K. Making the additional assumption that T$_{\rm rot}$=T$_{\rm vib}$, we derived a total beam averaged
A+E CH$_3$CHO column density of 4$\times$10$^{15}$~cm$^{-2}$ in the $\nu$=0 ground state. In this case, 
the lines of the $^{13}$C isotopologue were not detected. 
Like for other oxygenated molecules, we assumed that the source is extended
relative to our beam. Although the firm detections in the 218-221~GHz frequency range are only a few, this model is 
fully confirmed by the good match to the data published by FU05 (see Sect. 5).

\subsection{Formic acid: HCOOH}
Two lines of HCOOH, one at 218938~MHz and the other at 220038~MHz were detected in our spectrum. However their 
line intensities were not consistent with thermalized optically thin emission. Since the two lines have very similar
E$_u$, this discrepancy cannot be due to a wrong rotation temperature. Spatial filtering cannot be the cause of this
discrepancy, because such an intense line should have been detected in the 30m spectrum.  However the Einstein  
spontaneous emission coefficient, A$_{ij}$,  of the 220038~MHz  line is 10 times lower than that of the  218938~MHz line
which results in a higher opacity of the former for the same physical conditions. We propose that
the high opacity of the line at 220038~MHz causes the anomalous line intensity ratio.
The detection of H$^{13}$COOH at 219341.849~MHz also argues in favor of a high column density of HCOOH in this source.
We can reasonable fit both lines of HCOOH and that of H$^{13}$COOH assuming T$_{\rm rot}$=100~K, a total HCOOH column density
of 9.5$\times$10$^{18}$~cm$^{-2}$ and a source size of $\sim$0.1$\arcsec$.
The profile of the  H$^{13}$COOH line is wider than expected (see Fig. A.2), suggesting some contribution from 
the outflow or partial blending with other line. Because of this possible blending, the integrated intensity of this line 
is uncertain by a factor of 2. We selected the solution that best fitted the 218938~MHz line.
The fit to the HCOOH line at 220038~MHz is not perfect but this is not worrying because our model is not
adequate to reproduce the profile of very optically thick lines. In fact the intensity of this line is not sensitive to the HCOOH
column density in the range of column densities we are dealing with. Some HCOOH lines were also observed by FU05. As seen in Fig. 5, the
lines are reasonably reproduced with our values.
We have searched for the DCOOH and HCOOD lines in our spectrum and found one transition of HCOOD
at 218541~MHz from which we derive N(HCOOD)/N(HCOOH)$\approx$0.01 as long as both species
are coming from the same region. Since we have only one line of HCOOD and it is at 4$\sigma$, we indicated this detection
as tentative in Table~3.

\subsection{Dimethyl ether: CH$_3$OCH$_3$}
We detected four unblended lines of CH$_3$OCH$_3$, at 218492, 219301, 220847, and 220893~MHz. 
The other lines of this species are blended with intense lines. Assuming that the source is filling 
the beam, we derived T$_{\rm rot}$=500~K and N(CH$_3$OCH$_3$)=2$\times$10$^{16}$~cm$^{-2}$. We considered that this apparently 
high rotation temperature was an opacity effect and obtained a better fit
assuming T$_{\rm rot}$=200~K, N(CH$_3$OCH$_3$)=4$\times$10$^{18}$~cm$^{-2}$ and a source diameter of 0.1$\arcsec$.
These parameters are consistent with the non-detection of the  $^{13}$C isotopologues and  we adopted them.


\subsection{Ethylene glycol: aGg'-(CH$_2$OH)$_2$, gGg'-(CH$_2$OH)$_2$ }
The aGg' conformer of ethylene glycol, together with methyl formate, is one of those with the highest number of lines in our spectrum.
We detected twelve lines, and seven were unblended. We used these seven unblended lines to derive a rotation 
temperature of 145~K and
a beam averaged column density of  N(aGg'-(CH$_2$OH)$_2$)=2.3$\times$10$^{15}$~cm$^{-2}$ (see Fig.~2
and Table~2).
One of the lines, that at  218576~MHz was removed from the final fit since it was very
far from the intensity traced by the other lines (see Fig.~2) and is certainly blended with something else. 
In this case, we did not have information about the transitions of the $^{13}$C-isotopologues and
deuterated species. The emission is compatible with the beam filling assumption although we cannot
discard a smaller size. 

We also searched for gGg'-(CH$_2$OH)$_2$ in our spectrum. We have a good match with
8 lines that can be assigned to this species (218712, 219097, 219389, 219410, 220763, 220448, 220887,
221596~MHz). Modeling the spectrum, we obtain that 5 additional  lines are also present in our data but blended with 
other species. However there are 2 lines, at 218144~MHz and 220249~MHz, 
that should have been detected and do not appear in the spectrum. These lines have similar energies as the
previous ones and their absence cannot be explained with a simple excitation argument. This species has not
been detected in any other source thus far. Waiting for confirmation, 
we do not include it in the model.

\subsection{Ketene: H$_2$CCO}
We detected one o-H$_2$CCO line at 220177~MHz and one p-H$_2$CCO at 221545~MHz with similar intensity. The first
transition is at E$_u$= 62.4~K and the second one E$_u$= 894.8~K. It is not possible to fit both lines assuming reasonable
physical conditions and optically thin emission. One possibility is that the line at 221545~MHz is not ketene. In this case,
we can fit the other line assuming a rotation temperature of T$_{\rm rot}$=200 K and a beam averaged ortho+para H$_2$CCO column
density of 6.7$\times$10$^{14}$~cm$^{-2}$, where we have assumed an ortho-to-para ratio of 3. Another possibility is that
the emission is optically thick in both lines. In this case, we need a ketene column density of 4.0$\times$10$^{18}$~cm$^{-2}$
in a compact $\sim$0.1\arcsec region. There is no line of the $^{13}$C isotopologue in the observed frequency range that
would allow us to determine the opacities. The deuterated compounds have no intense transitions in the observed
wavelength range either. We selected the optically thin case adopting a conservative approach. Taking into account the uncertainty
in its column density, we do not use this species in our comparion with other sources. 

\subsection{Glycolaldehyde: CH$_2$OHCHO, Formamide: NH$_2$CHO}
There are thirteen intense lines of glycolaldehyde in the sampled wavelength range. From these, six are not blended
with other features and do not have any other likely identification. Assuming T$_{\rm rot}$=200~K, we 
estimated a beam averaged column density of 1.0$\times$10$^{15}$~cm$^{-2}$. Only one line of NH$_2$CHO lies in
our spectrum. It is at 218459~MHz and is blended with a line of T-ethanol. Our fit to T-ethanol failed to predict
the high intensity of the observed feature suggesting some contribution from NH$_2$CHO. 
The total intensity would be consistent with a beam averaged formamide column density of 3.0$\times$10$^{14}$~cm$^{-2}$.
%
%
\subsection{Vinyl cyanide: CH$_2$CHCN}
We detected two rotational lines, at 218675~MHz and 220345~MHz, that could correspond to CH$_2$CHCN. The intense line at 220345~MHz is 
overlapped with one of CH$_3$CN. Some features of the $\nu_{11}$=1 vibrational state at an energy of 328.4~K could have also been detected 
but are blended with other more intense lines. 
Since we could not determine the rotation temperature, we assumed the fixed value of
T$_{\rm rot}$=200~K and the same source size as the chemically related species CH$_3$CN, i.e., a diameter of $\sim$0.1$\arcsec$. 
But our fit was very bad. The intensities of the lines at 220345~MHz and 218675~MHz cannot 
be accounted by LTE models. We were not able to obtain a better fit by varying the rotation temperature and/or the source size.
One possibility is that the intensity of one of the lines is wrong because of baseline uncertainties or spatial filtering. The second possibility
is that our identification is false. Since we do not have a better candidate for the 220345~MHz line, 
we derived an upper limit to the vinyl cyanide column density using its line intensity and fixing the rotation temperature to 200~K.  
This compound is quite abundant in massive hot cores (see (L\'opez et al. 2014), and this upper limit is interesting for comparison with this
kind of objects.  
%
\begin{table*}
\caption{Comparison of rotation temperatures between hot cores and hot corinos}
\begin{tabular}{lllll} \hline \hline 
Hot core/corino         &  L$_\odot$ &  T$_{\rm rot}$ (CH$_3$CN)  & T$_{\rm rot}$ (X)   &   ref. \\ \hline
IRAS~16293-2422         &    6.9$-$23           & $\sim$150-390~K         & $\sim$60~K ( CH$_3$CHO)       & Cazaux et al. (2003), Bisschop et al. (2008) \\ 
                        &               &                         &                               & Pineda et al. (2012) \\ 
IC~1396~N               &  $<$300       &    $\sim$100~K          &                               & Neri et al. (2007), Fuente et al. (2009) \\ 
IRAS~22198+6336         &  370          &    $\sim$100~K          &  $\sim$120~K (CH$_3$CH$_2$OH)       & S\'anchez-Monge et al. (2010), Palau et al. (2011)  \\ 
NGC 7129 FIRS 2         &  480          &    $\sim$400~K          &  $\sim$120~K (aGg'-(CH$_2$OH)$_2$)        & This work \\
AFGL~5142               &  2300         &                         &  $\sim$140$-$210~K (CH$_3$CH$_2$OH)  & Palau et al. (2011) \\   
Orion KL                &  $\sim$10$^4$ &    400$-$600~K          &  $\sim$100-200~K (CH$_3$OCH$_3$,OCS) & Tercero et al. (2010), Beuther et al. (2011),\\         
                        &               &                         &                                       & Brouillet et al. (2013), Bell et al. (2013) \\         
\hline \hline
\end{tabular}

\end{table*}
\section{The hot core model}
Based on the calculations explained in Sect.~4, we propose the hot core model shown in Table~3. This model is not unique
and some parameters, especially source sizes, are quite uncertain. However it cannot be done better. Higher spatial resolution
are required to better constrained the spatial
structure. Fortunately, the total number of molecules in our beam is well constrained, especially when we
have detected the $^{13}$C isotopologue or the emision is optically thin, which is the case for most species. 
Our model is self-consistent, i.e., once a molecule is included, its 
isotopologues and deuterated compounds are also included. For isotopologues, we assume $^{12}$C/$^{13}$C=65, $^{16}$O/$^{18}$O=650, 
$^{32}$S/$^{34}$S=22 and $^{34}$S/$^{33}$S=6 (Chin et al. 1995, Milam et al. 2005). The column densities of the deuterated species are  
calculated independently to determine the deuterium fractions. When possible we directly compare the deuterated compound 
with the $^{13}$C isotopologue, that is expected to have similar fractional abundance, in order to avoid opacity effects in
the derived column density ratio.
The obtained values of deuterium fractions are: $\sim$0.06 for methyl formate, $\sim$0.02 for methanol, $\sim$0.008 for the doubly deuterated formaldehyde,  
$\sim$0.01 for DC$_3$N, and  $<$0.01 for HCOOD (see Table 3).
Of course these are average values in the hot core since the hydrogenated and deuterated species
could have different spatial distributions.
Using MADEX, we have synthesized the total spectrum assuming LTE, the parameters shown in Table~3 and convolving with a circular beam of $\sim$1.4$\arcsec$.The obtained 
spectrum is compared with observations in Figs.~A.1-4. Our fit is reasonably good taking into account the simplicity of our model. 
There are, however, two frequency ranges 218.0-219.12 GHz and
221.46-221.62 where the agreement is poor. One reason is possible observational errors. Note that the second region is at the end of
the observed frequency range where the passband calibration is more uncertain. But the existence of intense "U" lines suggests
the existence of a population of  COMs that have not been identified yet.

An important test for our model is to reproduce the interferometric observations published by FU05 with a
higher angular resolution of 0.63$\arcsec$$\times$0.46$\arcsec$. We synthesized the spectrum assuming the same parameters as in Table~3 
but convolving with a smaller beam, $\sim$0.52$\arcsec$, and compared with FU05 observations (see Fig.~A.5). 
We find a remarkably good agreement for most of
the species which confirmed that our model was not far from reality. 

We identified the line at 231143~MHz in FU05 as H$_2$$^{13}$CS providing an estimate of the H$_2$CS abundance in the hot
core (see Table~3). However we have only one line of the $^{13}$C isotopologue and the identification could be wrong.
A similar situation occurs for S$^{18}$O. We have detected one line at 228272~MHz but it is blended with intense CH$_3$OCHO 
and CH$_3$OCOD features. From this line we estimate a S$^{18}$O column density of 8$\times$10$^{15}$~cm$^{-2}$ within the
$\sim$0.52$\arcsec$ ($\sim$650 AU) beam. Because of the blending, this estimate is model-dependent and suffers from larger uncertainty.
   
An important discrepancy between our model and FU05 observations comes from G+ethanol.
There are two predicted G+ethanol lines, 228491~MHz and 228560~MHz that do not appear in the FU05 spectrum while all 
the T$-$ethanol lines  are correctly reproduced. 
One possibility is that we have a lower kinetic temperature than assumed since the G+ethanol is 60~K higher in energy than T$-$ethanol. But the
temperature would have to be unreasonably low, $<$50~K, to reproduce the observations.These lines are close to the edge of the spectral band 
and could suffer from some instrumental effects.

In Table~A.2 we give a
list with our new identifications of the data published by FU05. In this work we have identified more
lines and found some misidentifications. The lines previously identified as c-C$_3$D are instead
carried by CH$_3$CHO. In fact, CH$_3$CHO is the carrier of most of the lines in the band centered at 231.3~GHz.
We have a handful of lines around the $^{13}$CS frequency without any
plausible identification. One possibility is that we are observing high velocity gas emitting in the $^{13}$CS 5$\rightarrow$4 line, 
but this would imply an outflow with velocities as high as 50~km~s$^{-1}$. We have not seen such high velocities in the  
CO single-dish and interferometric spectra (Fuente et al. 2001, FU05). 
Therefore, we think that there are still some unidentified lines at these 
frequencies. The intense lines at 228307~MHz, 228363~MHz, 228427~MHz and 228467~MHz, although identified, 
are poorly reproduced by our model. We have
not found any simple way to improve the agreement between our model and observations. Either these lines are far from 
the LTE approximation or we are still missing important species. Additional observations could help to discern their
origin.  

The observations presented in
this paper have significantly contributed to the understanding of the spectra published by FU05 proving that
observations in a large range of frequencies are needed for the correct line identification.

\begin{figure*}
\includegraphics{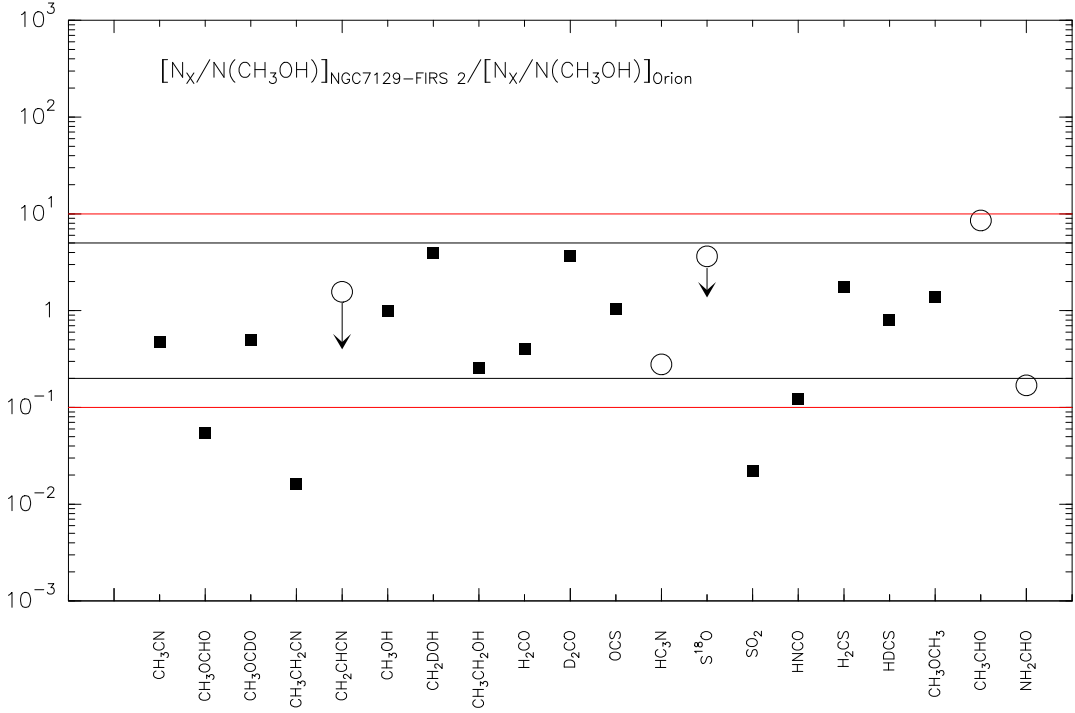}
\vspace{11.0cm}
      \caption{Comparison of the molecular abundances in FIRS 2 and the Orion hot core (see Table~6).
      All the molecular abundances have been normalized to that of CH$_3$OH.
      Black horizontal lines indicate the loci of the FIRS~2 values that differ by
      less than a factor of 5 from those of Orion. Red lines indicate the same but for a factor of 10. Empty circles
      indicate doubtful values.}
         \label{Fig 5}
 \end{figure*}
\begin{table}
\caption{Average fractional abundances$^1$}
\begin{tabular}{lcc} \hline \hline 
                                           &      T$_{dust}$=50~K            &  T$_{dust}$=400~K       \\ \hline
M$_{gas+dust}$ (M$_{\odot}$)               &      1.8                        &    0.46                  \\
N$_{\rm H_2}$      (cm$^{-2}$)                 &      2.0$\times$10$^{24}$       &    5.0$\times$10$^{23}$   \\ \hline
CH$_3$CN ($^{13}C$)$^2$            &   6$\times$10$^{-9}$    &  3$\times$10$^{-8}$ \\
CH$_3$OCHO                         &   2$\times$10$^{-8}$    &  7$\times$10$^{-8}$ \\
CH$_3$CH$_2$CN                     &   4$\times$10$^{-10}$   &  2$\times$10$^{-9}$  \\
CH$_2$CHCN                         &  $<$3$\times$10$^{-9}$  &  $<$2$\times$10$^{-8}$  \\
CH$_3$OH ($^{13}C$)                &   1$\times$10$^{-6}$    &  5$\times$10$^{-6}$ \\
CH$_2$DOH                          &   7$\times$10$^{-9}$    &  3$\times$10$^{-8}$ \\  
CH$_3$CH$_2$OH                     &   1$\times$10$^{-8}$    &  4$\times$10$^{-8}$ \\ 
CH$_3$COCH$_3$                     &   5$\times$10$^{-10}$   &  2$\times$10$^{-9}$    \\
H$_2$CO ($^{13}C$)                 &   2$\times$10$^{-8}$    &  8$\times$10$^{-8}$ \\
D$_2$CO                            &   2$\times$10$^{-10}$   &  7$\times$10$^{-10}$  \\
OCS ($^{13}C$)                     &   5$\times$10$^{-8}$    &  2$\times$10$^{-7}$ \\ 
HC$_3$N                            &   2$\times$10$^{-10}$   &  8$\times$10$^{-10}$ \\ 
SO$_2$                             &   3$\times$10$^{-9}$    &  1$\times$10$^{-8}$  \\
HNCO                               &   8$\times$10$^{-10}$   &  3$\times$10$^{-9}$  \\
H$_2$CS ($^{13}C$)                 &   7$\times$10$^{-9}$    &  3$\times$10$^{-8}$  \\
HCOOH ($^{13}C$)                   &   3$\times$10$^{-8}$    &  1$\times$10$^{-7}$ \\   
CH$_3$OCH$_3$                      &   1$\times$10$^{-8}$    &  6$\times$10$^{-8}$  \\   
CH$_3$CHO                          &   2$\times$10$^{-9}$    &  1$\times$10$^{-8}$    \\   
aGg'-(CH$_2$OH)$_2$                &   1$\times$10$^{-9}$    &  4$\times$10$^{-9}$ \\  
CH$_2$OHCHO                        &   5$\times$10$^{-10}$   &  2$\times$10$^{-9}$    \\ 
\hline \hline
\end{tabular}

\noindent
$^1$ Average fractional abundance in a disk with a diameter of $\sim$1.4\arcsec.

\noindent
$^2$ Derived from the $^{13}$C isotopologue assuming $^{12}$C/$^{13}$C=65.
\end{table}

\begin{table*}
\caption{Comparison between FIRS~2 and Orion KL}
\begin{tabular}{llll} \hline \hline 
\multicolumn{1}{c}{}  & \multicolumn{1}{c}{FIRS~2$^1$} & \multicolumn{1}{c}{Orion KL$^{2}$} &  \multicolumn{1}{c}{Refs$^3$}  \\ 
\multicolumn{1}{c}{}  & \multicolumn{1}{c}{N (cm$^{-2}$)} &  \multicolumn{1}{c}{N (cm$^{-2}$)} &  \multicolumn{1}{c}{} \\ \hline
{\tiny CH$_3$CN}      & {\tiny 1.3$\times$10$^{16}$}   ($^{13}C$)$^4$    & 
{\tiny 1.2$\times$10$^{16}$}    ($^{13}C$) &  Bell et al. (2014) \\
{\tiny CH$_3$OCHO}    & {\tiny 3.7$\times$10$^{16}$}                 & 
{\tiny 2.9$\times$10$^{17}$}  ($^{13}C$)   &  Carvajal et al. (2009), Haykal et al. (2014)   \\
{\tiny CH$_3$OCDO}        & {\tiny 1.1$\times$10$^{15}$}                 &  
{\tiny 1.0$\times$10$^{15}$} &  Margul{\`e}s et al. (2010) \\
{\tiny CH$_3$CH$_2$CN}    &  {\tiny 8.8$\times$10$^{14}$}     & 
{\tiny 2.3$\times$10$^{16}$} ($^{13}C$) &  Daly et al. (2013)       \\
{\tiny CH$_2$CHCN}    &  {\tiny $<$6.9$\times$10$^{15}$}     & 
{\tiny 2.4$\times$10$^{15}$} &  L\'opez et al.(2014)       \\
{\tiny CH$_3$OH}          &  {\tiny 2.5$\times$10$^{18}$} ($^{13}C$)     & 
{\tiny 1.1$\times$18$^{18}$} ($^{13}C$)   &   Kolesnikov\'a et al. (2014) \\  
{\tiny CH$_2$DOH}         &    {\tiny 1.4$\times$10$^{16}$}     &  
{\tiny 1.5$\times$10$^{15}$}  &  Neill et al., (2013)  \\ 
{\tiny CH$_3$CH$_2$OH}    &   {\tiny 2.2$\times$10$^{16}$}     & 
{\tiny 3.6$\times$10$^{16}$}   &  Kolesnikov\'a et al. (2014) \\ 
%
%
{\tiny H$_2$CO }         &   {\tiny 4.1$\times$10$^{16}$} ($^{13}C$)    & 
{\tiny 4.3$\times$10$^{16}$} ($^{13}C$) &  Neill et al. (2013)  \\
{\tiny D$_2$CO}         &  {\tiny 3.4$\times$10$^{14}$}   & 
{\tiny 4.2$\times$10$^{13}$} &  Turner, B.E. (1990)     \\
{\tiny OCS}              &  {\tiny 1.0$\times$10$^{17}$}  ($^{13}C$)  & 
{\tiny 4.2$\times$10$^{16}$}   ($^{13}C$)   &  Tercero et al. (2010)     \\ 
{\tiny HC$_3$N}          &  {\tiny 4.1$\times$10$^{14}$}   & 
{\tiny 7.7$\times$10$^{15}$}    ($^{13}C$)      &  Esplugues et al. (2013b)      \\ 
{\tiny S$^{18}$O}                            &  {\tiny $<$1.5$\times$10$^{15}$}         &  
{\tiny 2.2$\times$10$^{14}$}                    & Esplugues et al. (2013a)  \\   
{\tiny SO$_2$}                        & {\tiny 6.0$\times$10$^{15}$}   & 
{\tiny 1.1$\times$10$^{17}$}    &  Esplugues et al. (2013a)  \\
{\tiny HNCO}                          & {\tiny 1.6$\times$10$^{15}$}          & 
{\tiny 5.5$\times$10$^{15}$}   ($^{13}C$)     &  Marcelino et al. (2009) \\
{\tiny H$_2$CS}                       &  {\tiny 1.4$\times$10$^{16}$}  ($^{13}C$)     &      
{\tiny 3.3$\times$10$^{15}$}   ($^{13}C$)      &  Tercero et al. (2010)  \\
{\tiny HDCS}                       &  {\tiny 1.8$\times$10$^{14}$}                    &      
{\tiny 9.5$\times$10$^{13}$}        &  Tercero et al. (2010)  \\
{\tiny HCOOH}                     &  {\tiny 6.9$\times$10$^{16}$}    ($^{13}C$)                  & 
{\tiny 4.9$\times$10$^{13}$}          &  Turner, B.E. (1991) \\   
%
%
%
{\tiny CH$_3$OCH$_3$}             &  {\tiny 2.9$\times$10$^{16}$}   & 
{\tiny 9.0$\times$10$^{15}$}                     &   Comito et al. (2005) \\   
{\tiny CH$_3$CHO}                &  {\tiny 5.0$\times$10$^{15}$}   & 
{\tiny 2.8$\times$10$^{14}$}                     &   Turner, B. E. (1991) \\   
%
%
%
{\tiny NH$_2$CHO}$^*$                & {\tiny  3.0$\times$10$^{14}$}       & 
{\tiny 9.6$\times$10$^{14}$}   &  Motiyenko et al. (2012)  \\ 
\hline \hline
\end{tabular}

\noindent
$^1$ Average molecular column densities in a disk with a diameter of 1.4$\arcsec$. It is calculated with the expression $N=N_{c} \times (D(\arcsec)/1.4\arcsec)^2 + N_{ex}$ 
where $N_{c}$ and $D$ are the column density and size (diameter) of the compact component, respectively, and $N_{ex}$, the column density of the beam filling component as
shown in Table ~3.

\noindent
$^2$ Average molecular column densities considering a region of 15$\arcsec$ of diameter towards Orion KL. We assume a beam filling factor of 0.44 for
the hot core component (size=10$\arcsec$).

\noindent
$^3$ References for the Orion column densities.

\noindent
$^4$ Derived from the $^{13}$C isotopologues assuming $^{12}$C/$^{13}$C=65 for FIRS~2 and $^{12}$C/$^{13}$C=50 for Orion.

\end{table*}

\section{Discussion}
\subsection{Physical conditions}
It is known that hot cores are heterogenous objects where different species are coming from different regions with
different physical conditions (see. e.g. Beuther et al. 2011, Brouillet et al. 2013, Bell et al. 2013). 
For this reason,  we only compare rotation temperatures measured
with the same molecular species.

Methyl cyanide is easily thermalized for the densities prevailing in hot cores and hence a good thermometer of the molecular gas. 
The rotational diagram shows that the temperature in the
FIRS~2 hot core is $\approx$400~K. This temperature is similar to that measured in massive hot cores.
Beuther et al. (2011) proposed the existence of  a warm component
of 600$\pm$200~K in the Orion hot core based on submillimeter CH$_3$CN lines as observed with the 
Submillimeter Array (SMA). More recently,  Bell et al. (2013) estimated a rotation temperature of $\sim$220~K towards IRc2 
and $\sim$400~K in the hot spot located $\sim$14$\arcsec$ NE using data from the 30m telescope. 
Lower temperatures, however, are found in hot corinos. Bisschop et al. (2008) derived an upper limit of $<$390~K 
in IRAS~16293-2422 A and B based on interferometric SMA data. Fuente et al. (2009) derived a rotation temperature of
97$\pm$25~K in the hot core located in the IM protocluster IC~1396~N. A similar rotation temperature, $\sim$100~K, was derived
by S\'anchez-Monge et al. (2010) in the hot core associated with IRAS~22198+6336. Although the luminosity of FIRS~2
is similar, our observations suggest the existence of a warmer gas component ($\sim$400~K) in FIRS~2 than in  IC~1396~N
and  IRAS~22198+6336. IC~1396~N
is a protocluster with three young stellar objects (YSOs). Although the total luminosity is similar to that of FIRS~2, 
the luminosity of each YSO is lower and probably closer to that of an Ae star, i.e., $<$100 L$_\odot$. 
IRAS~22198+6336 is morphologically similar to FIRS~2.
S\'anchez-Monge et al. (2010) detected only the K-lines with  E$_u$=100$-$300~K, while we used lines with E$_u$$>$600~K in this
study. It is possible that a warmer component remains hindered in IRAS~22198+6336 and would require higher sensitivity observations of
transitions with higher E$_u$ to be detected. 
In fact, FU05 determined a gas temperature of 50$-$100~K in FIRS~2 from lower sensitivity observations of the  
CH$_3$CN 5$_K$$\rightarrow$4$_K$ ladder, from lines with E$_u$=13$-$154 K.

The rotational diagrams of the oxygenated species give lower rotation temperatures than with CH$_3$CN.
We derived rotation temperatures between 100~K and 200~K from the CH$_2$DOH, CH$_3$CHO and aGg'-(CH$_3$OH)$_2$ 
lines and between 200~K and 300~K from CH$_3$OCHO  and HNCO. Palau et al. (2011) derived rotation temperatures of $\sim$100$-$150~K
in the IM hot cores in IRAS 22198+6336 and AFGL 5142 from the ethanol lines. Rotation temperatures of around 100~K
are also derived in Orion from CH$_3$OCH$_3$, methyl formate, deuterated methanol and OCS (Tercero et al. 2010, Peng et al. 2012, 
Brouillet et al. 2013). Therefore, we do not detect significant differences between the rotation temperatures measured with these
O-bearing species between IM and massive hot cores. Lower rotation temperatures, $\sim$50~K, are measured with CH$_3$OCHO  in  
IRAS~16293-2422 (Pineda et al. 2012). 

In Table~4, we show a summary of the rotation temperatures derived from different
molecules and for different hot cores. The comparison is not totally fair since CH$_3$CN and the O-bearing molecules could come from different regions. 
This is the case of IRAS~16293-2422,  where CH$_3$CN is coming mainly from the southern component (A) and the O-bearing molecules
from the northern one (B) (Bisschop et al. 2008). As commented above, different regions are associated with O- and N-bearing molecules in
Orion. Towards the other hot cores, the spatial resolution provided by the existing observations is lower,
preventing us from resolving regions with different chemistries. Nevertheless, 
it seems that there is a weak trend with the 
maximum gas temperature increasing from low to massive hot cores as measured with CH$_3$CN. There is no trend for the O-bearing
molecules. 

This suggests a scenario in which the O-bearing molecules are more abundant in extended regions with temperatures
$\sim$100~K in both hot cores and corinos. Instead, CH$_3$CN  probes the hottest region of the core, and its rotation temperature 
increases with the luminosity. Of course, there is a problem with the spatial scales that are different for low-mass and massive stars.
But, taking into account that massive stars are more distant, this would favor the interpretation of higher temperatures in the latter.
Therefore, within the limitations of our analysis, we can conclude that there is a trend of increasing
temperature from hot corinos to massive hot cores. It is not so clear between IM and massive stars. As we discuss, there
is no obvious difference between FIRS~2 and Orion.

Another important parameter is the total amount of gas in the hot core. 
One possibility is to calculate it from 
the dust continuum emission assuming a dust temperature and a dust emissivity value. FU05
adopted a dust temperature of T$_{dust}$=100~K and $\kappa_\nu$=0.015 (1300/$\lambda$ ($\mu$m))~cm$^2$~g$^{-1}$ and
derived a total gas+dust mass of 2~M$_\odot$ in the hot core FIRS~2. More recently, Palau et al. (2013) derived a similar value 
(1.8~M$_\odot$) assuming T$_{dust}$=50~K and $\kappa_{1.3mm}$=0.00899~cm$^2$~g$^{-1}$. 
Our results suggest that at least a fraction of the dust must be at T$_{dust}$=400~K. 
Assuming T$_{dust}$=400~K and $\kappa_{1.3mm}$=0.01~cm$^2$~g$^{-1}$,
we obtain a lower limit to the gas+dust mass of 0.46~M$_\odot$. Therefore, the total gas+dust mass of the FIRS~2 hot core in
uncertan in a factor of $\sim$4.
In Table~5 we show the average molecular abundances in a region of 1.4$\arcsec$ assuming M=1.8~M$_\odot$ and 0.46~M$_\odot$
for the hot core mass, respectively. When detected, we used the $^{13}C$ isotopologue to determine the 
total average molecular column density which is more reliable. As discussed in Sect. 4, the uncertainties in the molecular column
densities would add an error of a factor $\leq$4$-$5 to these values.

\subsection{Comparison with the Orion hot core}
The hot core associated with FIRS~2 is extraordinarily rich in complex molecules. 
We have detected glycolaldehyde and tentatively formamide that had been detected only in a few objects thus far. 
Glycolaldehyde was detected in Sgr B2 (Hollis et al. 2000, Halfen et al. 2006, Requena-Torres et al. 2008), 
the massive hot core G31.41+0.31 (Beltr\'an et al. 2009) and more recently towards the hot corino IRAS~16293-2422B (Jorgensen et al. 2012,
Zapata et al. 2013). Formamide has been detected in Orion (Motiyenko et al. 2012), Sgr~B2 (Belloche et al. 2013), a handful of 
massive prototypical hot cores (Adande et al. 2013) and IRAS~16293-2422 (Kahane et al. 2013). 
In Figs.~A.1-4, we compare the FIRS~2 spectrum with that of the Orion hot core as observed with the 30m telescope (Tercero et al.  2010).  
Surprisingly, there is a  good match between both spectra, both in number of lines and 
relative line intensities, suggesting that both hot  cores have similar chemical characteristics.  
Of course, this is a qualitative comparison and we need to put it on quantitative grounds to establish firm conclusions.

In Table~6 and Fig. 3 we compare the molecular column densities in FIRS~2 and Orion KL. For this comparison, in FIRS 2,  we
have used the average column densities in a region of 1.4$\arcsec$. The reason is that 
the sizes of the emitting regions are not determined for most of the species (see Table 3). One needs to know the rotation
temperature and the line opacities independently, to be able to determine the source size. In many cases we do not
have detected the $^{13}$C isotopologue that would allow us to derive the line opacities. In others, the number of 
unblended lines is not enough to determine the rotation temperature and we had to assume a reasonable value. 
On the contrary, 
the number of molecules per  beam is a quite well determined parameter as long as the emission is optically thin which is
a reasonable assumption for many complex species.
When detected, we used the $^{13}C$ isotopologue to determine the total average column density.
 
Towards Orion KL, the molecular emission
is formed of 4 physically different components: hot core, plateau, compact ridge and extended ridge. The plateau
has a chemistry characteristic of a shocked region and is considered to be associated with the shocks produced
by a past eruptive episode. The extended ridge is the parent cloud that hosts the cluster of
young stars. The compact ridge is a small ($\sim$15$\arcsec$) $``$U" shaped feature that is characterized for being very rich
in complex O-bearing molecules. The hot core is a very compact source (6$\arcsec$-10$\arcsec$) particularly rich
in complex N-bearing molecules. We cannot resolve different regions within the FIRS~2 hot core, therefore we consider 
that the FIRS~2 hot core is comparable to the hot core (HC) and the compact ridge (CR) Orion components together. 
In Table~6 we show the average column densities of the HC+CR Orion components
in a region of 15$\arcsec$. To estimate this value, we have summed up the column densities of the HC and CR weighted by
their respective filling factors: 0.44 for HC (assuming a size of 10$\arcsec$) and 1 for the CR. When possible, we
used the $^{13}$C isotopologues to determine the Orion column densities (see Table~6). We assume $^{12}$C/$^{13}$C=50
for Orion (Comito et al. 2005, Tercero et al. 2010).
Some authors, like Neill et al. (2013) and Comito et al. (2005) used a size of 10$\arcsec$ for the CR. We have multiplied their column densities
by 0.44 to account for the different adopted source sizes. For CH$_3$CHO we have adopted the  value of Turner (1991). 
This author did not distinguish among the different Orion components (hot core, plateau, compact ridge and extended ridge) and we have
taken the total value. For this reason this point is drawn as an empty circle in Fig. 3. In the case of D$_2$CO, Turner (1990) assumed a 
source size of 15$\arcsec$$\times$25$\arcsec$ that is consistent with ours.

In Fig. 3 we compare the column densities of FIRS 2 and Orion normalized to methanol. Normalization is required if one wants
to compare the relative abundances of the different species in the warm gas. Methanol is commonly used
to make this kind of normalization in hot cores ({\"O}berg et al. 2011). Instead we could have used methyl cyanide
for the normalization but we would have obtained essentially the same trend. As it is clearly seen in Table~6, it would simply produce
an increase in all the ratios by factor of 2.

Most of the relative molecular abundances in FIRS~2 agree with those in Orion 
within a factor of 5, which is quite good taking into account the uncertainty due to the unknown
spatial structure of the molecular emission and opacity effects. Only a few molecules have abundances significantly different in both
hot cores: CH$_3$OCHO, CH$_3$CH$_2$CN, HNCO and SO$_2$ have fractional abundances more than 10 times lower
in FIRS~2 than in Orion. In the case of CH$_3$OCHO, we have not detected the $^{13}$C isotopologues in FIRS~2 and our column density
could be slightly underestimated. As discussed in Sect. 4.3, our estimate is accurate within a factor of 4. Increasing the methyl formate
by a factor of 4  would push the methyl formate back to the region within the two black lines in Fig. 3, although still in the lower end.
In the case of HNCO, the lines of the main isotologue are blended with those of the $^{13}$C isotopologue preventing us from a direct estimate 
of the line opacities.
Looking at the top of the Figure, there is a group of molecules that are more 
abundant in FIRS~2 by more than a factor of 4$-$5. These molecules are: CH$_2$DOH, D$_2$CO and CH$_3$CHO. As commented above, 
the value of CH$_3$CHO in Orion is uncertain.
However the differences observed in CH$_2$DOH and D$_2$CO could be significant.
In the following Section, we discuss possible scenarios to explain these differences.


\begin{figure*}
\includegraphics{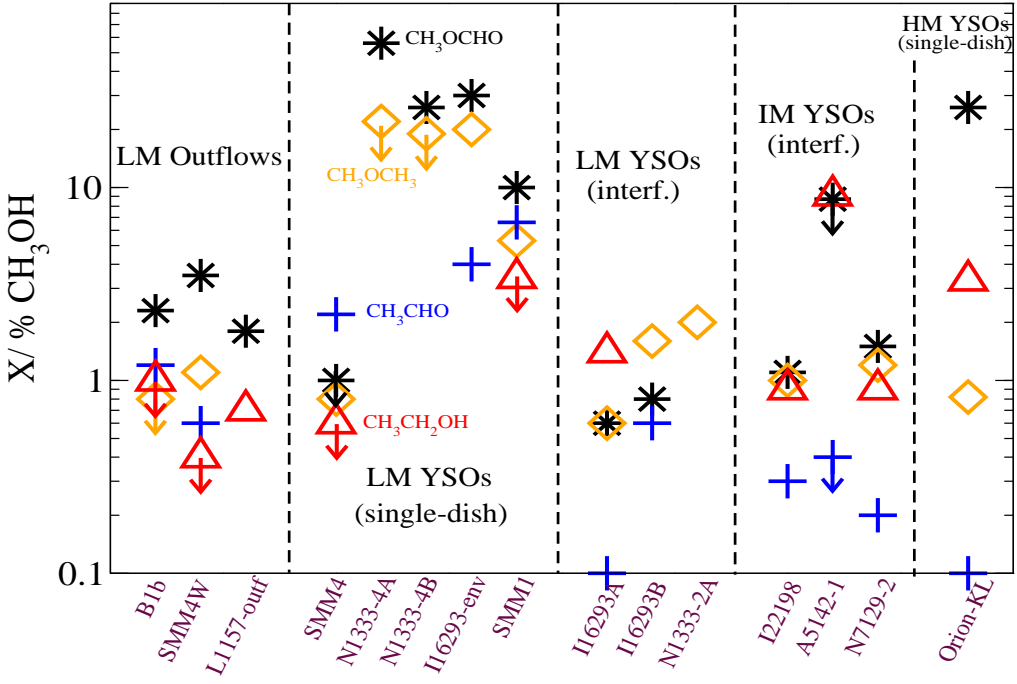}
\vspace{11.0cm}
      \caption{Abundance of complex organic molecules relative to CH$_3$OH for 
different types of sources, after {\"O}berg et al. (2011): from left to 
right, low-mass outflows, low-mass YSOs observed with single-dish 
(tracing envelopes), low-mass YSOs observed with interferometers 
(tracing the inner envelope/disk at $\sim$500 AU spatial scale), 
intermediate-mass YSOs observed with interferometers (also at ~500 AU 
spatial scale, this work and Palau et al. 2011), and Orion (references
in Table~6). Symbols with cold colors correspond to HCO-rich molecules (black: 
CH$_3$OCHO; blue: CH$_3$CHO), and symbols with warm colors correspond to 
CH$_{3/2}$-rich molecules (orange: CH$_3$OCH$_3$; red: CH$_3$CH$_2$OH).
}

         \label{Fig 5}
 \end{figure*}

\subsection{Sequential formation of complex molecules in hot cores/corinos}

Although gas phase reactions can play a role, surface chemistry dominates the formation of COMs.
The large sensitivity of surface chemistry on physical parameters, especially the dust temperature, can also
explain the observed chemical differentiation within a hot core and among different hot cores.
For example, Caselli et al. (1993) found that complex N-bearing species are more easily 
formed if the dust temperature (T$_{dust}$) is about 40~K during the collapse phase. This 
is primarily due to the higher mobility of carbon and the shortage of hydrogen (H, H$_2$) on
the grain surfaces. A similar argument has been used by {\"O}berg et al. (2011) to explain
the over-abundance of CH$_{2/3}$-rich molecules in hot cores and corinos compared with molecular ouflows.
{\"O}berg et al. (2011) studied a number of cores 
associated with outflows, low-mass cores observed with single-dish 
telescopes (tracing envelopes at $\sim5000$~AU spatial scales), low-mass 
cores observed with interferometers (tracing warmer material at 
$\sim500$~AU), and an average of high-mass hot molecular cores from 
Bisschop et al. (2007), and they found that HCO-rich molecules are in general 
more abundant than CH$_{3/2}$-rich molecules in outflows and envelopes of 
low-mass YSOs, while the situation is reversed for low-mass inner 
envelopes/disks and high-mass YSOs.
They proposed a sequential formation of complex molecules, starting with HCO-rich molecules as long 
as CO ice is abundant, followed by CH$_{3/2}$-rich molecules at higher ice temperatures.
 
In Fig.~4 we present an updated version of the figure by {\"O}berg et al. (2011) 
complemented with abundances of three 
intermediate-mass hot cores observed with interferometers (at $\sim500$~AU scales, 
from this work and Palau et al. 2011), and the new Orion data. 
In the Figure, symbols in warm colors (orange, red) correspond to CH$_{3/2}$-rich molecules, while 
symbols in cold colors (blue/black) correspond to HCO-rich molecules. 
Our new results for the intermediate-mass hot core FIRS~2 are in line with 
previous measurements towards IM hot cores at disk-scales ($\sim500$ AU). 
Putting all these data together, we do not find a clear trend for the abundances of the CH$_{3/2}$-rich and HCO-rich molecules 
with the stellar mass. Considering only interferometric
observations, the CH$_3$CHO abundance seems to decreases with the stellar mass. However,
the abundance of methyl formate, the only other HCO-rich molecule considered, is larger in 
Orion than in IM hot cores and hot corinos. 
Orion is the only source in which we have been able to use the $^{13}$C isotopologue to derive the methyl formate
abundance. If the methyl formate lines were typically optically thick in hot cores/corinos, this would have produced an underestimate 
of the CH$_3$OCHO abundance in the rest of objects. Using the upper limit to the emission of the $^{13}$C isotopologue,
we have estimated that the methyl formate abundance in FIRS~2 is accurate within a factor of 4. Even considering an
uncertainty of a factor of 4 in the CH$_3$OCHO abundances towards low mass and IM YSOs, the [CH$_3$OCHO]/[CH$_3$OH] ratio 
would be larger in Orion than in these objects. Besides, the [CH$_3$OCHO]/[CH$_3$OH] ratio seems to increase from the low-mass to the IM sample.
We can only compare two HCO-rich molecules in Fig.~4. Therefore we cannot know if methyl formate is
a special case (see also Taquet et al. 2012). 
Further observations in IM and massive hot cores are required to have a deeper insight into the chemistry of this molecule and other 
HCO-rich species.


We have yet to discuss the deuterium fractions 
in FIRS 2 and Orion. Deuterated species whose deuteration requires surface chemistry, 
such as D$_2$CO, present higher abundances in the warm regions associated with low-mass protostars. 
Loinard et al. (2003) searched for the doubly deuterated form of formaldehyde (D$_2$CO) in a 
large sample of young stellar objects. D$_2$CO was detected in all low-mass protostars, 
with [D$_2$CO]/[H$_2$CO] ratios of 2$-$40\%. On the other hand, no detection was obtained towards 
massive protostars (where [D$_2$CO]/[H$_2$CO]$<$0.5\%). This is consistent with	
the value reported by Turner (1990) in Orion. 
If the hot cores associated with massive stars are older and/or significantly denser than those 
surrounding low-mass objects, gas phase chemistry could have had the time to reset the deuterium 
fractions to values close to the cosmic D/H ratio. The longer time of the evaporated molecules 
in gas phase in would also affect the abundances of the COMs that can be formed in 
warm gas-phase chemistry (e.g. CH$_2$CHCN). 

An alternative explanation could be that the
temperature of the envelope material accreting onto the high mass protostars is larger than 30~K 
(Fontani et al. 2002). In this case the deuterium fractionation efficiency in the gas phase is 
strongly reduced before the hot core phase as the standard H$_2$D$^+$ route 
(H$_3^+$+HD $\rightarrow$ H$_2$D$^+$ + H$_2$) also starts to proceed from left to right . 
Moreover, at T$_{dust}$$>$30~K there is no way that deuteration can be enhanced on the grain surfaces.  
The reason is that H and D are easily evaporated before reacting on the surface (see
e.g. Cazaux et al. 2011).  The higher temperature during the collapse would also affect the formation
of COMs in grain mantles. As CO does not efficiently freezes-out at these values of T$_{dust}$, the formation
of HCO-rich molecules would be strongly diminished too. As commented above, methyl formate could
be an exception of this rule.

\subsection{Carbon budget}
We have used the column densities shown in Table~3 to estimate the Carbon
gas phase budget in the FIRS~2 hot core. Apart from CO and its isotopologues, most
of the Carbon  is locked in methanol with a total column density of 
$\sim$3.4$\times$10$^{20}$~cm$^{-2}$ in the compact component of the hot core. Taking into account 
the rest of molecules, the Carbon column density reaches to 
$\sim$4$\times$10$^{20}$~cm$^{-2}$ in this small region. One problem to derive absolute abundances
is that we do not know the total column density of gas in the inner R$<$0.05$\arcsec$ region. 
Based on higher angular resolution observations, FU05 derived
a size of 0.72$\arcsec$$\times$0.52$\arcsec$ for the continuum source that account for 75\% of the
continuum emission and a point source that contributes with 25\% of the flux. Adopting this value
for the fraction of continuum flux coming from the compact component of the hot core and a total gas 
mass of 0.46 M$_{\odot}$ (see Section 6.1), 
the total column density of molecular hydrogen would be 2$\times$10$^{25}$~cm$^{-2}$ and 
the fractional abundance of the Carbon locked in COMs relative
to H nuclei, $\sim$1.0$\times$10$^{-5}$. This value is 36 times lower than the solar value of 
the Carbon abundance (C/H$\sim$3.6$\times$10$^{-4}$, Anders \& Grevesse 1989) and suggests that
methanol and COMs (at least those considered in this paper) are not the main reservoirs of Carbon in hot cores.


It is also useful to compare methanol with C$^{18}$O. Again we have the problem of the unknown 
spatial distribution of the C$^{18}$O emission within our interferometric beam.  
We can derive a lower limit to the C$^{18}$O
abundance assuming that the emission is optically thin and fills the beam. 
Adopting T$_{\rm rot}$=400~K, we derive N(C$^{18}$O)=1.5$\times$10$^{17}$~cm$^{-2}$ and
N(CO)=9.4$\times$10$^{19}$~cm$^{-2}$. This solution, however, is not compatible with the non-detection
of the  C$^{18}$O 9$\rightarrow$8 line by Fuente et al. (2012) and we cannot reconcile the PdBI and Herschel
results by lowering the temperature down to T$_{\rm rot}$=200~K. The optically thick solution seems more reasonable.
Assuming again that the gas temperature is 400~K and that the C$^{18}$O emission is optically thick
(T$_B$$\approx$T$_{\rm rot}$), we derive an effective diameter of 0.15$\arcsec$ for the  C$^{18}$O emitting region. In this case, 
our results are compatible with the Herschel upper limit to the
C$^{18}$O 9$\rightarrow$8 line and the column density of CO would be $>$10$^{22}$~cm$^{-2}$ in the compact component of
the hot core.

\section{Conclusions}
FIRS~2 is probably the youngest IM protostar studied thus far and, as such, an excellent template upon which to base interpretations of  
other IM and massive star forming regions. The interferometric observations presented in this paper proved that this IM protostar hosts a 
hot core extraordinarily rich in complex molecules.The gas kinetic temperature as measured with methyl cyanide, is around 400 K, similar to that 
of the Orion hot core and higher than typical rotation temperatures in hot corinos. A detailed comparison of the chemistry of FIRS~2 with
Orion shows that the fractional abundances of most molecules relative to that of CH$_3$OH agree within a factor of 5, that is reasonable
taking into account the uncertainties of our column density estimates.
Only CH$_3$CH$_2$CN, HNCO, SO$_2$, CH$_2$DOH, D$_2$CO and CH$_3$CHO present a significant disagreement, with CH$_3$CH$_2$CN, HNCO and SO$_2$
being more abundant in Orion, and CH$_2$DOH, D$_2$CO and CH$_3$CHO, in FIRS~2. 
Since the physical conditions are similar in both hot cores, only different initial conditions 
(warmer pre-collapse and collapse phase in the case of Orion) and/or different crossing time of the gas in the hot core can explain this behavior.
FIRS~2 is a YSO in which most of the accreting envelope still maintains an overall temperature lower than 30~K, allowing deuteration
to proceed unhindered, while complex N-bearing molecules like CH$_3$CH$_2$CN are not efficiently formed, unlike the Orion hot core.
%


\begin{acknowledgements}
We thank the Spanish MINECO for funding support from grants CSD2009-00038, AYA2009-07304, and AYA2012-32032.
DJ is supported by the National Research Council of Canada 
and by a Natural Sciences and Engineering Research Council of Canada (NSERC) Discovery Grant.
A.P. is supported by the Spanish MICINN grant AYA2011-30228-C03-02 
(co-funded with FEDER funds), and by the AGAUR grant 2009SGR1172 (Catalonia). 
\end{acknowledgements}

\newpage
\begin{appendix} 
\section{Tables and Figures}

\onecolumn
\begin{longtable}{l|l|l}
\multicolumn{3}{c}{Table A.1.Line identifications} \\
\hline \hline
Freq (MHz) & \multicolumn{1}{c}{Molecular lines$^1$}  &  \multicolumn{1}{c}{Comments$^2$}\\
\hline 
\endfirsthead
\hline
Freq (MHz) & \multicolumn{1}{c}{Molecular lines$^1$}  &  \multicolumn{1}{c}{Comments$^2$}\\
\hline
\endhead
\hline \multicolumn{3}{l}{\textit{Continued on next page}} \\
\endfoot
\hline \hline
\endlastfoot
 218132     & EE$-$(CH$_3$)$_2$CO {\scriptsize 20$_{2,18}$-19$_{3,17}$,20$_{3,18}$-19$_{3,17}$,20$_{2,18}$-19$_{2,17}$, 20$_{3,18}$-19$_{2,17}$},  &       \\
             & aGg'-(CH$_2$OH)$_2$, HCOOH  &       \\
 218143     & U                                               &      \\
 218157     & U                                               &       \\
 218163     &  AA$-$(CH$_3$)$_2$CO {\scriptsize 20$_{2,18}$-19$_{3,17}$},CH$_2$OHCHO    &    \\
 218181     &  aGg'-(CH$_2$OH)$_2$ {\scriptsize 18$_{4,0}$-18$_{3,1}$,18$_{5,0}$-18$_{4,1}$}                  &       \\
 218199     &  O$^{13}$CS {\scriptsize 18-17}                 &      \\
 218222     &  p-H$_2$CO  {\scriptsize 3$_{0,3}$-2$_{0,2}$}   &       \\
 218238     &  aGg'-(CH$_2$OH)$_2$ {\scriptsize 22$_{17,5}$-21$_{17,4}$,22$_{17,6}$-21$_{17,5}$}     &       \\
 218259     &  A-CH$_3$OCHO  {\scriptsize 31$_{9,23}$-31$_{8,24}$}, CH$_2$OHCHO                   &       \\
 218282     &  E-CH$_3$OCHO  {\scriptsize 17$_{-3,14}$-16$_{-3,13}$}                                &        \\
 218298     &  A-CH$_3$OCHO  {\scriptsize 17$_{3,13}$-16$_{3,13}$}    &           \\
 218316     &  CH$_2$DOH {\scriptsize 5$_{2,4}$-5$_{1,5}$}     &            \\
 218324     &  HC$_3$N {\scriptsize 24-23}, CH$_2$OHCHO   &           \\
 218336     &   U          &                \\
 218371     &   aGg'-(CH$_2$OH)$_2$ {\scriptsize 22$_{4,19}$-21$_{4,18}$}  &      \\
 218380     &   aGg'-(CH$_2$OH)$_2$ {\scriptsize 22$_{15,7}$-21$_{15,6}$,22$_{15,8}$-21$_{15,7}$}  &     \\
 218389     &   CH$_3$CH$_2$CN {\scriptsize 24$_{3,21}$-23$_{3,20}$}, CH$_3$OCOD  &                \\
 218404     &  A-CH$_2$DOCOH {\scriptsize 18$_{5,14}$-17$_{5,13}$}, S-CH$_2$DOCOH, CH$_3$OCOD    &                \\
 218440     &  E-CH$_3$OH {\scriptsize 4$_{2}$-3$_{1}$}    &         \\
 218460     &  NH$_2$CHO {\scriptsize 10$_{1,9}$-9$_{1,8}$}, T-CH$_3$CH$_2$OH &                    \\
 218475     &  p-H$_2$CO {\scriptsize 3$_{2,2}$-2$_{2,1}$}                  &                        \\
 218493     &  CH$_3$OCH$_3$ {\scriptsize 23$_{3,21}$-23$_{2,22}$}          &                        \\
 218519     &  E-CH$_3$OCOD {\scriptsize 18$_{6,13}$-17$_{6,12}$}           &                     \\
 218526     &  A-CH$_3$OCOD {\scriptsize 18$_{6,13}$-17$_{6,12}$}          &                      \\
 218541     &  HCOOD {\scriptsize 10$_{3,7}$-9$_{3,6}$}                    &                           \\
 218554     &  T-CH$_3$CH$_2$OH {\scriptsize 21$_{5,16}$-21$_{4,17}$}     &                         \\
 218576     &  aGg'-(CH$_2$OH)$_2$ {\scriptsize 22$_{13,10}$-21$_{13,9}$,22$_{13,9}$-21$_{13,8}$}  &           \\
 218585     &  A-CH$_3$OCHO  {\scriptsize 36$_{9,28}$-36$_{8,29}$}  &                         \\
 218594     &  A-CH$_3$OCHO  {\scriptsize 27$_{7,21}$-27$_{5,22}$}  &                         \\
 218611     &  U     &                        \\
 218632     &  EE$-$(CH$_3$)$_2$CO {\scriptsize 12$_{9,4}$-11$_{8,3}$}, CH$_3$OCHO  {\scriptsize 27$_{7,21}$-27$_{-5,22}$} &            \\
 218655     &  T-CH$_3$CH$_2$OH {\scriptsize 7$_{2,5}$-6$_{1,6}$}, CH$_3$OCHO  $\nu_t$=1   &                 \\
 218672     &  A-CH$_3$CHO $\nu_t$=1 {\scriptsize 28$_{7,22}$-28$_{5,23}$}  &      \\
 218681     &  E-CH$_3$CHO $\nu_t$=1 {\scriptsize 33$_{-9,25}$-33$_{-8,26}$}  &        \\
 218688     &  CH$_2$DOH {\scriptsize 17$_{4,13}$-16$_{5,12}$}      &                 \\
 218696     &  CH$_2$DOH {\scriptsize 17$_{4,14}$-16$_{5,11}$}       &                 \\
 218707     &  aGg'-(CH$_2$OH)$_2$ {\scriptsize 22$_{12,10}$-21$_{12,9}$, 22$_{12,11}$-21$_{12,10}$}  &              \\
 218730     &  S-CH$_2$DOCOH  {\scriptsize 19$_{4,16}$-18$_{4,15}$}, wing of C$_3$H$_2$ {\scriptsize 6$_{0,6}$-5$_{1,5}$}?        &                  \\
 218738     &  CH$_3$CH$_2$CN  $\nu_{13}$=1 /$\nu_{21}$=1 {\scriptsize 24$_{3,22}$-23$_{3,21}$,24$_{3,21}$-23$_{3,20}$}, CH$_3$CHO $\nu_t$=1          &                 \\
 218760     &  p-H$_2$CO  {\scriptsize 3$_{2,1}$-2$_{2,0}$}                &                            \\
 218774     &  AE$-$(CH$_3$)$_2$CO {\scriptsize 12$_{9,4}$-11$_{8,3}$}     &                 \\ 
 218782     &  U                        &              \\
 218796     &  CH$_2$DOH {\scriptsize 20$_{1,20}$-19$_{2,17}$}, G-CH$_3$CH$_2$OH, CH$_3$OCOD    &                   \\
 218815     &  E-CH$_3$OH $\nu_t$=2 {\scriptsize 16$_{-8}$-15$_{-9}$}, AA$-$(CH$_3$)$_2$CO &                           \\
 218830     &  E-CH$_3$OCHO  $\nu_t$=1 {\scriptsize 18$_{13,5}$-17$_{13,4}$}       &          \\
 218850     &  U                    &                             \\
 218860     &  HC$_3$N $\nu_7$=1 {\scriptsize 24$_{-1}$-23$_{-1}$}          &                \\
 218873     &  aGg'-(CH$_2$OH)$_2$ {\scriptsize 22$_{11,12}$-21$_{11,11}$, 22$_{11,11}$-21$_{11,10}$ }    &         \\
 218903     &  OCS   {\scriptsize 18-17}  &                \\
 218914     &  U &         \\
 218922     &  U                 &                   \\
 218938     &  HCOOH {\scriptsize 8$_{1,8}$-7$_{0,7}$}      &                   \\
 218955     &  A-$^{13}$CH$_3$OCHO  {\scriptsize 40$_{8,33}$-39$_{9,30}$}  &                            \\
 218967     &  E-CH$_3$OCHO  $\nu_t$=1 {\scriptsize 18$_{12,6}$-17$_{12,5}$}, CH$_2$OHCHO, CH$_2$DOCOH,   &         \\
            &  $^{13}$CH$_3$OCHO   &         \\
 218980     &  HNCO {\scriptsize 10$_{1,10}$-9$_{1,9}$}, HN$^{13}$CO    &                             \\
 219007     &  S-CH$_2$DOCOH {\scriptsize 32$_{9,23}$-32$_{8,24}$}  &     \\ 
 219019     &  S-CH$_2$DCH$_2$CN  {\scriptsize 27$_{1,27}$-26$_{1,26}$} &                \\ 
 219037     &  CH$_2$OHCHO {\scriptsize 39$_{10,30}$-39$_{9,31}$}      &               \\ 
 219067     &  CH$_2$OHCHO {\scriptsize 27$_{2,25}$-27$_{2,26}$,27$_{2,25}$-27$_{1,26}$}, CH$_3$OCHO  $\nu_t$=1, $^{13}$CH$_3$OCHO, &   \\  
            &  CH$_2$DCH$_2$CN &   \\  
 219078     &  E-CH$_3$OCHO  $\nu_t$=1 {\scriptsize 28$_{3,25}$-28$_{2,26}$}, AE-(CH$_3$)$_2$CO, EA-(CH$_3$)$_2$CO        \\ 
 219090     &  aGg'-(CH$_2$OH)$_2$ {\scriptsize 22$_{10,13}$-21$_{10,12}$,22$_{10,12}$-21$_{10,11}$},CH$_3$OCHO            \\
 219098     &  A-CH$_2$DOCOH {\scriptsize 18$_{6,12}$-17$_{6,11}$}, $^{13}$CH$_3$OCHO, CH$_2$OHCHO, HCOOH   &   \\
 219108     &  A-CH$_3$OCHO   {\scriptsize 34$_{7,28}$-34$_{5,29}$}               \\ 
 219115     &  U                                          \\ 
%
 219123     &  CH$_2$OHCHO {\scriptsize 29$_{3,26}$-29$_{2,27}$}     &                \\ 
 219132     &  S-CH$_2$DOCOH {\scriptsize 19$_{8,12}$-18$_{8,11}$,19$_{8,11}$-18$_{8,10}$}        &         \\ 
 219143     &  U                    &                   \\ 
 219154     &  A-CH$_3$OCHO  $\nu_t$=1 {\scriptsize 18$_{11,7}$-17$_{11,6}$, 10$_{4,6}$-9$_{3,6}$ }    &                     \\ 
 219171     &  OCS $\nu_2$=1 {\scriptsize 18$_{-1}$-17$_{-1}$ },  T-CH$_3$CH$_2$OH, HC$_3$N $\nu_7$=1  &     \\ 
 219188     &  EE$-$(CH$_3$)$_2$CO {\scriptsize 5$_{5,1}$-5$_{4,1}$}   &         \\ 
 219196     &  U               &     \\ 
 219204     &  CH$_2$DOH {\scriptsize 20$_{5,16}$-19$_{6,13}$,20$_{5,15}$-19$_{6,14}$}     &                               \\ 
 219220     &  AE-(CH$_3$)$_2$CO {\scriptsize 21$_{1,20}$-20$_{2,19}$,21$_{2,20}$-20$_{1,19}$}, EA-(CH$_3$)$_2$CO,   &     \\ 
            &  CH$_3$CH$_2$CN $\nu_{13}$/$\nu_{21}$=1   &     \\ 
 219228     &  CH$_2$OHCHO   {\scriptsize 31$_{5,27}$-31$_{3,28}$}                   &           \\ 
 219242     &  A-CH$_3$OCOD {\scriptsize 21$_{0,21}$-20$_{1,20}$, 21$_{1,21}$-20$_{1,20}$, 21$_{0,21}$-20$_{0,20}$, 21$_{1,21}$-20$_{0,20}$ } , 
EE-(CH$_3$)$_2$CO            &           \\ 
 219264     &  AA-(CH$_3$)$_2$CO {\scriptsize 21$_{1,20}$-20$_{2,19}$,  21$_{2,20}$-20$_{1,19}$} , CH$_3$OCHO            &            \\ 
 219276     &  SO$_2$  {\scriptsize 22$_{7,15}$-23$_{6,18}$}                 &        \\ 
 219295     &  S-CH$_2$DCH$_2$CN  {\scriptsize 26$_{3,24}$-25$_{3,23}$} , CH$_3$CHO       &          \\
 219305     &  CH$_3$OCH$_3$  {\scriptsize 31$_{6,26}$-30$_{7,23}$}, CH$_2$OHCHO          &          \\ 
 219311     &  AA-(CH$_3$)$_2$CO  {\scriptsize 12$_{9,4}$-11$_{8,3}$}                     &        \\
 219321     &  CH$_3$CH$_2$CN  $\nu_{13}$=1 /$\nu_{21}$=1 {\scriptsize 25$_{1,25}$-24$_{1,24}$, 25$_{2,25}$-24$_{2,23}$} , CH$_3$OCHO  $\nu_t$=1     &                          \\ 
 219330     &  E-CH$_3$OCHO  $\nu_t$=1  {\scriptsize 18$_{-15,4}$-17$_{-15,3}$}                &       \\
 219341     &  H$^{13}$COOH  {\scriptsize 10$_{0,10}$-9$_{0,9}$}                &         \\
 219355     &  $^{34}$SO$_2$  {\scriptsize 11$_{1,10}$-10$_{0,10}$}   &        \\ 
 219370     &  S-CH$_2$DCH$_2$CN  {\scriptsize 26$_{12,14}$-25$_{12,13}$,  26$_{12,15}$-25$_{12,14}$}                 &         \\
 219386     &   aGg'-(CH$_2$OH)$_2$ {\scriptsize 22$_{9,14}$-21$_{9,13}$,  22$_{9,13}$-21$_{9,12}$}  &         \\
 219389     &   U &         \\
 219398     &   OCS  $\nu_2$=1   {\scriptsize 18-17}         &          \\
 219411     &   E-CH$_3$OCHO  $\nu_t$=1  {\scriptsize 18$_{0,8}$-17$_{0,7}$} &                   \\
 219441     &   U                &                \\
 219467     &   CH$_3$OCH$_3$ {\scriptsize 28$_{5,24}$-27$_{6,21}$}, CH$_3$CH$_2$CN, AE-(CH$_3$)$_2$CO,
 EA-(CH$_3$)$_2$CO   &       \\
 219479     &  CH$_3$OCHO  $\nu_t$=1 {\scriptsize 18$_{-14,5}$-17$_{-14,4}$},CH$_3$OCHO   &          \\
 219490     &  DC$_3$N   {\scriptsize 26-25}                                              &           \\
 219506     &  CH$_3$CH$_2$CN   {\scriptsize 24$_{2,22}$-23$_{2,21}$}                       &          \\
 219512     &   U                         &          \\  
 219541     &   aGg'-(CH$_2$OH)$_2$  {\scriptsize 22$_{2,21}$-21$_{2,20}$}         &          \\  
 219550     &   HNCO   {\scriptsize 10$_{4,6}$-9$_{4,5}$, 10$_{4,7}$-9$_{4,6}$}, CH$_3$OH               &      Partially overlapped with C$^{18}$O  \\
 219560     &   C$^{18}$O  {\scriptsize 2-1}     &                        \\
 219570     &  A-CH$_3$OCHO  $\nu_t$=1  {\scriptsize 18$_{15,4}$-17$_{15,3}$, 18$_{15,3}$-17$_{15,2}$,
18$_{16,2}$-17$_{16,1}$, 18$_{16,3}$-17$_{16,2}$ }     &       Partially overlapped with  C$^{18}$O  \\
 219580     &  aGg'-(CH$_2$OH)$_2$  {\scriptsize 22$_{1,21}$-21$_{1,20}$} , CH$_3$OCHO               &               \\
 219584     &  A-CH$_3$OCHO  $\nu_t$=1 {\scriptsize 18$_{13,6}$-17$_{13,5}$, 18$_{13,5}$-17$_{13,4}$} , CH$_3$OCHO                    &                 \\
 219593     &  E-CH$_3$OCHO  {\scriptsize 28$_{-9,19}$-28$_{-8,20}$}, CH$_3$OCOD,CH$_2$OHCHO   &                  \\
 219600     &  E-CH$_3$OCHO  {\scriptsize 30$_{9,22}$-30$_{8,23}$}             &                \\
 219607     &  E-CH$_3$OCHO  {\scriptsize 30$_{5,26}$-30$_{-3,27}$}, EE-(CH$_3$)$_2$CO             &              \\
 219622     &  A-CH$_3$OCHO  $\nu_t$=1 {\scriptsize 18$_{12,6}$-17$_{12,5}$, 18$_{12,7}$-17$_{12,6}$ }                &                  \\
 219642     &  E-CH$_3$OCHO  $\nu_t$=1 {\scriptsize 18$_{-13,6}$-17$_{-13,5}$}            &                  \\
 219657     &  HNCO {\scriptsize 10$_{3,7}$-9$_{3,6}$}, HN$^{13}$CO          &                 \\
 219674     &  A-CH$_3$OCHO   {\scriptsize 30$_{5,26}$-30$_{3,27}$}                    &                 \\
 219696     &  A-CH$_3$OCHO  $\nu_t$=1 {\scriptsize 18$_{11,8}$-17$_{11,7}$, 18$_{11,7}$-17$_{11,6}$ }      &                   \\
 219705     &  A-CH$_3$OCHO  $\nu_t$=1 {\scriptsize 18$_{4,15}$-17$_{4,14}$}        &                  \\
 219720     &  E-CH$_3$OCHO  $\nu_t$=1 {\scriptsize 32$_{-9,24}$-32$_{-8,25}$, 28$_{-4,25}$-28$_{2,26}$}        &                  \\
 219735     &  HNCO {\scriptsize 10$_{2,9}$-9$_{2,8}$, 10$_{2,8}$-9$_{2,7}$}, HN$^{13}$CO  &                 \\
 219751     &   AA-(CH$_3$)$_2$CO {\scriptsize 33$_{8,25}$-33$_{3,26}$, 33$_{9,25}$-33$_{7,26}$ }       &             \\
 219764     &   E-CH$_3$OCHO  $\nu_t$=1 {\scriptsize 18$_{9,9}$-17$_{9,8}$},  aGg'-(CH$_2$OH)$_2$ {\scriptsize 20$_{4,16}$-19$_{4,15}$}        &             \\
 219780     &   A-CH$_3$CHO $\nu_t$=2  {\scriptsize 11$_{1,10}$-10$_{1,9}$}, CH$_3$OCHO  $\nu_t$=1          &             \\
 219798     &   HNCO  {\scriptsize 10$_{1,10}$-9$_{0,9}$}              &                  \\
 219804     &   aGg'-(CH$_2$OH)$_2$  {\scriptsize 22$_{8,15}$-21$_{8,14}$}, HN$^{13}$CO   &                  \\     
 219823     &   A-CH$_3$OCHO  $\nu_t$=1 {\scriptsize 18$_{10,9}$-17$_{10,8}$, 18$_{10,8}$-17$_{10,7}$ } , AE-(CH$_3$)$_2$CO, EA-(CH$_3$)$_2$CO,  &       \\
            &   CH$_3$CHO   &       \\
 219863     &    U                 &                  \\
 219876     &    U                 &        \\
 219894     &    U                 &                \\
 219900     &  U     &   CH$_2$CHCN {\scriptsize 10$_{2,9}$-9$_{1,8}$}    \\
 219908     &  o-H$_2$$^{13}$CO  {\scriptsize 3$_{1,2}$-2$_{1,1}$}    &                 \\
 219916     &  U                   &     E-CD$_3$CN {\scriptsize 14$_{5}$-13$_{5}$}                \\
 219949     &  SO {\scriptsize 5$_{6}$-4$_{5}$}                             &                  \\
 219965     &  U            &    A-CD$_3$CN {\scriptsize 14$_{3}$-13$_{3}$}                \\
 219980     &  U            &    E-CD$_3$CN {\scriptsize 14$_{2}$-13$_{2}$}                  \\
 219991     &  U            &    A-CD$_3$CN {\scriptsize 14$_{0}$-13$_{0}$}                     \\
 220030     &  CH$_3$OCHO  $\nu_t$=1  {\scriptsize 18$_{9,10}$-17$_{9,9}$, 18$_{9,9}$-17$_{9,8}$}     &                  \\
 220037     &  HCOOH  {\scriptsize 10$_{0,10}$-9$_{0,9}$}, CH$_3$OCHO  $\nu_t$=1 &                \\  

 220054     &  CH$_2$OHCHO  {\scriptsize 35$_{10,26}$-35$_{9,27}$},                &                \\
 220077     &  E-CH$_3$OH   {\scriptsize 8$_{0,8}$-7$_{1,6}$}, CH$_2$DOH                  &                  \\
 220094     &   aGg'-(CH$_2$OH)$_2$  {\scriptsize 24$_{11,13}$-24$_{10,15}$,  24$_{11,14}$-24$_{10,14}$} &   \\
 220140     &   U                  &                    \\
 220155     &  T-CH$_3$CH$_2$OH  {\scriptsize 24$_{3,22}$-24$_{2,23}$}        &                 \\
 220167     &  E-CH$_3$OCHO   {\scriptsize 17$_{-4,13}$-16$_{-4,12}$}, AA-(CH$_3$)$_2$CO             &               \\
 220178     &  o-H$_2$CCO  {\scriptsize 11$_{1,11}$-10$_{1,10}$}  &                                 \\
 220190     &  A-CH$_3$OCHO  {\scriptsize 17$_{4,13}$-16$_{4,12}$}          &                   \\
 220196     &  CH$_2$OHCHO  {\scriptsize 7$_{6,2}$-6$_{5,1}$, 7$_{6,1}$-6$_{5,2}$}               &                   \\
 220203     &  CH$_2$OHCHO   {\scriptsize 11$_{4,7}$-10$_{3,8}$}             &                   \\
 220225     &   U                                &                   \\
 220235     &   E-CH$_3$CN  {\scriptsize 12$_{11}$-11$_{11}$}             &                   \\
 220243     &   U                                   &            \\
 220258     &   E-CH$_3$OCHO  $\nu_t$=1 {\scriptsize 18$_{8,10}$-17$_{8,9}$} , CH$_2$DOCOH       &                   \\
 220270     &   U                  &                   \\
 220279     &   U                  &                   \\
 220300     &   A-CH$_2$DOCOH  {\scriptsize 20$_{1,19}$-19$_{1,18}$}  , CH$_3$$^{13}$CN  &        \\
 220307     &   E-CH$_3$OCHO  $\nu_t$=1 {\scriptsize 18$_{-10,9}$-17$_{-10,8}$}      &                    \\
 220321     &   A-$^{13}$CH$_3$OH {\scriptsize 17$_{7,11}$-18$_{6,12}$,17$_{7,10}$-18$_{6,13}$}, CH$_3$CN        &                    \\
 220332     &   U                   &                                \\ 
 220345     &   U                   & CH$_2$CHCN  {\scriptsize 9$_{4,5}$-10$_{3,8}$}, G+CH$_3$CH$_2$OH          \\
            &                       & CH$_2$DOCOH             \\
 220355     &  AE-(CH$_3$)$_2$CO {\scriptsize 22$_{1,22}$-21$_{1,21}$,22$_{0,22}$-21$_{0,21}$}, EA-(CH$_3$)$_2$CO                          &        \\
 220361     &  EE-(CH$_3$)$_2$CO {\scriptsize 22$_{0,22}$-21$_{1,21}$,22$_{1,22}$-21$_{1,21}$,22$_{1,22}$-21$_{0,21}$,22$_{0,22}$-21$_{0,21}$} &  \\
 220368     &  A-CH$_3$OCHO  $\nu_t$=1  {\scriptsize 18$_{8,11}$-17$_{8,10}$, 18$_{8,10}$-17$_{8,9}$} ,  AA-(CH$_3$)$_2$CO, EA-(CH$_3$)$_2$CO      &                  \\
 220390     & H$^{13}$CC$_2$N {\scriptsize 25-24}             &        Partially blended with $^{13}$CO  \\
 220398     & $^{13}$CO {\scriptsize 2-1}, CH$_3$OH                             &            \\
 220403     & A-CH$_3$CN {\scriptsize 12$_9$-11$_9$}, aGg'-(CH$_2$OH)$_2$                         &         Partially blended with $^{13}$CO  \\
 220409     & E-CH$_3$OCHO  $\nu_t$=1 {\scriptsize 18$_{-4,15}$-17$_{-4,14}$}   &          Partially blended with $^{13}$CO  \\
 220417     & A-CH$_3$OCHO  $\nu_t$=1 {\scriptsize 18$_{3,16}$-17$_{2,15}$}  &                  \\
 220427     &   U    &        \\
 220446     &   A-CH$_3$CHO {\scriptsize 21$_{4,18}$-22$_{1,21}$}  &                           \\
 220466     &   CH$_2$OHCHO {\scriptsize 20$_{2,18}$-19$_{3,17}$},  AE-(CH$_3$)$_2$CO         &                  \\
 220476     &   E-CH$_3$CN {\scriptsize 12$_8$-11$_8$}             &                  \\
 220486     &   A-CH$_3$$^{13}$CN  {\scriptsize 12$_6$-11$_6$}         &                 \\
 220498     &   aGg'-(CH$_2$OH)$_2$ {\scriptsize 22$_{7,15}$-21$_{7,14}$}        &                  \\
 220504     &  $^{33}$SO$_2$  {\scriptsize 27$_{1}$-27$_{28}$, 27$_{1}$-27$_{27}$}  &   \\
 220525     &   A-CH$_3$OCHO  $\nu_t$=1 {\scriptsize 10$_{4,6}$-9$_{3,7}$} &   \\
 220532     &   E-CH$_3$$^{13}$CN  {\scriptsize 12$_{5}$-11$_{5}$}            &                   \\ 
 220540     &   E-CH$_3$CN {\scriptsize 12$_{7}$-11$_{7}$}                   &                   \\
 220551     &   CH$_2$DOH {\scriptsize 17$_{1,16}$-17$_{0,17}$}              &                \\
 220570     &   E-CH$_3$$^{13}$CN {\scriptsize 21$_{4,18}$-22$_{1,21}$}    &                \\
 220585     &   HNCO {\scriptsize 10$_{1,9}$-9$_{1,8}$}            &                \\
 220595     &   A-CH$_3$CN {\scriptsize 12$_{6}$-11$_{6}$}               &          \\
 220602     &   T-CH$_3$CH$_2$OH  {\scriptsize 13$_{1,13}$-12$_{0,12}$}           &           \\
 220606     &  E-CH$_3$OCHO   $\nu$=1  {\scriptsize 18$_{-3,16}$-17$_{2,15}$}           &                         \\
 220622     &  E-CH$_3$$^{13}$CN {\scriptsize 12$_{2}$-11$_{2}$}      &                         \\
 220640     &  E-CH$_3$CN {\scriptsize 12$_{5}$-11$_{5}$}                  &                        \\
 220647     &  E-CH$_3$OCHO  $\nu_t$=1  {\scriptsize 18$_{-9,10}$-17$_{-9,9}$}               &                 \\
 220662     &  CH$_3$CH$_2$CN {\scriptsize 25$_{2,24}$-24$_{2,23}$}, EE- (CH$_3$)$_2$CO           &              \\
 220679     &  E-CH$_3$CN {\scriptsize 12$_{4}$-11$_{4}$}        &                                         \\
 220695     &  AE-(CH$_3$)$_2$CO {\scriptsize 29$_{5,25}$-29$_{4,26}$}, {\scriptsize 29$_{4,25}$-29$_{3,26}$}  &   CH$_2$CHCN  $\nu$=1 {\scriptsize 10$_{3,7}$-10$_{2,8}$}    \\
 220701     & CH$_2$OHCHO {\scriptsize 40$_{9,32}$-40$_{8,33}$}  &  \\ 
 220709     &  A-CH$_3$CN  {\scriptsize 12$_{3}$-11$_{3}$}                    &                 \\
 220730     &  E-CH$_3$CN  {\scriptsize 12$_{2}$-11$_{2}$}                  &                 \\
 220745     &  A-CH$_3$CN {\scriptsize 12$_{0}$-11$_{0}$}, E-CH$_3$CN {\scriptsize 12$_{1}$-11$_{1}$}                   &                 \\
 220764     &  EE-(CH$_3$)$_2$CO  {\scriptsize 11$_{11,9}$-10$_{10,0}$}    &                \\
 220785     &  E-CH$_3$OCHO  {\scriptsize 28$_{-3,25}$-28$_{3,26}$}, CH$_3$OCHO  $\nu_t$=1              &                 \\
 220793     &  U                                                                          &          \\            
 220814     &  E-CH$_3$OCHO  {\scriptsize 28$_{-3,25}$-28$_{-2,26}$}  &    \\
 220847     &  CH$_3$OCH$_3$  {\scriptsize 24$_{4,20}$-23$_{5,19}$}               &                \\ 
 220857     &  U       &              \\
 220866     &  CH$_3$OCHO  $\nu_t$=1 {\scriptsize 19$_{2,17}$-18$_{3,16}$}, CH$_3$OCHO , G-CH$_3$CH$_2$OH, CH$_3$CHO $\nu_t$=1           &             \\
 220887     &  A-CH$_3$OCHO {\scriptsize 18$_{17,1}$-17$_{17,0}$, 18$_{17,2}$-17$_{17,1}$}                   &            \\
 220892     &  CH$_3$OCH$_3$ {\scriptsize 23$_{4,20}$-23$_{3,21}$}, CH$_3$OCHO      &             \\
 220901     &  E-CH$_3$OCHO   {\scriptsize 18$_{-17,1}$-17$_{-17,0}$}     &              \\
 220913     &  CH$_3$OCHO  $\nu_t$=1 {\scriptsize 18$_{7,12}$-17$_{7,11}$}, CH$_3$OCHO             &              \\
 220926     &  A-CH$_3$OCHO  {\scriptsize 18$_{16,3}$-17$_{16,2}$, 18$_{16,2}$-17$_{16,1}$},CH$_2$DOCOH,     &             \\
            &  AE-(CH$_3$)$_2$CO, EA-(CH$_3$)$_2$CO      &             \\
 220936     &  E-CH$_3$OCHO  {\scriptsize 18$_{-16,2}$-17$_{-16,1}$}             &              \\
 220946     &  A-CH$_3$OCHO  $\nu$=1 {\scriptsize 18$_{7,11}$-17$_{7,10}$}, CH$_3$OCHO           &              \\ 
%
 220961     &  AA-(CH$_3$)$_2$CO {\scriptsize 11$_{11,1}$-10$_{10,0}$,11$_{11,0}$-10$_{10,1}$},CH$_3$OCHO ,CH$_3$NC             &              \\
 220978     &  A-CH$_3$OCHO  {\scriptsize 18$_{15,4}$-17$_{15,3}$, 18$_{15,3}$-17$_{15,2}$},CH$_3$CN $\nu_8$=1,EE-(CH$_3$)$_2$CO & \\
 220985     &  E-CH$_3$OCHO  $\nu_t$=1 {\scriptsize 18$_{7,11}$-17$_{7,10}$},CH$_3$OCHO             &             \\
 220989     &  S-CH$_2$DOCOH {\scriptsize 19$_{6,13}$-18$_{6,12}$}     &              \\
 220999     &  G-CH$_3$CH$_2$OH {\scriptsize 13$_{0,13}$-12$_{0,12}$},CH$_3$OCHO         &            \\
 221008     &  aGg'-(CH$_2$OH)$_2$ {\scriptsize 21$_{4,18}$-20$_{4,17}$},CH$_2$OHCHO  &       \\
 221013     &  E-CH$_3$NC   {\scriptsize 11$_{5}$-10$_{5}$}      &              \\
 221026     &  A-CH$_3$OCHO  {\scriptsize 36$_{6,30}$-36$_{5,31}$}      &              \\
 221038     &  aGg'-(CH$_2$OH)$_2$ {\scriptsize 22$_{6,17}$-21$_{6,16}$}  &                 \\
 221049     &  A-CH$_3$OCHO  {\scriptsize 18$_{14,5}$-17$_{14,4}$,18$_{14,4}$-17$_{14,3}$}   &              \\ 
 221059     &  CH$_3$CN $\nu_8$=1 {\scriptsize 12$_{8,1}$-11$_{8,1}$},CH$_3$NC            &             \\
 221066     &  E-CH$_3$OCHO  {\scriptsize 37$_{9,29}$-37$_{8,30}$}    &              \\  
 221077     &  A-CH$_3$OCHO  {\scriptsize 29$_{9,21}$-29$_{8,22}$}, CH$_3$CN $\nu_8$=1 &          \\
 221087     &  E-CH$_3$OCHO  {\scriptsize 29$_{9,21}$-29$_{8,22}$}             &              \\ 
 221092     &  A-CH$_3$NC {\scriptsize 11$_{3}$-10$_{3}$}                    &             \\
 221100     &  aGg'-(CH$_2$OH)$_2$ {\scriptsize 22$_{5,18}$-21$_{5,17}$}     &              \\
 221112     &  E-CH$_3$OCHO  $\nu_t$=1 {\scriptsize 18$_{-8,11}$-17$_{-8,10}$}, $^{34}$SO$_2$             &              \\ 
 221119     &  E-CH$_3$NC {\scriptsize 11$_{2}$-10$_{2}$}   &              \\ 
 221132     &  E-CH$_3$NC {\scriptsize 11$_{1}$-10$_{1}$},CH$_3$CN $\nu_8$=1, CH$_3$OCHO  $\nu_t$=1 &                   \\ 
 221138     &  A-CH$_3$NC {\scriptsize 11$_{0}$-10$_{0}$} & \\
 221141     &  A-CH$_3$OCHO  {\scriptsize 18$_{13,6}$-17$_{13,5}$,18$_{13,5}$-17$_{13,4}$}, E-CH3OCHO  {\scriptsize 18$_{-13,5}$-17$_{-13,4}$}      &   \\ 
 221157     &  E-CH$_3$OCHO  {\scriptsize 18$_{13,6}$-17$_{13,5}$}, CH$_2$DOH, CH$_3$OCOD       &              \\
 221165     &  A-CH$_3$OCHO  $\nu$=1 {\scriptsize 37$_{8,30}$-37$_{7,31}$} &  \\
 221178     &  CH$_2$DOH {\scriptsize 6$_{2,5}$-6$_{1,6}$}, HDCS {\scriptsize 7$_{1,6}$-6$_{1,5}$}                &              \\ 
 221191     &  p-D$_2$CO {\scriptsize 4$_{1,4}$-3$_{1,3}$}        &              \\
 221198     &  CH$_3$OCH$_3$ {\scriptsize 27$_{5,22}$-27$_{4,23}$},CH$_3$CN $\nu_8$=1    &              \\   
 221210     &  CH$_3$CN $\nu_8$=1 {\scriptsize 12$_{8,2}$-11$_{8,2}$}     &              \\  
 221216     &  E-CH$_3$CHO $\nu_t$=2  {\scriptsize 11$_{2,10}$-10$_{2,9}$}                  &              \\  
 221222     &  A-CH$_3$CHO $\nu_t$=1  {\scriptsize 22$_{1,21}$-22$_{0,22}$}              &           \\
 221253     & CH$_3$CN $\nu_8$=1 {\scriptsize 12$_{5,1}$-11$_{5,1}$}           &          \\
 221266     & A-CH$_3$OCHO {\scriptsize 18$_{12,7}$-17$_{12,6}$,18$_{12,6}$-17$_{12,5}$},CH$_3$CN $\nu_8$=1  {\scriptsize 12$_{7,2}$-11$_{7,2}$}   &     \\
 221273     & CH$_2$DOH {\scriptsize 5$_{1,5}$-4$_{1,4}$}           &          \\
 221285     & E-$^{13}$CH$_3$OH {\scriptsize 8$_{-1,8}$-7$_{0,7}$}, CH$_3$OCHO  &          \\ 
 221298     & A-CH$_3$CHO {\scriptsize 13$_{3,10}$-13$_{2,11}$}, CH$_3$CN $\nu_8$=1 {\scriptsize 12$_{4,1}$-11$_{4,1}$}, CH$_3$OCHO           &          \\
 221312     & CH$_3$CN $\nu_8$=1 {\scriptsize 12$_{6,2}$-11$_{6,2}$}, NH$_2$CHO          &         \\
 221326    &  U                 &      \\
 221338    &  aGg'-(CH$_2$OH)$_2$ {\scriptsize 23$_{3,21}$-22$_{3,20}$},CH$_3$CN $\nu_8$=1 {\scriptsize 12$_{3,1}$-11$_{3,1}$}, aGg'-(CH$_2$OH)$_2$  &     \\
 221350    &  CH$_3$CN $\nu_8$=1 {\scriptsize 12$_{5,2}$-11$_{5,2}$}         &         \\
 221367    &  CH$_3$CN $\nu_8$=1 {\scriptsize 12$_{2,1}$-11$_{2,1}$}             &          \\
 221382    &  CH$_3$CN $\nu_8$=1 {\scriptsize 12$_{4,2}$-11$_{4,2}$}                     &          \\
 221392    &  CH$_2$DOH {\scriptsize 10$_{1,10}$-9$_{0,9}$},CH$_3$CN $\nu_8$=1 {\scriptsize 12$_{1,1}$-11$_{1,1}$}                &          \\
 221404    &  CH$_3$CN $\nu_8$=1 {\scriptsize 12$_{3,2}$-11$_{3,2}$}, CH$_2$DOCOH,CH$_3$OCHO  &          \\
 221423    &  A-$^{13}$CH$_3$OH $\nu_t$=1  {\scriptsize 6$_ {1, 6}$ - 7$_{ 2, 5}$}, CH$_3$OCHO , CH$_3$CN $\nu_8$=1     &         \\
 221433    &  A-CH$_3$OCHO  {\scriptsize 18$_{11,8}$-17$_{11,7}$,18$_{11,7}$-17$_{11,6}$}, CH$_3$CH$_2$CN $\nu_{13}$/$\nu_{21}$=1           &          \\
 221445    &  E-CH$_3$OCHO  {\scriptsize 18$_{11,8}$-17$_{11,7}$}, CH$_3$CH$_2$CN $\nu_{13}$/$\nu_{21}$=1          &          \\
 221465    &  CH$_2$OHCHO {\scriptsize 34$_{10,25}$-34$_{9,26}$}              &           \\
 221480    &  E-CH$_3$CHO $\nu_t$=1 {\scriptsize 23$_{8,15}$-24$_{7,17}$}, CH$_3$CHO $\nu_t$=2            &        \\
 221494    &  aGg'-(CH$_2$OH)$_2$ {\scriptsize 12$_{5,8}$-11$_{4,8}$}    &     \\
 221500    &  EE-(CH$_3$)$_2$CO {\scriptsize 27$_{2,25}$-27$_{2,26}$,27$_{2,25}$-27$_{1,26}$,27$_{3,25}$-27$_{2,26}$,27$_{3,25}$-27$_{1,26}$ }   &  \\
 221519    &  U   &             \\
 221526    &  EA-(CH$_3$)$_2$CO  {\scriptsize 31$_{20,11}$-31$_{17,14}$}              &            \\
 221534    &   U                &        \\
 221547    &   U                &       \\
 221560    &  A-CH$_2$DOCOH {\scriptsize 7$_{6,2}$-6$_{5,1}$,7$_{6,1}$-6$_{5,2}$}, S-CH$_2$DOCOH {\scriptsize 21$_{3,18}$-20$_{4,17}$,50$_{9,41}$-50$_{8,42}$} &      \\
 221566    &  U                 &      \\
 221582    &  aGg'-(CH$_2$OH)$_2$  {\scriptsize 18$_{4,15}$-17$_{3,14}$}   &        \\
 221597    &  U   &          \\
 221605    &  U                  &            \\
 221627    &  CH$_3$CN $\nu_8$=1 {\scriptsize 12$_{1,2}$-11$_{-1,2}$}             &            \\ 
 221649    &  E-CH$_3$OCHO  {\scriptsize 18$_{-10,8}$-17$_{-10,7}$}  &         \\
 221660    &  A-CH$_3$OCHO  {\scriptsize 18$_{10,9}$-17$_{10,8}$,18$_{10,8}$-17$_{10,7}$},  E-CH$_3$OCHO  {\scriptsize 18$_{4,15}$-17$_{4,14}$} &     \\
\end{longtable}
\noindent
$^1$ Lines are ordered from major to minor contribution to the total integrated flux according with our model.

\noindent
$^2$ The lines in this column correspond to doubtful identifications.


\begin{table*}
\begin{tabular}{l|l|l} 
\multicolumn{3}{c}{Table A.2.Line identifications (FU05)} \\ \hline \hline
 (MHz)        &  Molecular lines$^1$             & Comments$^2$ \\ \hline
228140  & G-CH$_3$CH$_2$OH {\scriptsize 18$_{5,14}$-18$_{4,15}$}     &   \\
228151  & OC$^{33}$S {\scriptsize 19-18}    &   \\
228167  & AE-(CH$_3$)$_2$CO {\scriptsize 34$_{9,26}$-34$_{8,27}$,34$_{8,26}$-34$_{7,27}$}, 
EA-(CH$_3$)$_2$CO {\scriptsize 34$_{9,26}$-34$_{8,27}$,34$_{8,26}$-34$_{7,27}$},CH$_3$CH$_2$CN  &   \\
228186  & U      &   \\
228199  & HCOOH {\scriptsize 21$_{6,16}$-22$_{5,17}$}     &   \\
228211  & CH$_3$OCHO  $\nu_t$=1 {\scriptsize 18$_{3,15}$-17$_{3,14}$}   &   \\
228231  & CH$_2$DOH  {\scriptsize 11$_{2,9}$-11$_{1,10}$}        &   \\
228246  & CH$_2$DOH  {\scriptsize 15$_{2,13}$-15$_{1,14}$}        &   \\
228257  & U     &  \\
228272  & A-CH$_3$OCHO  {\scriptsize 24$_{9,16}$-24$_{8,17}$}, S$^{18}$O,CH$_3$OCOD        &   \\
228293  & A-CH$_3$OCOD {\scriptsize 18$_{4,14}$-17$_{4,13}$}                     &   \\
228307  & HC$_3$N $\nu_7$=1 {\scriptsize 25$_{1}$-24$_{1}$}, EE-CH$_3$COCH$_3$   &   \\
228336  & U   & CH$_2$CHCN {\scriptsize 16$_{3,14}$-16$_{2,15}$}   \\
228359  & E-CH$_3$OCHO  {\scriptsize 24$_{-9,15}$-24$_{8,17}$},CH$_3$OCHO  $\nu_t$=1,CH$_2$DCH$_2$CN, aGg'-(CH$_2$OH)$_2$      &   \\
224374  & U                         &   \\
228384  & U                         &   \\
228410  & H$^{13}$COOH {\scriptsize 10$_{2,8}$-9$_{2,7}$}              &   \\
228436  & CH$_3$OCH$_3$ {\scriptsize 26$_{3,23}$-25$_{4,21}$, 26$_{3,24}$-25$_{4,21}$ }    &   \\
228439  & A-CH$_2$DOCOH {\scriptsize 19$_{9,10}$-18$_{9,9}$,19$_{9,10}$-18$_{9,10}$,19$_{9,11}$-18$_{9,9}$,19$_{9,11}$-18$_{9,10}$},CH$_3$OCOD, 
HCOOH                        &          \\
228448  & A-CH$_3$OCOD {\scriptsize 20$_{2,18}$-19$_{2,17}$}, AA-(CH$_3$)$_2$CO  &   \\
228464  & E-CH$_3$OCOD {\scriptsize 19$_{-9,10}$-18$_{-9,9}$}, CH$_3$OCHO  $\nu_t$=1    &   \\
228483  & CH$_3$CH$_2$CN {\scriptsize 25$_{2,23}$-24$_{2,22}$},CH$_3$OCOD    &   \\
228490  & G+CH$_3$CH$_2$OH {\scriptsize 8$_{5,3}$-8$_{4,4}$, 8$_{5,4}$-8$_{4,5}$}            & Undetected \\
228500  & AE-(CH$_3$)$_2$CO {\scriptsize 12$_{10,3}$-11$_{9,2}$}         &   \\
228512  & G-CH$_3$CH$_2$OH {\scriptsize 15$_{5,10}$-15$_{4,11}$}, AE-(CH$_3$)$_2$CO, EA-(CH$_3$)$_2$CO  &   \\
228526  & G-CH$_3$CH$_2$OH {\scriptsize 17$_{5,13}$-17$_{4,14}$} &   \\
228544  & HCOOH {\scriptsize 10$_{2,8}$-9$_{2,7}$}   &   \\
228559  & G+CH$_3$CH$_2$OH {\scriptsize 7$_{5,2}$-7$_{4,3}$, 7$_{5,3}$-7$_{4,4}$}  & Undetected  \\
231060   & OCS {\scriptsize 19-18}  &   \\
231077   & U    & OCS wing?, CH$_3$OCHO  $\nu_t$=1  \\
231102   & U    &   \\
231114   & E-CH$_3$CHO {\scriptsize 9$_{3,6}$-9$_{2,7}$}      &    \\
231127   & aGg'-(CH$_2$OH)$_2$ {\scriptsize 23$_{7,16}$-22$_{7,15}$}    &   \\
231144   & o-H$_2$$^{13}$CS {\scriptsize 7$_{3,5}$-6$_{3,4}$,7$_{3,4}$-6$_{3,3}$}, CH$_3$OCHO  $\nu_t$=1, CH$_3$CHO $\nu_t$=1  &    \\
231167   & E-CH$_3$CHO {\scriptsize 10$_{10,2}$-11$_{10,1}$}   &    \\
231188   & H$^{13}$COOH {\scriptsize 10$_{1,9}$-9$_{1,8}$},CH$_3$OCHO , CH$_3$CHO & contaminated with $^{13}$CS wings?  \\
231200   & A-CH$_3$OCHO  {\scriptsize 21$_{9,12}$-21$_{8,13}$},E-CH$_3$OCHO  {\scriptsize 21$_{-9,12}$-21$_{-8,13}$},E-CH$_3$CHO & 
contaminated with $^{13}$CS wings?  \\
231220   & $^{13}$CS  {\scriptsize 5-4} &   \\
231226   & A-CH$_3$CHO {\scriptsize 12$_{9,3}$-11$_{9,2}$,12$_{9,4}$-11$_{9,3}$},E-CH$_3$CHO {\scriptsize 12$_{7,5}$-11$_{7,4}$} &   \\
231232   & A-CH$_3$CHO {\scriptsize 12$_{8,4}$-11$_{8,3}$,12$_{8,5}$-11$_{8,4}$},E-CH$_3$CHO {\scriptsize 12$_{8,5}$-11$_{8,4}$},
CH$_3$OCHO  $\nu_t$=1 & contaminated with $^{13}$CS wings?  \\
231245   & A-CH$_3$CHO {\scriptsize 12$_{7,6}$-11$_{7,5}$,12$_{7,5}$-11$_{7,4}$},CH$_3$OCHO  $\nu_t$=1 &                \\
231255   & U                   &    \\
231268   & A-CH$_3$CHO {\scriptsize 12$_{6,7}$-11$_{6,6}$,12$_{6,6}$-11$_{6,5}$}, E-CH$_3$CHO {\scriptsize 12$_{7,6}$-11$_{7,5}$} &    \\
231280   & A-CH$_3$OH {\scriptsize 10$_{2,-1}$-9$_{3,-1}$}, CH$_3$CHO  &   \\
231293   & U  &   \\
231311   & CH$_3$CH$_2$CN {\scriptsize 26$_{1,25}$-25$_{1,24}$,27$_{0,27}$-26$_{1,26}$,24$_{2,23}$-23$_{1,22}$}, CH$_3$CHO   &   \\
231329   & A-CH$_3$CHO {\scriptsize 12$_{5,8}$-11$_{5,7}$,12$_{5,7}$-11$_{5,6}$}   &   \\
231344   & OCS $\nu_2$=1 {\scriptsize 19$_{-1}$-18$_{-1}$}  &   \\
231369   & E-CH$_3$CHO {\scriptsize 12$_{5,7}$-11$_{5,6}$,12$_{5,8}$-11$_{5,7}$}    &   \\
231383   & E-CH$_3$CHO $\nu$=1 {\scriptsize 12$_{5,8}$-11$_{5,7}$}    &    \\
231410   & o-D$_2$CO {\scriptsize 4$_{0,4}$-3$_{0,3}$}       &    \\        
231417   & E-CH$_3$COOH $\nu$=1 {\scriptsize 19$_{-16,4}$-18$_{-16,3}$}     &   \\
231454   & A-CH$_3$CHO {\scriptsize 12$_{4,9}$-11$_{4,8}$}  &   \\
231466   & A-CH$_3$CHO {\scriptsize 9$_{3,6}$-9$_{2,7}$}  &   \\
231483   & E-CH$_3$CHO {\scriptsize 12$_{4,8}$-11$_{4,7}$}    &   \\
231506   & HCOOH {\scriptsize 10$_{1,9}$-9$_{1,8}$}, CH$_3$CHO    &    \\
231525?  & aGg'-(CH$_2$OH)$_2$ {\scriptsize 23$_{6,18}$-22$_{6,17}$} & \\
231533?  & U  &        \\
231560   & T-CH$_3$CH$_2$OH {\scriptsize 20$_{5,16}$-20$_{4,17}$}    &    \\ \hline \hline
\end{tabular}

\noindent
$^1$ Lines are ordered from major to minor contribution to the total integrated flux according with our model.

\noindent
$^2$ The lines in this column correspond to doubtful identifications.

\end{table*}

\newpage
\begin{figure*}
\includegraphics{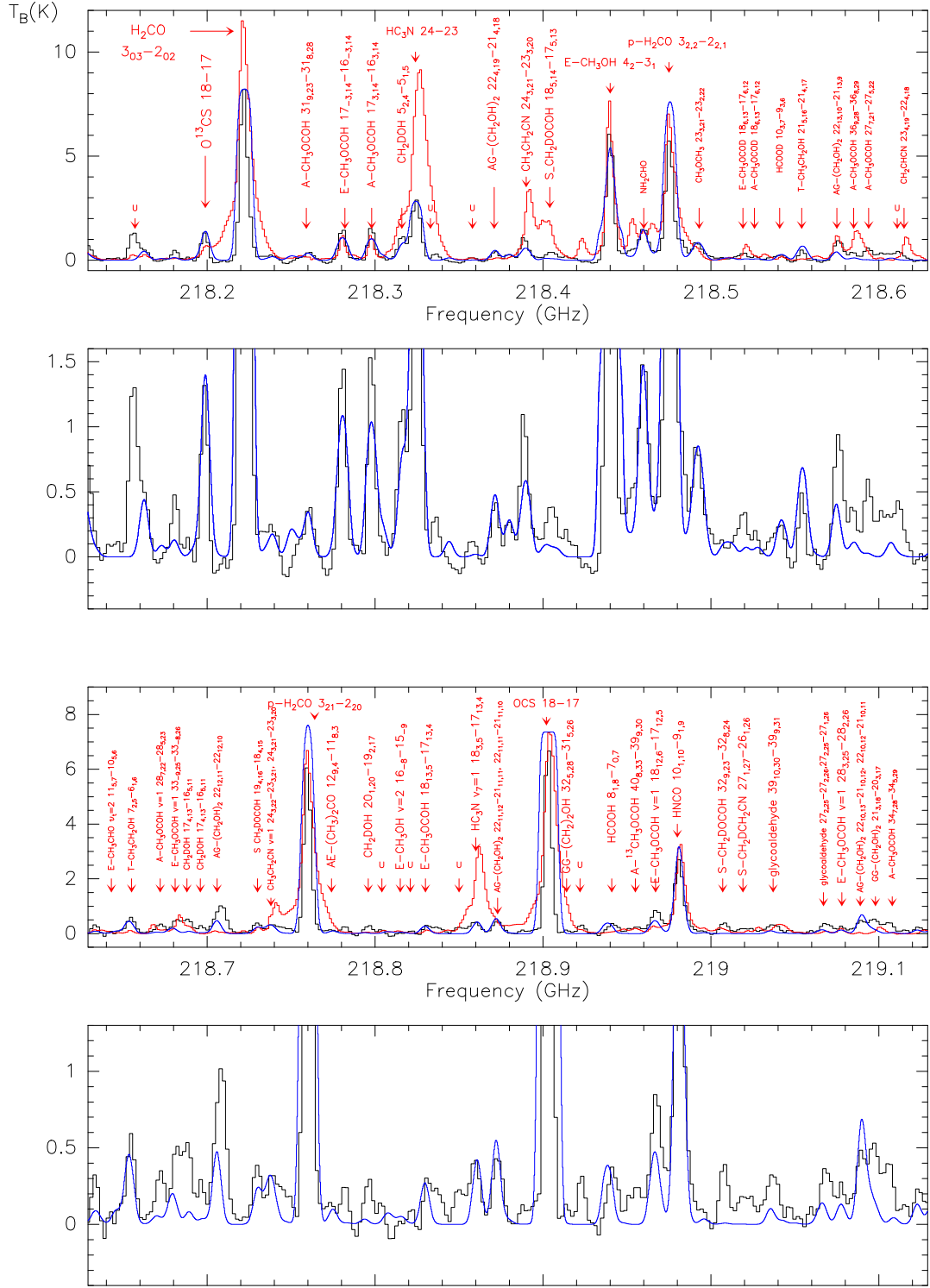}
\vspace{22.0cm}
      \caption{Comparison of the interferometric spectra towards NGC 7129 - FIRS 2 (black line) with the single-dish spectrum of Orion KL (red line). 
       The Orion spectra have been provided by B. Tercero and J. Cernicharo and multiplied by a factor 0.43 for an easier comparison.  
       The spectrum was observed towards: $5^{\mathrm{h}}35^{\mathrm{m}}14\fs5$, $-5\degr22\arcmin30\farcs0$ (J2000) and is centered at a velocity
       of 7~km~s$^{-1}$. Lines are labelled in red. When several lines are blended, only the most intense ones are indicated.
       In blue, the spectra synthesized with the parameters shown in Table~3. The intensity scale is brightness temperature.}
 \end{figure*}

\begin{figure*}
\includegraphics{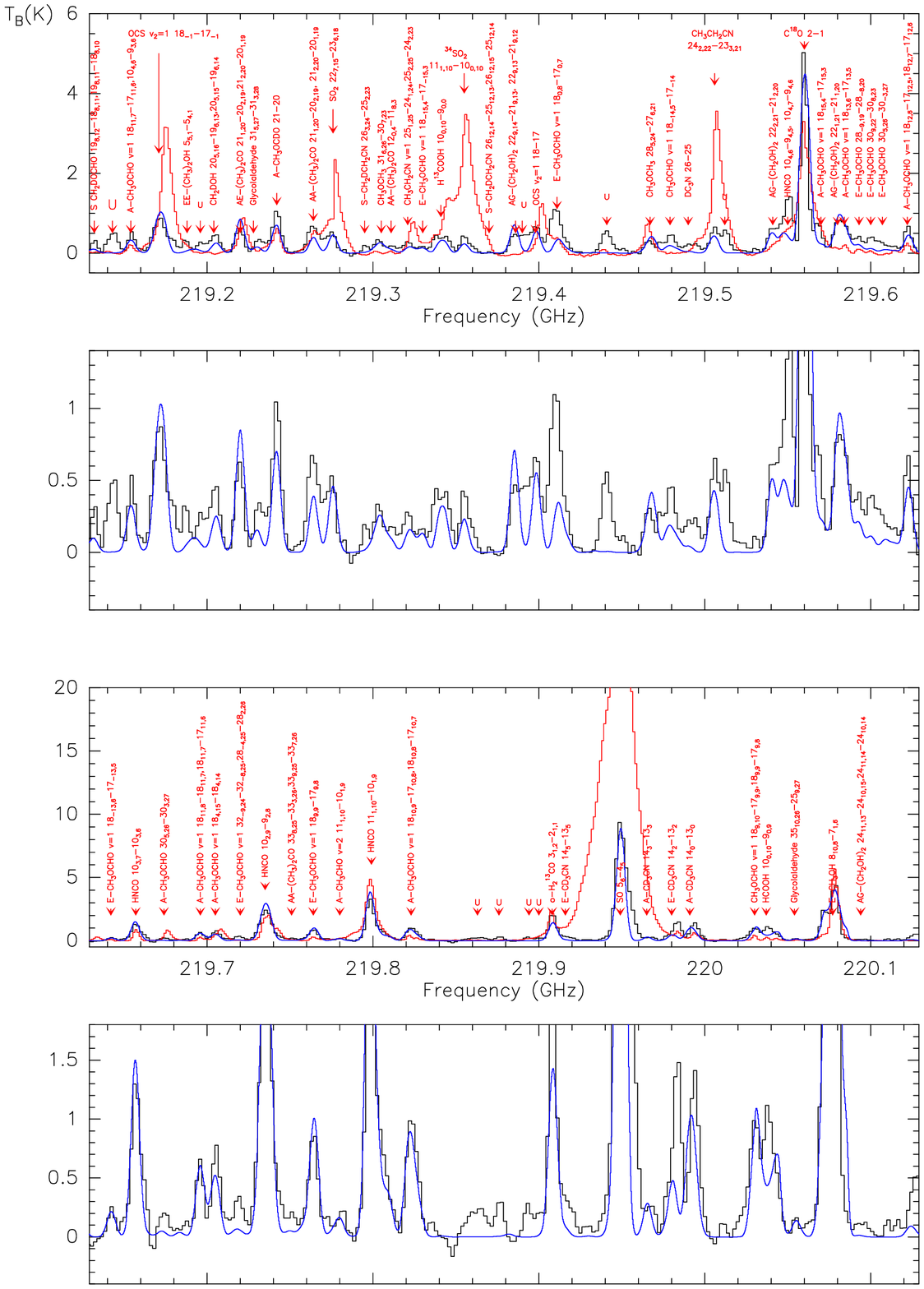}
\vspace{22.0cm}
      \caption{The same as Fig. A.1. }
 \end{figure*}

\begin{figure*}
\includegraphics{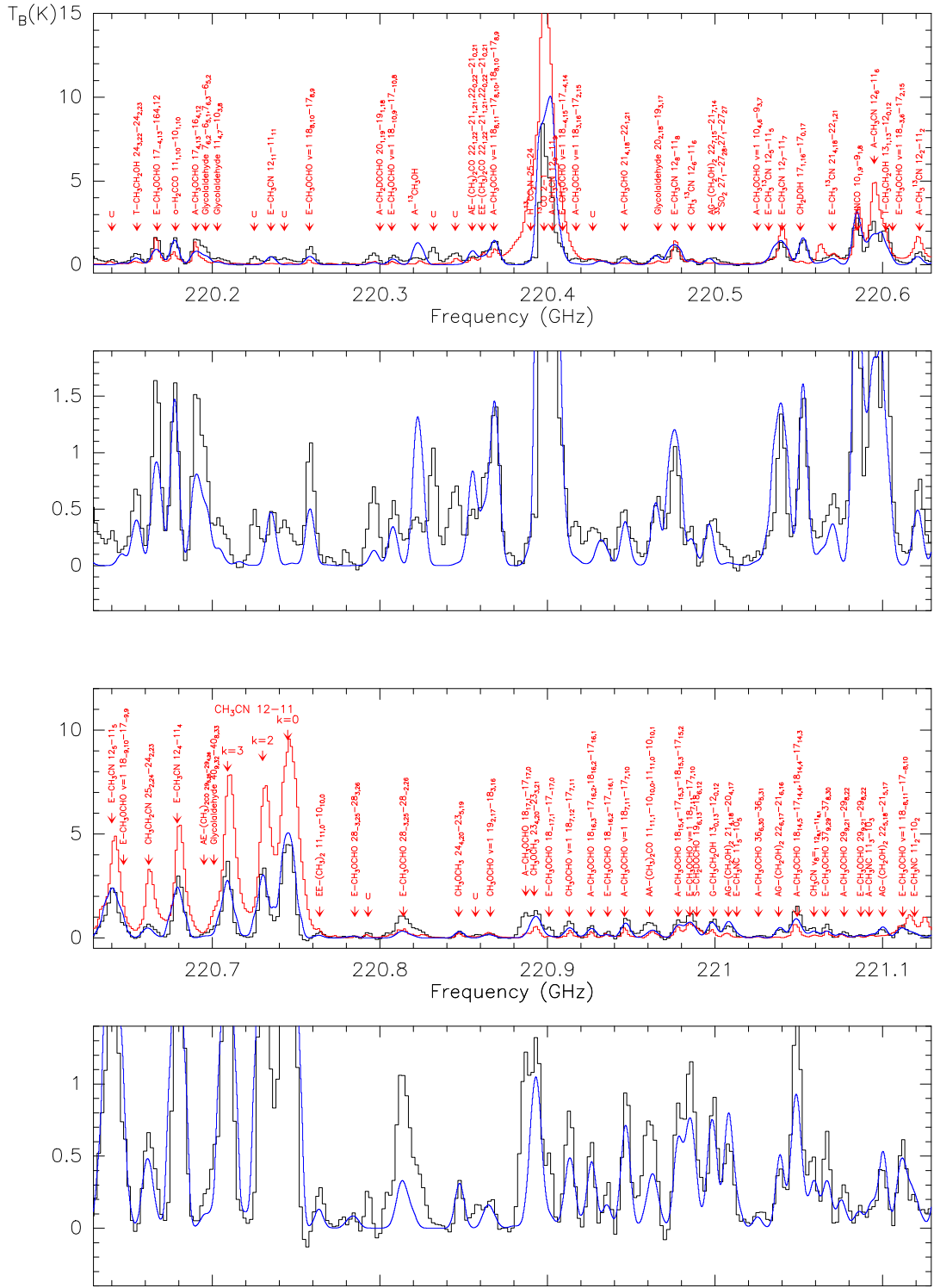}
\vspace{22.0cm}
      \caption{The same as Fig. A.1. }
\end{figure*}

\begin{figure*}
\includegraphics{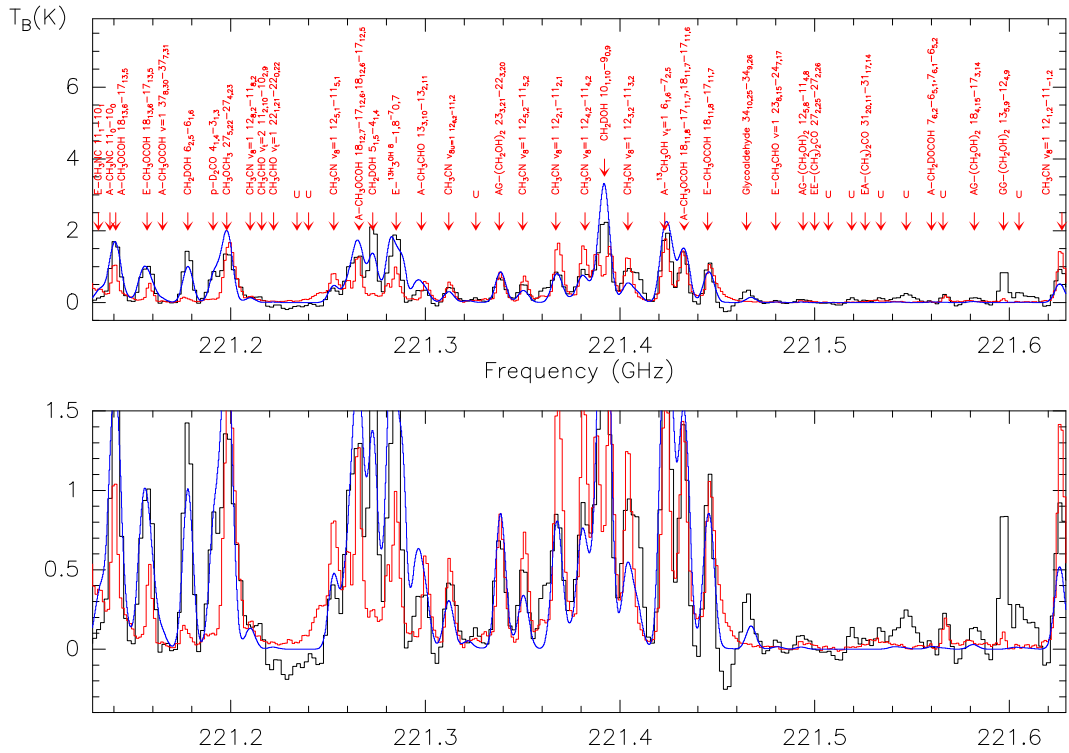}
\vspace{22.0cm}
      \caption{The same as Fig. A.1. }
 \end{figure*}

\begin{figure*}
\includegraphics{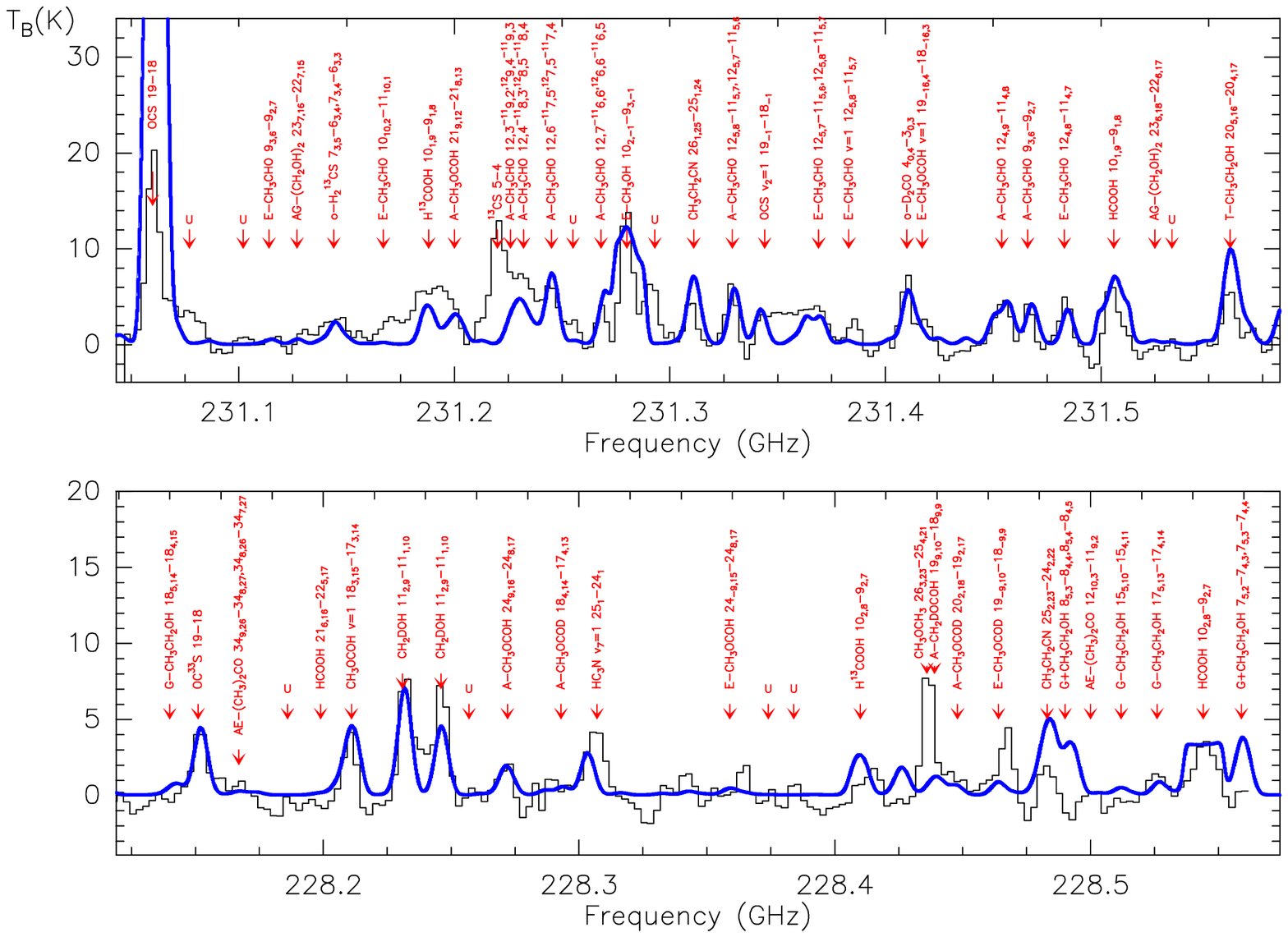}
\vspace{13.0cm}
      \caption{In blue, our synthetic spectrum superposed to the interferometric spectra published by FU05.}
 \end{figure*}
\vskip 4cm

\newpage
\vskip 4cm
\begin{table*}
\begin{tabular}{llllllllllllllcl} 
\multicolumn{16}{c}{Table A.3. CH$_3$CN~$\nu$=0 line parameters$^1$} \\ \hline \hline
N$^2$  & \multicolumn{8}{c}{Quantum numbers: up-low$^3$} &  $\nu$(MHz)   &  E$_u$(K)$^4$        &   A$_{ul}$ (s$^{-1}$)$^5$ &   Sij$^6$  &   gu$^7$ &  W (K km s$^{-1}$)$^8$ &  \\ \hline 
 1  & 12  &  11  &  & & 11  &  11 & & &  220235.047  &   923.4  &  1.469E-04 &  1.92  &  25 & 1.95(1.95)  & \\
 2  & 12  &  10  &  & & 11  &  10 & & &  220323.643  &   774.0  &  2.813E-04 &  3.67  &  25 & 4.73(1.38)  &  Blended with A-$^{13}$CH$_3$3OH\\
 3  & 12 &   8   &  & & 11  &   8 & & &  220475.814  &   517.6  &  5.125E-04 &  6.67  &  25 & 7.57(1.17)  & \\
 4  & 12 &   7   &  & & 11  &   7 & & &  220539.328  &   410.6  &  6.091E-04 &  7.92  &  25 & 10.18(6.0)  & \\
 5  & 12 &   4   &  & & 11  &   4 & & &  220679.287  &   175.1  &  8.223E-04 &  10.7  &  25 & 27.05(10.63) &   \\
 6  & 12 &   2   &  & & 11  &   2 & & &  220730.259  &   89.4   &  9.000E-04 &  11.7  &  25 & 29.82(6.74)  & \\
 7  & 12 &  1    &  & & 11  &  1  & & &  220743.008  &   68.0   &  9.195E-04 &  11.9  &  25 & 58.83(7.66)  & \\
 7  & 12 &  0    &  & & 11  &  0  & & &  220747.259  &   68.9   &  9.259E-04 &  12.0  &  25 &   &  \\
 8  & 12 &   6   &  & & 11  &   6 & & &  220594.426  &   325.9  &  3.465E-04 &  9.00  &  50 & 19.37(6.83)  & \\
 9  & 12 &   3   &  & & 11  &  3  & & &  220709.015  &   133.2  &  4.338E-04 &  11.   &  50 & 31.35(10.23) & \\ \hline
\end{tabular}

\noindent
$^1$Exponential notation: 1.469E-04=1.469$\times$10$^{-4}$; $^2$Index to enumerate the line features in the spectrum (blended lines share the same index); 
$^3$Quantum numbers of the upper and lower leves of the transitions; $^4$Energy of the upper level; $^5$ Einstein coefficient of spontaneous emission;
$^6$Line strength defined as $|\mu_{ij}|^2= S_{ij} \mu^2$ where $\mu$ is the dipole moment; $^7$Degeneracy of the upper level; $^8$Velocity integrated intensity      
\end{table*}

\begin{table*}
\begin{tabular}{llllllllllllllcl} 
\multicolumn{16}{c}{Table A.4. CH$_3$CN~$\nu_8$=1 (E$_{vib}$=525.17~K) line parameters $^1$} \\ \hline \hline
 N$^2$  & \multicolumn{8}{c}{Quantum numbers: up-low$^3$} &  $\nu$(MHz)   &  E$_{R_u}$(K)$^4$  &   A$_{ul}$ (s$^{-1}$)$^5$  &   Sij$^6$ &   gu$^7$  &  W (K km s$^{-1}$)$^8$   &  \\ \hline

  1 &   12  &   8  &   1  &  12  &  11  &   8   &  1  &  11   &  221059.437   &  637.3  &  1.906E-03  & 24.6  &  25   & 3.63(1.57)  & \\ 
  1 &   12  &   8  &   1  &  13  &  11  &   8   &  1  &  12   &  221059.797   &  637.3  &  1.923E-03  & 26.8  &  27   &  & \\
  1 &   12  &   8  &   1  &  11  &  11  &   8   &  1  &  10   &  221059.828   &  637.3  &  1.904E-03  & 22.6  &  23   &  & \\
%
%
  2  &     12 &   -1 &    2  &  11  &  11  &  1   &  2  &  10  &  221199.140   &   68.1  & 1.705E-03  &  20.2   &  23 & 17.62(8.89) & Blended with CH$_3$OCH$_3$\\ 
  2  &     12 &   -1 &    2  &  12  & 11   &  1   &  2  &  11  &  221199.140   &   68.1  & 1.708E-03  &  22.0   &  25 &  &  \\  
  2  &     12 &   -1 &    2  &  13  & 11   &  1   &  2  &  12  &  221199.156   &  68.1   & 1.718E-03  &  23.9   &  27 &  &  \\ 
  3   &    12 &    8 &    2  & 12  &  11  &   8  &  2  &  11   &  221209.781   &  423.9 & 9.551E-04  & 12.3  &  25 & 2.72 (2.0) & \\
  3   &    12 &    8 &    2  & 13  &  11  &   8  &  2  &  12   &  221210.140   &  423.9 & 9.634E-04  & 13.4  &  27 &  & \\ 
  3   &    12 &    8 &    2  & 11  &  11  &   8  &  2  &  10   &  221210.172   &  423.9 & 9.537E-04  & 11.3  &  23 &  & \\ 
  4  &     12 &    5 &    1  &  12  &  11  &   5  &   1  &  11  &   221252.937   &  319.3 &  2.844E-03  & 36.6  &  25  & 5.65(2.27) & \\
  4  &     12 &    5 &    1  &  13  &  11  &   5  &   1  &  12  &   221252.937   &  319.3 &  2.863E-03  & 39.8  &  27  & & \\
  4  &     12 &    5 &    1  &  11  &  11  &   5  &   1  &  10  &   221252.937   &  319.3 &  2.846E-03  & 33.7  &  23  & & \\
  5  &     12 &    7 &    2  &  12  &  11   &  7   &  2  &  11   &  221265.547   &  330.5  & 2.269E-03 &  29.2  &   25  & 19.98(2.08) & Blended with CH$_3$OCHO \\ 
  5  &     12 &    7 &    2  &  13  &  11   &  7   &  2  &  12   &  221265.547   &  330.5  & 2.288E-03 &  31.8  &   27  & &  \\  
  5  &     12 &    7 &    2  &  11  &  11   &  7   &  2  &  10   &  221265.547   &  330.5  & 2.272E-03 &  26.9  &   23  & &  \\
  6  &     12 &    4 &    1  &  12  &  11   &  4   &  1 & 11  &  221299.875  & 241.8  & 1.532E-03 & 19.7  &  25 & 4.19(4.0) & Blended with CH$_3$OCHO\\ 
  6  &     12 &    4 &    1  &  13  &  11   &  4   &  1 & 12  &  221299.875  & 241.8  & 1.540E-03 & 21.4  &  27 &  &   \\  
  6  &     12 &    4 &    1  &  11  &  11   &  4   &  1 & 10  &  221299.875  & 241.8  & 1.529E-03 & 18.1  &  23 &  &   \\ 
  7  &     12 &    6 &    2  &  12  &  11  &   6  &  2 & 11  &  221311.953   & 251.2  & 1.291E-03  & 16.6   &  25  & 3.35(3.0) & \\  
  7  &     12 &    6 &    2  &  13  &  11  &   6  &  2 & 12  &  221311.953   & 251.2  & 1.303E-03  & 18.1   &  27  & & \\  
  7  &     12 &    6 &    2  &  11  &  11  &   6  &  2 & 10  &  221311.953   & 251.2  & 1.293E-03  & 15.3   &  23  & & \\ 
  8  &     12 &    3 &    1  &  12  &  11  &   3  &   1 & 11  &  221338.218   &  178.5  & 1.618E-03 & 20.8  &  25  & 6.28(5.34) & Blended  with aGg'-(CH$_2$OH)$_2$\\ 
  8  &     12 &    3 &    1  &  11  &  11  &   3  &   1 & 10  &  221338.218   &  178.5  & 1.615E-03 & 19.1  &  23  &  &  \\  
  8  &     12 &    3 &    1  &  13  &  11  &   3  &   1 & 12  &  221338.218   &  178.5  & 1.628E-03 & 22.6  &  27  &  & \\  
  9  &    12 &    5 &    2  &  12  &  11 &    5   &  2 & 11  &  221350.375  &  186.1 &  1.424E-03  & 18.3  &   25  & 4.52(2.42) & \\  
  9  &    12 &    5 &    2  &  13  &  11 &    5   &  2 & 12  &  221350.375  &  186.1 &  1.433E-03  & 19.9  &   27  & & \\  
  9  &     12 &    5 &    2  &  11  &  11 &    5   &  2 & 10  &  221350.375  &  186.1 &  1.421E-03  & 16.8  &   23  & & \\  
  10 &     12 &    2 &    1  &  12  &  11  &   2  &  1 &  11  &  221367.672   &  129.5 & 3.354E-03  & 43.1  &   25  & 9.47(4.30) & Blended with A-$^{13}$CH$_3$OH\\  
  10 &     12 &    2 &    1  &  11  &  11  &   2  &  1 &  10  &  221367.672   &  129.5 & 3.349E-03  & 39.6  &   23  & & \\  
  10 &     12 &    2 &    1  &  13  &  11  &   2  &  1  & 12  &  221367.672   &  129.5 & 3.379E-03  & 46.9  &   27  & & \\  
  11  &    12 &    4 &    2  &  12  &  11  &   4  &  2  & 11  &  221380.734  &  135.3 & 3.066E-03  & 39.4  &   25  & 16.77(1.81)& \\ 
  11  &    12 &    4 &    2  &  11  &  11  &   4  &  2  & 10  &  221380.734  &  135.3 & 3.062E-03  & 36.2  &   23  & & \\  
  11  &    12 &    4 &    2  &  13  &  11  &   4  &  2  & 12  &  221380.734  &  135.3 & 3.084E-03  & 42.8  &   27  & & \\   
  12  &    12  &   0   &  2   & 11 &  11  &   0  &   2  & 10  &  221394.156  &  74.3  & 1.726E-03 & 20.4  &   23  & 18.99(1.15) & Blended with CH$_2$DOH \\  
  12  &    12  &   0   &  2   & 12 &  11  &   0  &   2  & 11  &  221394.156  &  74.3  & 1.728E-03 & 22.2  &   25  & &  \\  
  12  &    12  &   0   &  2   & 13 &  11  &   0  &   2  & 12  &  221394.156  &  74.3  & 1.737E-03 & 24.1  &   27  & & \\ 
  13  &   12  &   3   &  2   &  12  &  11  &   3 &  2  & 11  &  221403.812  &  98.7   & 1.619E-03 & 20.8  &   25 & 7.49(2.09) &  Blended \\  
  13  &   12  &   3   &  2   &  11  &  11  &   3 &  2  & 10  &  221403.812  &  98.7   & 1.616E-03 & 19.1  &   23 & &  \\ 
  13  &   12  &   3   &  2   &  13  &  11  &   3 &  2  & 12  &  221403.812  &  98.7   & 1.629E-03 & 22.6  &   27 & &  \\ 
  14  &    12 &    2   &  2  &  12  &  11  &   2 &  2  & 11  &  221422.375  & 76.3  & 1.674E-03  & 21.5  &   25 & 15.61(7.19) &   Blended with A-$^{13}$CH$_3$OH \\ 
  14  &    12 &    2   &  2  &  11  &  11  &   2 &  2  & 10  &  221422.375  & 76.3  & 1.676E-03  & 19.8  &   23 &  &  \\ 
  14  &    12 &    2   &  2  &  13  &  11  &   2 &  2  & 12  &  221422.375  & 76.3  & 1.687E-03  & 23.4  &   27 &  &  \\  
  15  &    12 &    1   &  2  &  11  &  11 &  -1 &  2  & 10  &  221625.906   &  68.2 & 1.715E-03  & 20.2   &  23 & 9.94(2.67)&  \\ 
  15  &    12 &    1   &  2  &  12  &  11 &  -1 &  2  & 11  &  221625.906   & 68.2  & 1.718E-03  & 22.0   &  25 & &  \\  
  15  &    12 &    1   &  2  &  13  &  11 &  -1 &  2  & 12  &  221625.906   & 68.2  & 1.728E-03  & 23.9   &  27 & &  \\ 
 \hline
\end{tabular}
\noindent

\noindent
$^1$Exponential notation: 1.469E-04=1.469$\times$10$^{-4}$; $^2$Index to enumerate the line features in the spectrum (blended lines share the same index); 
$^3$Quantum numbers of the upper and lower leves of the transitions; $^4$Rotational energy of the upper level; $^5$ Einstein coefficient of spontaneous emission;
$^6$Line strength defined as $|\mu_{ij}|^2= S_{ij} \mu^2$ where $\mu$ is the dipole moment; $^7$Degeneracy of the upper level; $^8$Velocity integrated intensity      
\end{table*}
%

\begin{table*}
\begin{tabular}{llllllllllllll} 
\multicolumn{14}{c}{Table A.5. CH$_3$OCHO line parameters$^1$} \\ \hline \hline
 N & \multicolumn{6}{c}{Quantum numbers: up-low}  &  $\nu$(MHz) &  E$_u$(K)  &  A$_{ul}$ (s$^{-1}$) &  Sij  &  gu  &  W (K km s$^{-1}$) &  \\ \hline
\multicolumn{14}{c} {$\nu_t$=0}\\
 1  &    17 &  3  & 14  &  16  &  3 &  13 &   218297.867 &   99.7  & 1.506E-04 & 16.4  &  35 & 10.21(8.88) & \\ 
 2  &    17 &  4  &  13 &  16 &   4 &  12 &   220190.266 &   103.1 & 1.522E-04 & 16.1  &  35 & 18.05(2.43) & \\   
 3  &    33 &   5 &  28 &  33 &   5 &  29 &   220432.932 &   357.4 & 5.710E-06 & 1.16  &  67 & 9.36(3.90)  & \\   
 4  &    17 &  -3 &  14 &  16 &  -3 &  13 &   218280.838 &    99.1 & 1.507E-04 & 43.6  &  35 & 9.96(2.90)  &  \\  
 5  &    17 &  -4 &  13 &  16 &  -4 &  12 &   220166.852 &   102.6 & 1.522E-04 & 42.9  &  35 & 13.96(0.37) & \\  
 6  &    18 &  13 &   6 &  17 &  13 &   5 &   221158.521 &  212.4  & 7.789E-05 & 22.9  &  37 & 10.09(3.66) & Blended with CH$_2$DOH  \\ 
 7  &    18 &  12 &   7 &  17 &  12 &   6 &   221280.890 &   195.8 & 9.063E-05 & 26.6  &  37 & 18.00(4.90) & Blended with $^{13}$CH$_3$OH   \\ 
 8  &    18 & -11 &   7 &  17 & -11 &   6 &   221424.615 &  180.6  & 1.024E-04 & 30.0  &  37 & 16.19(4.35) & Blended with CH$_3$CN $\nu_t$=8  \\  
 9  &    18 &  11 &   8 &  17 &  11 &   7 &   221445.622 &  180.6  & 1.025E-04 & 30.0  &  37 & 7.23(3.90)  & \\  
 10 &    18&  -10 &   8 &  17 & -10 &   7 &   221649.374 &   166.7 & 1.134E-04 & 33.1  &  37 & 6.51(0.65)  & \\  
\multicolumn{14}{c}{$\nu_t$=1}\\
  1 &   18 &   4 &  15 &  17 &   4  & 14  &  219704.898  &   299.5 &  1.507E-04  & 45.2 &   37  & 4.69(3.67) & \\  
  2 &   18 &   9 &   9 &  17 &   9  &   8 &  219764.090  &   342.7 &  1.198E-04  & 35.9 &   37  & 6.16(4.18) & \\ 
  3 &   18 &  10 &   9 &  17 &  10  &  8  &  219822.160  &   355.1 &  1.105E-04  & 33.1 &   37  & 8.99(1.76) & \\ 
  3 &   18 &  10 &   8 &  17 &   10 &   7 &  219822.161  &   355.1 &  1.105E-04  & 33.1 &   37  &            & \\ 
  4 &   18 &   9 &  10 &  17 &   9 &   9  &  220030.289  &  342.4  &  1.201E-04  & 35.9 &   37  & 7.60(7.00) & \\ 
  4 &   18 &   9 &   9 &  17 &   9 &   8  &  220030.339  &  342.4  &  1.201E-04  & 35.9 &   37  &            & \\ 
  5 &   18 & -10 &   9 &  17 & -10 &   8  &  220307.810  &  354.8  &  1.115E-04  & 33.1  &  37  & 3.65(0.60) & \\  
  6 &   18 &   7 &  12 &  17 &   7 &  11 &   220913.784  &  321.3  &  1.375E-04  & 40.6  &  37  & 8.33(4.37) & \\ 
  7 &   18 &  -8 &  11 &  17 &  -8 &  10  &  221111.015 &   330.7  &  1.308E-04  & 38.5  &  37  & 3.29 (2.88) & \\ \hline
\end{tabular}

\noindent
$^1$Same notation as Table A.5.    

\end{table*}

\begin{table*}
\begin{tabular}{llllllllllllllll} 
\multicolumn{16}{c}{Table A.6. CH$_2$DOH line parameters$^1$} \\ \hline \hline
 N  & \multicolumn{8}{c}{Quantum numbers: up-low} &  $\nu$(MHz)  &  E$_u$(K) &   A$_{ul}$ (s$^{-1}$) &  Sij  &  gu &  W (K km s$^{-1}$)  &  \\ \hline 
  1 &   5 &  2 &  4 &  1  & 5   & 1  &  5 &  1 &  218316.390  & 58.7  & 1.818E-05 & 0.83  &  11 & 7.07(1.27)  & \\ 
  2 &  20 &  5 & 15 &  1  & 19  & 6  & 14 &  0 &  219206.135  & 557.6 & 2.977E-05 & 4.99  &  41 & 2.05(1.16)  & \\
  3 &   5 &  1 &  5 &  0  & 4   & 1  & 4  & 0  &  220071.805  & 35.8  & 2.614E-05 & 2.89  &  11 & 16.05(0.60) & Blended with CH$_3$OH\\ 
  4 &  17 &  1 & 16 &  0  & 17  & 0  & 17 &  0 &  220552.586  & 335.9 & 7.545E-05 & 10.6  &  35 & 11.74(5.35) & \\ 
  5 &   5 &  1 &  5 &  2  & 4   & 1  &  4 &  2 &  221273.004  &  54.7 & 3.371E-05 & 3.66  &  11 & 8.25(3.23)  &  \\ 
  6 &  10 &  1 & 10 &  0  &  9  & 0  &  9 &  1 &  221391.766  & 120.2 & 5.674E-05 & 4.73  &  21 & 21.17(1.58) &  Blended with CH$_3$CN $\nu_8$=1 \\ 
  7 &  20 &  2 & 18 &  2  & 20  & 1  & 19 & 2 &   221154.575  & 486.0 & 7.322E-05 & 12.0  &  41 & 13.20(2.10) & Blended with methyl formate \\ 
  8 &  6 &   2 &  5 &  1  & 6   & 1  & 6  &  1&   221178.153  &   71.6 & 2.067E-05 & 1.07 &  13 & 12.48(5.20) & Blended with HDCS\\ \hline
\end{tabular}
\noindent
$^1$Same notation as Table A.5.    

\end{table*}

\begin{table*}
\begin{tabular}{llllllllllllll} 
\multicolumn{14}{c}{Table A.7. HNCO line parameters$^1$} \\ \hline \hline
 N   & \multicolumn{6}{c}{Quantum numbers: up-low}  &  $\nu$(MHz) &  E$_u$(K)  &  A$_{ul}$ (s$^{-1}$) &  Sii  &  gu &  W (K km s$^{-1}$)  &  \\ \hline
  1  &   10  &  1  &  10  &  9  &   1   &  9   &  218981.009   &  101.1  & 1.462E-04  & 9.79  &  21  &  22.71(1.42) & \\
  2  &   10  &  3  &  8   &  9  &   3   &  7   &  219656.769   &  433.0  & 1.235E-04  & 8.19  &  21  &  10.68(1.48) & \\
  2  &   10  &  3  &  7   &  9  &   3   &  6   &  219656.771   &  433.0  & 1.235E-04  & 8.19  &  21  &              &\\
  3  &   10  &  2  &  9   &  9  &   2   &  8   &  219733.850   &  228.3  & 1.383E-04  & 9.17  &  21  &  27.10(1.89) & \\
  3  &   10  &  2  &  8   &  9  &   2   &  7   &  219737.193   &  228.3  & 1.383E-04  & 9.17  &  21  &               &  \\
  4  &   10  &  0  & 10   &  9  &   0   &  9   &  219798.274   &   58.0  & 1.510E-04  & 1.00E+01a  &  21  &  27.85(4.74) & \\
  5  &   10  &  1  &  9   &  9  &   1   &  8   &  220584.751   &  101.5  & 1.494E-04  & 9.79E+00a  &  21  &  22.70(4.17) & \\ \hline
\end{tabular}

\noindent
$^1$Same notation as Table A.5.    

\end{table*}

\begin{table*}
\begin{tabular}{llllllllllllllll} 
\multicolumn{16}{c}{Table A.8. aGg'-(CH$_2$OH)$_2$ line parameters$^1$} \\ \hline \hline
 N  & \multicolumn{8}{c}{Quantum numbers: up-low} &  $\nu$(MHz)   &  E$_u$(K)          &   A$_{ul}$ (s$^{-1}$)     &   Sij  &   gu &  W (K km s$^{-1}$)  &  \\ \hline 
 1  &   22  &   4   &  19  &  0  &  21  &  4   &   18  &  1  &   218371.495  &  132.7  &   1.881E-04  & 145   & 405  & 3.07(3.00) & \\
 2  &   22  &  15   &  7   &  0  &  21  &  15  &   6   &  1  &   218379.983  &  234.8  &   1.355E-04  & 81.4  & 315  & 2.21(0.17) & \\
 2  &   22  &  15   &  8   &  0  &  21  &  15  &   7   &  1  &   218379.983  &  234.8  &   1.355E-04  & 105   & 405  &            & \\
 3  &   22  &  13  &   9   &  0  &  21   &   13    & 8   &  1   &  218574.680   &  207.3  &  1.652E-04  & 98.9 &  315  &  8.35(2.65) &  Unidentified line? \\
 3  &   22  &  13  &  10   &  0  &  21   &  13   &   9   &  1   &  218574.680   &   207.3 &  1.651E-04  & 127  & 405   &  & \\ 
 4   &   22  &  11  & 12   &  0  &  21  &  11  &   11   &  1  &   218872.112  &   183.8 &   1.911E-04  & 147  & 405  & 3.90(0.15) &   \\
 4   &   22  &  11  & 11   &  0  &  21  &  11  &   10   &  1  &   218872.112  &   183.8 &   1.911E-04  & 114  & 315  &            &  \\
 5   &   22  &  10  &  13  &   0  &  21 &  10  &  12   &  1  &   219089.720   &  173.5  &  2.027E-04  & 155  & 405  & 5.06(0.49) & Blended with CH$_3$OCHO \\
 5   &   22  &  10  &  12  &   0  &  21 &  10  &  11   &  1  &   219089.727   &  173.5  &  2.027E-04  & 121  & 315  &            &  \\
 6  &    26  &  11  &  16  &   1  &  26  &  10  &  16   &  0  &   219384.910   &  232.5  & 5.221E-06  & 71.4  & 371  & 4.03 (0.05)&  Well below the fit!  \\
 6  &    22  &  9   &  14  &   0  &  21  &   9  &  13   &  1  &   219385.177   &  164.3  & 2.136E-04  & 163   & 405  &  &  \\
 6  &    26  &  11  &  15  &   1  &  26  &  10  &  17   &  0  &   219385.324   &  232.5  & 5.222E-06  & 91.7  & 477  &  &  \\
 6  &    22  &   9   & 13  &   0  &  21  &   9  &  12   &  1  &   219385.425   &  164.3  & 2.136E-04  & 127   & 315  &  & \\
 7 &    22   &  1  &  21  &   1  &  21  &   1  &  20  &   0   &  219580.671   &   122.2  &  2.568E-04  &  195  &  405  & 12.63(0.56) & Blended with CH$_3$OCHO\\
 9  &   20  &   4   & 16   &  1  &  19   &  4  &  15  &   0   &  219764.925    &  114.4  &  2.453E-04  &  169   & 369   & 9.50(0.60) &  Blended with CH$_3$OCHO\\
 9  &   21  &   4   & 18   &  1  &  20   &  4  &  17  &   0  &   221007.823    &  122.6  & 2.657E-04   &  189    & 387  & 5.67(0.28) &   \\
 10  &   22   &  6   & 17  &   0  &  21   &  6  &  16   &  1  &   221038.799    &  142.6  & 2.395E-04  &  178     & 405  & 3.77(0.42) &   \\
 11  &   22   &  5  &  18   &  0   & 21  &   5  &  17  &   1  &  221100.315    &  137.4  & 2.368E-04  &  176     &  405  & 3.39(0.26)  \\
 12  &   23   &  3  &  21 &    0  &  22  &   3 &   20   &  1  &   221338.974   &  138.4  & 2.717E-04  &  164     & 329  & 7.19(0.075)$^*$ & Blended with CH$_3$CN $\nu_8$=1 \\
\hline
\end{tabular}
\noindent
$^1$Same notation as Table A.5.

\noindent
$^*$In Fig. 2, we have subtracted the CH$_3$CN $\nu_8$=1 contribution that accounts for $\sim$50\% of the flux\\
\end{table*}

\end{appendix}


\begin{thebibliography}{}

\bibitem[Adande et al.(2013)]{2013AsBio..13..439A} Adande, G.~R., Woolf, 
N.~J., \& Ziurys, L.~M.\ 2013, Astrobiology, 13, 439 

\bibitem[Alonso-Albi et 
al.(2010)]{2010A&A...518A..52A} Alonso-Albi, T., Fuente, A., Crimier, N., et al.\ 2010, \aap, 518, A52 

\bibitem[Anders \& Grevesse(1989)]{1989GeCoA..53..197A} Anders, E., \& Grevesse, N.\ 1989, \gca, 53, 197 

\bibitem[Bell et 
al.(2014)]{bell} Bell, T.A., Cernicharo, J., Viti, S. et al, 2013, A\&A accepted

\bibitem[Belloche et 
al.(2013)]{2013A&A...559A..47B} Belloche, A., M{\"u}ller, H.~S.~P., Menten, K.~M., Schilke, P., \& Comito, C.\ 2013, \aap, 559, A47 

\bibitem[Beltr{\'a}n et al.(2009)]{2009ApJ...690L..93B} Beltr{\'a}n, M.~T., 
Codella, C., Viti, S., Neri, R., \& Cesaroni, R.\ 2009, \apjl, 690, L93 

\bibitem[Beuther et al.(2005)]{2005ApJ...632..355B} Beuther, H., Zhang, Q., 
Greenhill, L.~J., et al.\ 2005, \apj, 632, 355 

\bibitem[Beuther et al.(2006)]{2006ApJ...636..323B} Beuther, H., Zhang, Q., 
Reid, M.~J., et al.\ 2006, \apj, 636, 323 

\bibitem[Bisschop et 
al.(2008)]{2008A&A...488..959B} Bisschop, S.~E., J{\o}rgensen, J.~K., Bourke, T.~L., Bottinelli, S., \& van Dishoeck, E.~F.\ 2008, \aap, 488, 959 

\bibitem[Blake et al.(1987)]{1987ApJ...315..621B} Blake, G.~A., Sutton, 
E.~C., Masson, C.~R., \& Phillips, T.~G.\ 1987, \apj, 315, 621 

\bibitem[Bottinelli et al.(2004)]{2004ApJ...615..354B} Bottinelli, S., 
Ceccarelli, C., Lefloch, B., et al.\ 2004b, \apj, 615, 354 

\bibitem[Bottinelli et al.(2004)]{2004ApJ...617L..69B} Bottinelli, S., 
Ceccarelli, C., Neri, R., et al.\ 2004a, \apjl, 617, L69 

\bibitem[Bottinelli et 
al.(2007)]{2007A&A...463..601B} Bottinelli, S., Ceccarelli, C., Williams, J.~P., \& Lefloch, B.\ 2007, \aap, 463, 601 

\bibitem[Brouillet et 
al.(2013)]{2013A&A...550A..46B} Brouillet, N., Despois, D., Baudry, A., et al.\ 2013, \aap, 550, A46 

\bibitem[Carvajal et 
al.(2009)]{2009A&A...500.1109C} Carvajal, M., Margul{\`e}s, L., Tercero, B., et al.\ 2009, \aap, 500, 1109 

\bibitem[Cazaux et al.(2003)]{2003ApJ...593L..51C} Cazaux, S., Tielens, 
A.~G.~G.~M., Ceccarelli, C., et al.\ 2003, \apjl, 593, L51 

\bibitem[Cazaux et al.(2011)]{2011ApJ...741L..34C} Cazaux, S., Caselli, P., 
\& Spaans, M.\ 2011, \apjl, 741, L34 

\bibitem[Cernicharo et 
al.(1988)]{1988A&A...189L...1C} Cernicharo, J., Kahane, C., Guelin, M., \& G\'omez-Gonzalez, J.\ 1988, \aap, 189, L1 

\bibitem[Cernicharo (2012)]{cerni12}Cernicharo, J. 2012, in Proceedings of the European Conference on Laboratory
Astrophysics, Eur. Astron. Soc. Publ. Ser, eds. C. Stehl\'e, C. Joblin, \& L. d'Hendecourt

\bibitem[Cernicharo et al.(2013)]{2013ApJ...771L..10C} Cernicharo, J., 
Tercero, B., Fuente, A., et al.\ 2013, \apjl, 771, L10 

\bibitem[Chin et al.(1995)]{1995yCat..33050960C} Chin, Y.-N., Henkel, C., 
Whiteoak, J.~B., Langer, N., \& Churchwell, E.~B.\ 1995, VizieR Online Data Catalog, 330, 50960 

\bibitem[Comito et al.(2005)]{2005ApJS..156..127C} Comito, C., Schilke, P., 
Phillips, T.~G., et al.\ 2005, \apjs, 156, 127 


\bibitem[Daly et al.(2013)]{2013ApJ...768...81D} Daly, A.~M., Berm{\'u}dez, 
C., L{\'o}pez, A., et al.\ 2013, \apj, 768, 81 

\bibitem[Demyk et 
al.(2010)]{2010A&A...517A..17D} Demyk, K., Bottinelli, S., Caux, E., et al.\ 2010, \aap, 517, A17 

\bibitem[Esplugues et 
al.(2013)]{2013A&A...556A.143E} Esplugues, G.~B., Tercero, B., Cernicharo, J., et al.\ 2013a, \aap, 556, A143 

\bibitem[Esplugues et 
al.(2013)]{2010A&A...517A..17D} Esplugues, G.B., Cernicharo, J., Viti, S. et al, 2013b, A\&A 559, 51

\bibitem[Fich et 
al.(2010)]{2010A&A...518L..86F} Fich, M., Johnstone, D., van Kempen, T.~A., et al.\ 2010, \aap, 518, L86 

\bibitem[Fuente et 
al.(2001)]{2001A&A...366..873F} Fuente, A., Neri, R., Mart{\'{\i}}n-Pintado, J., et al.\ 2001, \aap, 366, 873 

\bibitem[Fuente et 
al.(2005)]{2005A&A...433..535F} Fuente, A., Rizzo, J.~R., Caselli, P., Bachiller, R., \& Henkel, C.\ 2005b, \aap, 433, 535 

\bibitem[Fuente et 
al.(2005)]{2005A&A...444..481F} Fuente, A., Neri, R., \& Caselli, P.\ 2005a, \aap, 444, 481 (FU05)

\bibitem[Fuente et 
al.(2009)]{2009A&A...507.1475F} Fuente, A., Castro-Carrizo, A., Alonso-Albi, T., et al.\ 2009, \aap, 507, 1475 

\bibitem[Fuente et 
al.(2012)]{2012A&A...540A..75F} Fuente, A., Caselli, P., McCoey, C., et al.\ 2012, \aap, 540, A75 

\bibitem[Halfen et al.(2006)]{2006ApJ...639..237H} Halfen, D.~T., Apponi, 
A.~J., Woolf, N., Polt, R., \& Ziurys, L.~M.\ 2006, \apj, 639, 237 

\bibitem[Haykal et al.(2014)]{2006ApJ} Haykal, I., Carvajal, M., Tercero, B. et al., 2013, \aap, accepted

\bibitem[Hollis et al.(2000)]{2000ApJ...540L.107H} Hollis, J.~M., Lovas, 
F.~J., \& Jewell, P.~R.\ 2000, \apjl, 540, L107 

\bibitem[Johnstone et 
al.(2010)]{2010A&A...521L..41J} Johnstone, D., Fich, M., McCoey, C., et al.\ 2010, \aap, 521, L41 

\bibitem[Kahane et al.(2013)]{2013ApJ...763L..38K} Kahane, C., Ceccarelli, 
C., Faure, A., \& Caux, E.\ 2013, \apjl, 763, L38 

\bibitem[Kolesnikov\'a et al.(2014)]{2013ApJ...763L..38K} Kolesnikov\'a et al., 2014, \apjl, in press
(arXiv:1401.7810)

\bibitem[J{\o}rgensen et al.(2005)]{2005ApJ...632..973J} J{\o}rgensen, 
J.~K., Bourke, T.~L., Myers, P.~C., et al.\ 2005, \apj, 632, 973 

\bibitem[J{\o}rgensen et al.(2007)]{2007ApJ...659..479J} J{\o}rgensen, 
J.~K., Bourke, T.~L., Myers, P.~C., et al.\ 2007, \apj, 659, 479 

\bibitem[Loinard et al.(2003)]{2003cdsf.conf..351L} Loinard, L., Castets, 
A., Ceccarelli, C., et al.\ 2003, SFChem 2002: Chemistry as a Diagnostic of 
Star Formation, 351 

\bibitem[L\'opez et al. (2014)]{2007ApJ...659..479J} L\'opez, A., Tercero, B., Kisiel, Z. et al., submmitted to A\&A


\bibitem[Marcelino et al.(2009)]{2009ApJ...690L..27M} Marcelino, N., 
Cernicharo, J., Tercero, B., \& Roueff, E.\ 2009, \apjl, 690, L27 

\bibitem[Margul{\`e}s et al.(2010)]{2010ApJ...714.1120M} Margul{\`e}s, L., 
Huet, T.~R., Demaison, J., et al.\ 2010, \apj, 714, 1120 

\bibitem[Milam et al.(2005)]{2005ApJ...634.1126M} Milam, S.~N., Savage, C., 
Brewster, M.~A., Ziurys, L.~M., \& Wyckoff, S.\ 2005, \apj, 634, 1126 

\bibitem[Motiyenko et 
al.(2012)]{2012A&A...548A..71M} Motiyenko, R.~A., Tercero, B., Cernicharo, J., \& Margul{\`e}s, L.\ 2012, \aap, 548, A71 

\bibitem[M{\"u}ller et 
al.(2001)]{2001A&A...370L..49M} M{\"u}ller, H.~S.~P., Thorwirth, S., Roth, D.~A., \& Winnewisser, G.\ 2001, \aap, 370, L49 

\bibitem[M{\"u}ller et al.(2005)]{2005JMoSt.742..215M} M{\"u}ller, 
H.~S.~P., Schl{\"o}der, F., Stutzki, J., 
\& Winnewisser, G.\ 2005, Journal of Molecular Structure, 742, 215 

\bibitem[Neill et al.(2013)]{2013ApJ...777...85N} Neill, J.~L., Crockett, 
N.~R., Bergin, E.~A., Pearson, J.~C., \& Xu, L.-H.\ 2013, \apj, 777, 85 

\bibitem[Neri et 
al.(2007)]{2007A&A...468L..33N} Neri, R., Fuente, A., Ceccarelli, C., et al.\ 2007, \aap, 468, L33 

\bibitem[Palau et al.(2011)]{2011ApJ...743L..32P} Palau, A., Fuente, A., 
Girart, J.~M., et al.\ 2011, \apjl, 743, L32 

\bibitem[Palau et al.(2013)]{2013ApJ...762..120P} Palau, A., Fuente, A., 
Girart, J.~M., et al.\ 2013, \apj, 762, 120 

\bibitem[Parise et 
al.(2004)]{2004A&A...416..159P} Parise, B., Castets, A., Herbst, E., et al.\ 2004, \aap, 416, 159 

\bibitem[Peng et 
al.(2013)]{2013A&A...554A..78P} Peng, T.-C., Despois, D., Brouillet, N., et al.\ 2013, \aap, 554, A78 

\bibitem[Pickett (1991)]{pick91} Pickett, H.M. 1991, J. Molec. Spectroscopy 148, 371

\bibitem[Pickett et al.(1998)]{1998JQSRT..60..883P} Pickett, H.~M., 
Poynter, R.~L., Cohen, E.~A., et al.\ 1998, Journal of Quantitative Spectroscopy and Radiative Transfer, 60, 883 

\bibitem[Pineda et 
al.(2012)]{2012A&A...544L...7P} Pineda, J.~E., Maury, A.~J., Fuller, G.~A., et al.\ 2012, \aap, 544, L7 

\bibitem[{\"O}berg et al.(2011)]{2011ApJ...740...14O} {\"O}berg, K.~I., van 
der Marel, N., Kristensen, L.~E., 
\& van Dishoeck, E.~F.\ 2011, \apj, 740, 14 


\bibitem[Requena-Torres et al.(2008)]{2008ApJ...672..352R} Requena-Torres, 
M.~A., Mart{\'{\i}}n-Pintado, J., Mart{\'{\i}}n, S., 
\& Morris, M.~R.\ 2008, \apj, 672, 352 

\bibitem[S{\'a}nchez-Monge et al.(2010)]{2010ApJ...721L.107S} 
S{\'a}nchez-Monge, {\'A}., Palau, A., Estalella, R., et al.\ 2010, \apjl, 
721, L107 

\bibitem[Shevchenko 
\& Yakubov(1989)]{1989SvA....33..370S} Shevchenko, V.~S., \& Yakubov, S.~D.\ 1989, \sovast, 33, 370 

\bibitem[Taquet et al.(2012)]{2012A&A...538A..42T}
Taquet, V., Ceccarelli, C., \& Kahane, C.\ 2012, \aap, 538, A42

\bibitem[Tercero et 
al.(2010)]{2010A&A...517A..96T} Tercero, B., Cernicharo, J., Pardo, J.~R., \& Goicoechea, J.~R.\ 2010, \aap, 517, A96 

\bibitem[Tercero et 
al.(2011)]{2011A&A...528A..26T} Tercero, B., Vincent, L., Cernicharo, J., Viti, S., \& Marcelino, N.\ 2011, \aap, 528, A26 

\bibitem[Tercero et al.(2013)]{2013ApJ...770L..13T} Tercero, B., Kleiner, 
I., Cernicharo, J., et al.\ 2013, \apjl, 770, L13 

\bibitem[Turner(1990)]{1990ApJ...362L..29T} Turner, B.~E.\ 1990, \apjl, 
362, L29 

\bibitem[Turner(1991)]{1991ApJS...76..617T} Turner, B.~E.\ 1991, \apjs, 76, 
617 


\bibitem[Vastel et al.(2003)]{2003ApJ...593L..97V} Vastel, C., Phillips, 
T.~G., Ceccarelli, C., \& Pearson, J.\ 2003, \apjl, 593, L97 


\bibitem[Zapata et al.(2013)]{2013ApJ...764L..14Z} Zapata, L.~A., Loinard, 
L., Rodr{\'{\i}}guez, L.~F., et al.\ 2013, \apjl, 764, L14 

%
%


\end{thebibliography}
\end{document}